\pdfoutput=1
\newcommand*{\ATLASLATEXPATH}{}
\documentclass[cernpreprint,USenglish,texlive=2011,txfonts,texmf]{\ATLASLATEXPATH atlasdoc}
\usepackage{\ATLASLATEXPATH atlaspackage}
\usepackage[articletitle=false]{\ATLASLATEXPATH atlasbiblatex}
\usepackage{\ATLASLATEXPATH atlasphysics}
\graphicspath{{./}}

\usepackage{mydocument-defs}
\usepackage{multirow}

\usepackage[utf8]{inputenc}

\AtlasTitle{Measurements of \WZ production cross sections in $pp$ collisions at $\rts = 8\tev$ with the ATLAS detector and limits on anomalous
gauge boson self-couplings}

\author{The ATLAS Collaboration}

\AtlasRefCode{STDM-2014-02}
 \PreprintIdNumber{CERN-EP-2016-017}

\AtlasJournal{Phys.\ Rev.\ D.}
\AtlasAbstract{%
  This paper presents measurements of \wz production in $pp$ collisions
  at a center-of-mass energy of $8$~\TeV. The gauge
  bosons are reconstructed using their leptonic decay modes into  electrons and muons.
  The data were  collected in $2012$ by the ATLAS experiment  at the Large Hadron Collider,
  and correspond  to an integrated luminosity of {\mbox {$20.3$~fb$^{-1}$}}. The measured
  inclusive cross section in the detector fiducial region is
   $\sigma_{W^\pm Z \rightarrow \ell^{'} \nu\ \ell \ell} =  35.1~\pm$~0.9 (stat.)~$\pm~0.8$ (sys.)~$\pm~0.8$ (lumi.) fb,
   for one leptonic decay channel. In comparison, the next-to-leading-order Standard Model expectation is  $30.0  \pm 2.1$~fb.
  Cross sections for $W^+Z$ and $W^-Z$ production and their ratio are presented as well
  as differential cross sections for several kinematic observables.
  Limits on anomalous triple gauge boson couplings are derived
  from the transverse mass spectrum of the \wz  system.
  From the analysis of events with a $W$ and a $Z$ boson associated with two
  or more forward jets an upper limit at $95\%$ confidence level on  the \wz scattering cross section of $0.63$~fb, for each leptonic decay channel, is established, while the Standard Model prediction at next-to-leading order is $0.13$~fb.
Limits on anomalous  quartic gauge boson couplings are also extracted.
}

\hypersetup{pdftitle={ATLAS document},pdfauthor={The ATLAS Collaboration}}

\makeatletter
\DeclareFieldFormat{eprint:arxiv}{%
   arXiv\addcolon
  \ifhyperref
    {\href{http://arxiv.org/\abx@arxivpath/#1}{%
       \nolinkurl{#1}%
       \iffieldundef{eprintclass}
         {}
         {\addspace\texttt{\mkbibbrackets{\thefield{eprintclass}}}}}}
    {\nolinkurl{#1}
     \iffieldundef{eprintclass}
       {}
       {\addspace\texttt{\mkbibbrackets{\thefield{eprintclass}}}}}}
\makeatother

\begin{document}

\maketitle

\tableofcontents

% List of contributors - print here or after the Bibliography
%\AtlasPrintContribute{0.3}

%\clearpage
%-------------------------------------------------------------------------------
\section{Introduction}
\label{sec:intro}

The study of  \wz diboson production is an important test of the Standard Model (SM) for its sensitivity to the gauge
 boson self-interactions, related to the non-Abelian structure of the electroweak interaction.
 It provides the means to investigate vector boson scattering (VBS) processes, which directly probe
 the electroweak symmetry breaking sector of the SM, and to extract constraints on anomalous
 triple and quartic gauge boson couplings (aTGC and aQGC).  
Improved constraints can probe scales of
  new physics in the multi-\TeV\ range and provide a way to look for signals of new physics in a
  model-independent way.
Precise measurements of \wz\ production will also help to improve the existing QCD calculations of this process.

This paper presents measurements of  the \wz production cross section and limits
on the aTGC and aQGC obtained by analyzing proton--proton ($pp$) collisions at a center-of-mass energy
of $\sqrt{s}$ = $8$~\TeV. The leptonic decay modes of the $W$ and $Z$ bosons
are used and all quoted fiducial production cross sections include the branching ratio of the gauge bosons into
channels with electrons or muons.
The analyzed data sample was  collected in 2012 by the ATLAS experiment at the Large Hadron Collider (LHC),
  and corresponds  to an integrated luminosity of {\mbox {$20.3$~fb$^{-1}$}}.
Experimentally, \wz production has the advantage of a higher cross section than $ZZ$ production.
At the same time, with three charged leptons and the requirement that two of them originate from a $Z$ boson, the leptonic \wz final states
are easier to discriminate from the background than the leptonic $WW$ final states.

Measurements of the \wz production cross section have been reported in proton--antiproton collisions
at a center-of-mass energy of $\sqrt{s} = 1.96$~\TeV\  by
the CDF and D0 collaborations~\cite{CDF:2012WZ,D0:2012WZ} using integrated luminosities of $7.1$~fb$^{-1}$ and $8.6$~fb$^{-1}$, respectively, and for $\sqrt{s}= 7$~\TeV\  proton--proton collisions,
using an integrated luminosity of $4.6$~fb$^{-1}$, by the ATLAS
Collaboration~\cite{Aad:2012twa}.
Limits on anomalous charged-current gauge couplings were also reported previously by the LEP, Tevatron, and LHC
experiments~\cite{LEP_aTGC,Tevatron_aTGC,CMS_aTGC_v1}. In hadron collisions, the selection of \wz final states
allows direct access to the $WWZ$ gauge coupling without the need of disentangling it
from the $WW\gamma$ gauge coupling as in  $W^\pm W^\mp$ events from hadronic or $e^+e^-$ collisions.

Compared to the previously  published measurements, this  paper uses data collected at a higher center-of-mass energy with a four-fold increase in integrated luminosity and presents additional measurements.
  The production cross section is measured
  in a fiducial phase space inclusively and as single differential cross sections as a function of each of several kinematic variables:
  the transverse momentum \pt\ of the $W$ and $Z$ bosons, the jet multiplicity, the transverse mass of the $WZ$ system, $m_{\textrm{T}}^{WZ}$, and
  the \pt of the neutrino associated with the $W$ boson decay.
An interesting feature of this last distribution is its sensitivity to the polarization of the $W$ boson, similarly to the \pt of the lepton of the $W$ boson decay.
Finally, the distribution of the absolute difference between the rapidities of the $Z$ boson and
  the lepton from the $W$ boson decay is measured, which was proposed as an alternative variable to look for aTGC and was
also found to be sensitive to the approximately zero helicity amplitude that is
  predicted at leading order (LO) in the SM~\cite{Baur:1994aj,Accomando:2005xp}.
  Limits on aTGC are extracted from the $m_{\textrm{T}}^{WZ}$ distribution, which is found to be less sensitive to
  higher-order perturbative effects in QCD~\cite{Sapeta:2012} and electroweak
  (EW) theory~\cite{Bierweiler:2013dja, Baglio:2013toa} than other observables, e.g., $p_{\textrm{T}}^Z$.

 The ratio of $W^+Z$/$W^-Z$ integrated production cross sections, sensitive to the choice of
 parton distribution functions (PDF), is measured along with the evolution of this
 ratio as a function of the kinematic variables introduced above.
 Charge-dependent distributions may be helpful in investigating CP violation effects in the interaction between
 gauge bosons.
 This paper  also  includes a study of \wz vector boson scattering, characterized
 by the presence of at least two forward jets, which is sensitive to quartic gauge  couplings.
Events with a $W^\pm Z jj$ final state are used to set limits on the VBS cross section and on
aQGC in the $WZWZ$ vertex.

 The results  are compared with the SM cross-section predictions, which at present are
 fully calculated only up to the next-to-leading order (NLO) in QCD~\cite{Ohnemus:1991,Frixione:1992}.

The paper is organized as follows. The ATLAS detector is described in Section~\ref{sec:Detector}.
The definition of the fiducial phase space used in this paper  is presented
in Section~\ref{sec:FiducialPS}.
Section~\ref{sec:Theorypredictions} discusses the available theoretical predictions.
Section~\ref{sec:SimulatedSamples} provides details of the simulated samples used for the measurements.
A description of the data set and the selection criteria are given in Section~\ref{sec:EventSelection}.
Section~\ref{sec:BackgroundEstimation} presents the background estimation and Section~\ref{sec:RecoResults}
provides comparisons of observed and expected events and of kinematic distributions at reconstructed level.
The procedure used to correct for detector effects and for acceptance
is described in Section~\ref{sec:CrossSectionDefinition}.
The treatment of the systematic uncertainties is detailed in Section~\ref{sec:Systematics}.
Sections~\ref{sec:CrossSections},~\ref{sec:aTGC}, and~\ref{sec:aQGC} describe the
 combination procedure of the four leptonic \wz decay channels and discuss
the results. Finally, concluding remarks are presented in  Section~\ref{sec:Conclusion}.

%-------------------------------------------------------------------------------
\section{The ATLAS detector}
\label{sec:Detector}

The ATLAS detector~\cite{ATLAS_Detector} is a multipurpose detector with a cylindrical geometry and nearly $4\pi$ coverage in solid
angle.
The collision point is surrounded by inner tracking devices, which are followed in increasing distance from the center by a superconducting solenoid providing a $2$~T axial magnetic field, a calorimeter system, and a muon spectrometer.

 The inner tracker provides precise position and momentum measurements of charged particles in the pseudorapidity\footnote{ATLAS uses a right-handed coordinate system with its origin at the nominal
interaction point (IP) in the center of the detector and the $z$-axis along the beam direction.
The $x$-axis points from the IP to the center of the LHC ring, and the $y$-axis points upward.
Cylindrical coordinates $(r,\phi)$ are used in the transverse $(x,y)$ plane, $\phi$ being the
azimuthal angle around the beam direction. The pseudorapidity is defined in terms of the
 polar angle $\theta$ as $\eta = -\textrm{ln}\left[ \textrm{tan}(\theta/2) \right]$.
} range $|\eta| < 2.5$.
 It consists of three subdetectors arranged in a coaxial geometry around the beam axis: the silicon pixel detector, the silicon microstrip detector, and the transition radiation tracker.

Electromagnetic calorimetry in the region $|\eta|<3.2$ is based on a high-granularity, lead/liquid-argon (LAr) sampling technology. Hadronic calorimetry uses a steel/scintillating-tile detector covering the region $|\eta|<1.7$ and a copper/LAr detector in the region $1.5 < |\eta| < 3.2$. The most forward region of the detector $3.1 <|\eta| < 4.9$ is equipped with a dedicated forward calorimeter, measuring electromagnetic and hadronic energies using copper/LAr and tungsten/LAr modules.

The muon spectrometer comprises separate trigger
and high-precision tracking chambers to measure the deflection of muons in a magnetic field generated by three
large superconducting toroids arranged with an eightfold azimuthal coil symmetry around the calorimeters.
The high-precision chambers cover a range of $|\eta|< 2.7$.
The muon trigger system covers the range $|\eta|< 2.4$ with
resistive plate chambers in the barrel, and thin gap chambers in the endcap regions.

A three-level trigger system is used to select events in real time. A hardware-based Level-1 trigger uses a subset of detector information to reduce the event rate to a value of at most $75$~kHz. The rate of accepted events is then reduced to about $400$~Hz by two software-based trigger levels, Level-2 and the Event Filter.

%-------------------------------------------------------------------------------
\section{Phase-space definition}
\label{sec:FiducialPS}

The phase-space definition used in this paper relies on  final-state prompt leptons\footnote{A prompt lepton
 is a lepton that is not produced in the decay of a hadron or a $\tau$ or their descendants.}~\cite{TruthWS}
 associated with the $W$ and $Z$ boson decay, as explained in detail below.

  At particle level, the kinematics of final-state prompt electrons and muons is computed including the contributions from
final-state radiated photons
within a distance in the ($\eta,\phi$) plane of $\Delta R = \sqrt{(\Delta\eta)^2 + (\Delta\phi)^2} = 0.1$ around the direction of the charged lepton.

These dressed leptons and the final-state neutrinos that do not originate from
  hadron or $\tau$ decays  are associated with the $W$ and $Z$ boson decay products with an
  algorithmic approach, called ``resonant shape''.
This algorithm is  based on the value of an estimator expressing
   the product of the nominal line shapes of the $W$ and $Z$ resonances
\begin{equation}\label{eq:Pk}
   P = \left| \frac{1}{ m^2_{(\ell^+,\ell^-)} - \left(m_Z^{\textrm{PDG}}\right)^2 + i \; \Gamma_Z^{\textrm{PDG}} \; m_Z^{\textrm{PDG}} } \right|^2  \times \; \left| \frac {1} { m^2_{(\ell',\nu_{\ell'})} - \left(m_W^{\textrm{PDG}}\right)^2 + i \; \Gamma_W^{\textrm{PDG}} \; m_W^{\textrm{PDG}} } \right|^2 \, ,
 \end{equation}
where $m_Z^{\textrm{PDG}}$ ($m_W^{\textrm{PDG}}$) and $\Gamma_Z^{\textrm{PDG}}$ ($\Gamma_W^{\textrm{PDG}}$) are the world average mass and total width of the $Z$ ($W$) boson, respectively, as reported by the Particle Data Group~\cite{Agashe:2014kda}.
   The input to the estimator is the invariant mass $m$ of all possible pairs ($\ell^+,\ell^-$) and ($\ell',\nu_{\ell'}$)
   satisfying the fiducial selection requirements defined in the next paragraph.
The final choice of which leptons are assigned to the $W$ or $Z$ bosons corresponds to the configuration  exhibiting the highest  value of the estimator.
Using this specific association algorithm, the gauge boson kinematics can be computed using the kinematics of the associated leptons independently of any internal Monte Carlo generator details.

The integrated and differential cross-section measurements are performed in a fiducial phase space
defined at particle level by the following requirements:
the \pt\ of the leptons from the $Z$ boson decay is
greater than $15$~\GeV, the \pt  of the charged lepton from the $W$ decay is greater than $20$~\GeV,
the absolute value of the  pseudorapidity of the charged leptons from the $W$ and $Z$ bosons are less than $2.5$,
the invariant mass of the two leptons from the $Z$ boson decay differs at most by $10$~\GeV\ from the world average value of the $Z$ boson mass $m_Z^{\textrm{PDG}}$.
The $W$ transverse mass, defined as $m_{\textrm{T}}^{W} = \sqrt{2 \cdot p_{\rm{T}}^\nu \cdot p_{\rm{T}}^\ell \cdot [1 -\cos{\Delta \phi(\ell, \nu)}]}$, where $\Delta \phi(\ell, \nu)$
is the angle between the lepton and the neutrino in the transverse plane, is required to be greater than $30$~\GeV.
In addition, it is required that the angular distance $\Delta R$
between the charged leptons from $W$ and $Z$ decay is larger than $0.3$, and
that $\Delta R$ between the two leptons from the $Z$ decay is larger than $0.2$.

The  integrated cross section, measured in the fiducial region of the detector,
is extrapolated to a total phase space, defined by requiring that the
 invariant mass of the lepton pair associated with the $Z$ boson decay is in the range $66 < m_Z < 116$ \GeV, and extrapolating to all decay channels of the $W$ and $Z$ bosons.

In order to define the VBS fiducial region for the cross-section measurement, in addition to the inclusive
fiducial criteria, at least two jets with a \pt greater than $30$~\GeV\ and an absolute value of the pseudorapidity $\eta_j$ below $4.5$ are required.
These particle level jets are defined using the anti-$k_{t}$ algorithm with a radius parameter $R$ = $0.4$.
The angular distance between all selected leptons and jets, $\Delta R (j,\ell)$, is required to be greater than $0.3$.
If the $\Delta R (j,\ell)$ requirement is not satisfied, the jet is discarded.
The invariant mass of the two leading jets, $m_{jj}$, must be above  $500$~\GeV\ to enhance the sensitivity to VBS processes.

For setting limits on aQGC, the fiducial region definition was optimized to give the best expected limits.
Therefore, in addition to the criteria used for the VBS fiducial cross-section measurement, it is required
that the difference in azimuthal angle $| \Delta \phi(W,Z) |$ between the $W$ and $Z$ directions is greater than $2$ rad.
In addition, in order to increase the sensitivity to aQGC signals, the scalar sum of the transverse momenta of the three charged leptons associated with the $W$ and $Z$ bosons, $\sum |p_{\textrm T}^\ell |$,  is greater than $250$~\GeV.

 A summary of the phase-space definition used in this paper is given in Table~\ref{tab:PhasePS}.

\begin{table}[!htbp]
\begin{center}
\begin{tabular}{lllll}
\hline
Variable                &   Total  &  Fiducial and aTGC         &  VBS             & aQGC \\
\hline

Lepton $|\eta|$                                                & ---               & $< 2.5$  & $< 2.5$  & $< 2.5$  \\ 
\pt of $\ell_Z$,  \pt of  $\ell_W$ [GeV]             & ---             & $>  15$,~$>  20$   & $>  15$,~$> 20$ & $>  15$,~$>  20$ \\ 
$m_Z$ range [GeV]                                        &  $66-116$   & $|m_Z - m_Z^{\textrm{PDG}}| < 10$                        &  $|m_Z - m_Z^{\textrm{PDG}}| < 10$                     & $|m_Z - m_Z^{\textrm{PDG}}| < 10$                       \\ 
$m^{W}_{\textrm T}$ [GeV]                            &  ---		 & $> 30$    &  $> 30$   & $> 30$  \\
 $\Delta R (\ell^-_Z, \ell^+_Z)$,  $\Delta R (\ell_Z, \ell_W)$    &  ---	     & $> 0.2$,~$> 0.3$    &  $> 0.2$,~$> 0.3$   & $> 0.2$,~$> 0.3$  \\
   \pt  two   leading jets [GeV]                                                  & --- 	    & ---				     &  $> 30$		      & $> 30$  \\ 
  $|\eta_j|$  two  leading jets                                             &  ---	     & ---    &  $< 4.5$  & $< 4.5$  \\ 
  Jet multiplicity                                                                       &  --- 	    & ---    &  $\ge 2$   & $\ge 2$  \\
  $m_{jj}$ [GeV]                                                                       &  ---		  & ---    &  $> 500$   & $> 500$  \\ 
 $\Delta R (j,\ell)$                                                                   &  ---		 & ---    &  $> 0.3$   & $> 0.3$ \\
 $| \Delta \phi(W,Z) |$                                                             &  ---		 & ---    &  ---		& $> 2$ \\
 $\sum |p_{\textrm T}^\ell |$ [GeV]                                                                &  ---	     & ---    &  ---		    & $> 250$ \\
\hline
\end{tabular}
\end{center}
\caption{Phase-space definitions used for the total, fiducial, VBS cross-section measurements and for the extraction of
 limits on the aTGC and aQGC. The symbols $\ell_Z$ and $\ell_W$ refer to the
leptons associated with the $Z$ and $W$ boson, respectively. The symbol $m_Z^{\textrm{PDG}}$ refers to the mean experimental mass of the $Z$ boson from the Particle Data Group~\protect\cite{Agashe:2014kda}. The other symbols are defined in the text. }
\label{tab:PhasePS}
\end{table}

%-------------------------------------------------------------------------------
\section{Standard Model predictions for \wz production}
\label{sec:Theorypredictions}

The measured integrated  cross section is compared with the SM NLO prediction
from the \POWHEG event generator~\cite{powheg1:2004, powheg, Alioli:2010xd, Melia:2011tj}, interfaced with \PYTHIA~8.175~\cite{Sjostrand:2007gs} for parton
showering (PS) and hadronization.
The \powheg MC event generator implements the next-to-leading order QCD corrections to the production of 
electroweak vector boson pairs at hadron colliders, including the full spin and decay angle correlations~\cite{Dixon:1998py}.
This calculation is referred to as \powhegpythia later on.
At a center-of-mass energy of $\sqrt{s} = 8$~\TeV, in proton--proton collisions, the SM NLO cross section for \wz
production
in the fiducial phase space defined in Section~\ref{sec:FiducialPS}, estimated with \powhegpythia using factorization and dynamic renormalization scales $\mu_{\mathrm{R}}$ and $\mu_{\mathrm{F}}$ equal to $m_{WZ}/2$, where $m_{WZ}$ is the invariant mass of the $WZ$ system, and the CT10~\cite{CT10} PDF set, is
\begin{eqnarray}
\sigma_{W^{\pm}Z \rightarrow \ell^{'} \nu \ell \ell}^{\textrm{fid., th.}} & = & 30.0  \pm 0.8 \,\textrm{(PDF)} \pm 1.3 \,\textrm{(scale)} \; \textrm{fb} \nonumber.
\end{eqnarray}
The predicted cross sections for $W^+Z$ and $W^-Z$ inclusive production are
\begin{eqnarray}
\sigma_{W^{+}Z \rightarrow \ell^{'} \nu \ell \ell}^{\textrm{fid., th.}} & = & 18.8 \pm 0.5 \,\textrm{(PDF)} \pm 0.8 \,\textrm{(scale)} \; \textrm{fb} \nonumber,
\end{eqnarray}
\begin{eqnarray}
\sigma_{W^{-}Z \rightarrow \ell^{'} \nu \ell \ell}^{\textrm{fid., th.}} & = & 11.1  \pm 0.5 \,\textrm{(PDF)} \pm 0.5 \,\textrm{(scale)} \; \textrm{fb} \nonumber.
\end{eqnarray}
In these estimates, the $W$ and $Z$ decays in a single leptonic channel  with muons or electrons are considered.
 The uncertainty due to the PDF is computed using the eigenvectors of the CT10 PDF set scaled to $68$\% confidence level (CL) and  the envelope of the
 differences between the results obtained with CT10, MSTW 2008~\cite{Martin:2009iq}, NNPDF 3.0~\cite{Ball:2014uwa}, and ATLAS-epWZ12 NLO~\cite{Aad:2012sb} PDF sets.
The QCD scale uncertainty is estimated conventionally by varying $\mu_{\mathrm{R}}$ and $\mu_{\mathrm{F}}$ by factors of two around the nominal scale $m_{WZ}/2$ with the constraint $0.5 \leq \mu_{\mathrm{R}} /\mu_{\mathrm{F}} \leq 2$. 
A maximum variation of the cross section of $4\%$ is found.
However, the SM prediction, which is at NLO accuracy in perturbative QCD, is highly sensitive to the choice of renormalization scale $\mu_{\mathrm{R}}$.
For example, choosing a fixed renormalization scale of $\mu_{\mathrm{R}} = (m_W + m_Z)/2$ instead of a dynamic scale $\mu_{\mathrm{R}} = m_{WZ}$ increases the SM predicted cross section by $7\%$.
 The total uncertainty on the theoretical prediction  is estimated  as the linear sum of the PDF and QCD scale  uncertainties,
 following the recommendations in Ref.~\cite{Dittmaier:2011ti}.

  The differential distributions are compared to the predictions of the \powhegpythia, \mcatnlo 4.0~\cite{MCNLO},
interfaced with \HERWIG~\cite{Herwig} for PS and hadronization, and  \SHERPA 1.4.5~\cite{Sherpa, Sherpa1} event generators.
  The \SHERPA  predictions used in this paper are computed at LO and take into account the real emission  of up to three partons
  in the matrix element calculations. They are therefore expected to describe the jet multiplicity distribution and the event kinematics
  at higher jet multiplicity better than \powhegpythia where only the real emission of at most one parton is directly calculated at NLO.

The uncertainties on predicted differential cross sections arising from the PDF and the QCD scale uncertainties are estimated as described above.
 Recently, approximate next-to-next-to-leading-order ($\bar{n}$NLO) corrections
 have been computed and presented as $K$-factors for differential cross-section distributions~\cite{Sapeta:2012}.
 For a number of commonly used observables these corrections are sizable, of the order of  $30\%$ to $100\%$.
 The $\bar{n}$NLO correction to the $m_\textrm{T}^{WZ}$ distribution is smaller ($ < 10\%$) indicating that  this observable
 is less sensitive to higher order  perturbative contributions to the transition amplitude that appear at next-to-next-to-leading order (NNLO).
The approximate $\bar{n}$NLO calculation can only account for the dominant part of the NNLO QCD corrections and in certain regions of the phase space.

Electroweak quantum corrections at NLO to the \wz cross sections, including photon-quark-induced processes,
 have been computed~\cite{Bierweiler:2013dja, Baglio:2013toa}. These corrections have an impact mainly on differential
 cross sections.
 The  complete calculation is done in the zero-width approximation and the decays of vector bosons are not included.
It is therefore not included in the available Monte Carlo (MC) generators.
 For this reason,  the uncertainty on the differential distributions arising from  missing higher  orders in  the EW theory is
 included by taking the existing EW corrections
  at NLO as an additional theory uncertainty on the predictions from  \powhegpythia.
  The effect increases with increasing $p_\textrm{T}^Z$ and $m_\textrm{T}^{WZ}$.
At a center-of-mass energy of $8$~\TeV, they range from $-0.3\%$ 
  to a value of $3.2\%$ in the highest  $p_\textrm{T}^Z$  bin  considered  in this analysis and from $0.12\%$ to a value of $1.1\%$
   in the highest   $m_\textrm{T}^{WZ}$ bin  considered in this analysis~\cite{Baglio:2013toa}.
 The total uncertainty on the differential theoretical predictions  is estimated  as the linear sum of the PDF, QCD scale, and EW correction uncertainties~\cite{Dittmaier:2011ti}.

  The SM cross section of the VBS process is calculated at NLO in QCD with the Monte Carlo generator VBFNLO~\cite{Arnold:2008rz,Arnold:2011wj,Baglio:2014uba,Campanario:2010hp,Campanario:2010xn}.
 In proton--proton collisions with a center-of-mass energy of $8$~\TeV\ this cross section in the VBS fiducial phase space defined
 in Section~\ref{sec:FiducialPS} is $0.13 \pm 0.01$~fb.
 %0.52 fb.
This calculated cross section is for one single leptonic decay channel of the $W$ and $Z$ in  muons or electrons.
 The total uncertainty on the VBS theoretical prediction  is estimated  as the linear sum of the PDF and QCD scale uncertainties, each determined
 as described above.

%-------------------------------------------------------------------------------
\section{Simulated event samples}
\label{sec:SimulatedSamples}

Simulated event samples are used for estimates of the  irreducible background, for the correction of the signal yield for detector effects,
for the extrapolation from the fiducial to the total phase space, for the extraction of the gauge couplings, 
and for comparisons of the results to the theoretical expectations. 

The simulated samples are overlaid with additional proton--proton interactions (pileup) 
generated with \PYTHIA~8.1 using the MSTW2008 LO PDF set and the A2~\cite{ATL-PHYS-PUB-2012-003} set of tuned parameters.
The MC events are also reweighted to better reproduce the distribution of the mean number of interactions 
per bunch crossing and of the longitudinal position of the primary $pp$ collision vertex observed in the data.
All  generated events are passed through the ATLAS detector simulation~\cite{Aad:2010ah} based on GEANT4~\cite{Geant4} and processed using the same reconstruction software as for data.

Scale factors are applied to the simulated events to correct for the small differences from data in the trigger, reconstruction, and identification efficiencies
 for electrons and muons~\cite{ElectronPerf:2014,ATLAS-CONF-2014-032,MuonPerf:2014}. 
Furthermore, in simulated events the electron energy and the muon momentum  are smeared to account for the small differences in 
 resolution between data and simulation~\cite{Aad:2014nim,MuonPerf:2014}.
%

%-------------------------------------------------------------------------------
\subsection{Signal samples}
\label{sec:Signal}

The \wz\  SM production processes and subsequent  leptonic decays are generated at NLO in QCD
using the \POWHEG  MC event generator~\cite{powheg1:2004, powheg, Alioli:2010xd, Melia:2011tj}, interfaced with \PYTHIA~8.175~\cite{Sjostrand:2007gs} for PS, hadronization and the underlying event (UE) simulation.
This sample is used to correct for acceptance and detector effects.

Signal events with aTGC are generated at NLO with the \mcatnlo 4.0~\cite{MCNLO} Monte Carlo generator
interfaced with \HERWIG~\cite{Herwig} and \jimmy~\cite{Jimmy:1996} for the simulation of the PS, hadronization, and UE.

In all above-mentioned signal samples, the final-state radiation (FSR) resulting from the quantum electrodynamics (QED)
interaction is modeled with \PHOTOS~\cite{photos}.

For the VBS analysis, the \wz\ production associated with  at least two jets is generated at LO with
 \SHERPA 1.4.5~\cite{Sherpa, Sherpa1}, which uses the CKKW~\cite{CKKW} matching scheme and an internal
 model for QED radiation based on the YFS method~\cite{YFS}.  Signal events in the VBS analysis arise from
the processes that occur at zero-order in the strong coupling constant $\alpha_{\mathrm{s}}$ and are labeled $WZjj$-EW. The remaining
processes leading to \wz final states, associated with at least two jets, are called $WZjj$-QCD processes.

Events with aQGC are generated at LO using the \WHIZARD~\cite{Kilian:2007gr} MC generator.
A $K$-matrix unitarization method~\cite{Alboteanu:2008my,Chung:1995dx} is employed in order to ensure the unitarity of the scattering
amplitude, which would be violated for values of quartic gauge couplings different from the SM value.

The CT10~\cite{CT10} PDF set is used for all signal samples.

% 

%

%-------------------------------------------------------------------------------
\subsection{Background samples}
\label{sec:BackgroundSimu}

Backgrounds to the \wz\ signal come from events with two or more electroweak gauge bosons, top quarks
 and gauge bosons associated with jets ($V+j$, $V = W,Z$),
 and events from double parton scattering (DPS) processes where the \wz signature results  from collisions between two pairs of
 partons producing a single $W$ and a single $Z$ boson. In the VBS analysis the $WZjj$-QCD process is a background to the $WZjj$-EW
 production.
 Interference effects between $WZjj$-QCD and $WZjj$-EW processes are expected to be negligible and are therefore not considered.

 Monte Carlo simulation is used to compute the contribution from processes with at least three prompt leptons and for comparison with
 the data-driven estimation of the contribution from background processes with at least one misidentified  lepton.

  The $q\bar{q} \rightarrow ZZ^{(*)}$ processes are generated at NLO with \powheg interfaced with \PYTHIA~8.175 or at leading order (LO)
  with \SHERPA~1.4.5, which includes up to three partons in the matrix element calculation.
The first sample is used in the inclusive analysis, the second in the VBS analysis.
  The $gg \rightarrow ZZ^{(*)}$ process is simulated with \ggZZ at LO~\cite{Binoth:2008pr} interfaced
  with \HERWIG~\cite{Herwig} for the simulation of the PS and of the hadronization and \jimmy~\cite{Jimmy:1996} for the UE.
  Processes with three gauge bosons are simulated with   \MADGRAPH~\cite{madgraph} interfaced with \PYTHIA.
  The associated production of top pairs with  a weak gauge boson is simulated with \MADGRAPH interfaced with \PYTHIA and
  the associated production of a single top and a $Z$ boson is simulated with \SHERPA.
The total predictions of these MC samples are rescaled to match NLO predictions from Refs.~\cite{Campbell:2012dh,Garzelli:2012bn} and Ref.~\cite{tZj}, respectively.
  The contribution from DPS processes is estimated using \PYTHIA MC samples generated with two hard scatterings with single-boson production processes ($W$, $Z/\gamma^*$).
The cross section of the DPS
 samples is estimated using its factorization into the product of two single scattering cross sections~\cite{Gaunt:2010pi} and the effective area parameter for hard double-parton interactions recently measured by ATLAS~\cite{Aad:2013bjm}.

 \ALPGEN~\cite{AlpgenJimmy} interfaced with \jimmy~\cite{Jimmy:1996} and \SHERPA~\cite{Sherpa} samples are used to model the $W+j$ and $Z+j$ backgrounds, respectively.
  Top pair production  is simulated with \powhegpythia.
 The $WW$ diboson production  is modeled with \powhegpythia and \textsc{GG2WW+Herwig}. \ALPGEN and \SHERPA are used to model $W\gamma$ and ~$Z\gamma$ diboson production, respectively.

  The set of PDF used to generate \ALPGEN and  \MADGRAPH samples is CTEQ6L1~\cite{Pumplin:2002vw} while the CT10~\cite{CT10} PDF set
  is used to generate all the other background samples.

%-------------------------------------------------------------------------------
\section{Data sample and selections}
\label{sec:EventSelection}

The data set was collected in $2012$ during $pp$ collisions at $\sqrt{s} = 8$~\TeV.
It only includes data recorded with stable beam conditions and with all relevant subdetector systems operational,
and corresponds to a total integrated luminosity of $20.3$~\ifb.
The absolute luminosity scale is derived from beam-separation scans performed in November $2012$. 
The uncertainty on the integrated luminosity is $1.9\%$~\cite{LumiUncertainty_2012}.

Data events are selected by requiring at least one electron or muon candidate.
The electron and muon triggers impose a \pt\ threshold of  $24$~\GeV\ along with an isolation requirement on the lepton.
In order to increase the efficiency for high-\pt\ leptons, the electron and muon triggers are complemented
by single-electron or single-muon triggers with no isolation requirement and with a threshold of $60$~\GeV\ or $36$~\GeV, respectively.
Events are required to have at least one primary  vertex reconstructed from at least
three tracks, where the tracks must have a \pt\ greater than $400$~\MeV.

All final states with electrons, muons, and \met\ from \WZ\ leptonic decays are considered.
In the following, the different final states are referred to as $\mu^{\pm}\mu^{+}\mu^{-}$, $e^{\pm}\mu^{+}\mu^{-}$, $\mu^{\pm}e^{+}e^{-}$, and $e^{\pm}e^{+}e^{-}$.
No requirement on the number of jets is applied in the inclusive analysis, while jets are explicitly required in the dedicated analysis
 in order to enhance the contribution from the VBS process.
%
%
%-------------------------------------------------------------------------------
\subsection{Object reconstruction and selection}
\label{sec:ObjectReconstruction}

 Muon candidates are identified by tracks or track segments reconstructed in the
 muon  spectrometer system and matched to tracks reconstructed in the inner detector~\cite{MuonPerf:2014}.
The \pt\ of the muon must be greater than $15$~\GeV\ and its $|\eta|$
less than $2.5$.
 The ratio between the transverse impact parameter $d_0$ (with respect to the primary vertex) to its uncertainty ($d_0$  significance) must be smaller than $3$ and
the longitudinal impact parameter $|z_0\cdot \sin(\theta)|$ must be less than  $0.5$~mm.
Isolated muons are then selected with a requirement that the scalar sum of the \pt of the tracks
within a cone of size $\Delta R = 0.2$ around the muon, excluding the muon itself,
must be less than $15\%$ of the  muon \pt.

 Electron candidates are reconstructed from energy clusters in the calorimeter and matched to an
 inner detector track~\cite{ElectronPerf:2014}.
The lateral and transverse shapes of the cluster must be consistent with those
 of an electromagnetic shower.
The \pt\ of the electron must be greater than $15$~\GeV\ and
 the pseudorapidity of the cluster must be in the ranges $ |\eta| < 1.37$ or  $1.52 < |\eta| < 2.47$.
 The $d_0$ significance of the electron candidate  must be smaller than $6$ and the longitudinal impact
 parameter  $|z_0\cdot \sin(\theta)|$ must be less than $0.5$~mm.
To ensure that the electron candidate is isolated, the total transverse energy \et, corrected for pileup effects,
 in an isolation cone of $\Delta R = 0.2$ around the electron candidate and excluding the electron itself must be less than $14\%$ of
the electron \et. The scalar sum of the \pt of all tracks excluding the electron track itself within  the isolation cone
 must be less than $13\%$ of the electron \pt.
 If an electron overlaps with a muon candidate within  $\Delta R = 0.1$, the electron is rejected.
 This criterion mainly removes  photons from final-state radiation and jets misidentified as electrons.

Jets are reconstructed using the anti-$k_{t}$ algorithm~\cite{antikt}  with a radius parameter
$R=0.4$ using topological clusters of energy deposition in the calorimeter. Jets arising from detector noise
or noncollision events are rejected~\cite{JetCleaning:2013}.  Jets are calibrated and corrected for detector effects using a combination
 of simulated events and in situ methods~\cite{JetCleaning:2013,ATLAS-CONF-2015-002,ATLAS-CONF-2015-017}. The jet energies are also corrected  to account for energy arising
 from pileup~\cite{Aad:2015ina}.
In order to reject jets from pileup, the summed  scalar \pt of tracks associated with both the jet and the primary vertex
is required to be greater than $50\%$ of the summed scalar \pt of all the tracks associated with the jet~\cite{Aad:2015ina}.
This criterion is applied to jets with \pt smaller than $50$~\GeV\ and within $|\eta| < 2.4$.
%
%I
%
The presence of jets with \pt $> 30$~\GeV\ and a pseudorapidity $|\eta_j| < 4.5$, is explicitly required only in the VBS analysis.
Jets overlapping with an electron or muon candidate within $\Delta R =0.3$ are rejected.

The missing transverse momentum, \met\, in the event is calculated as the negative vector sum of the transverse
momentum of calibrated leptons, photons, and jets, and additional low-energy deposits in the calorimeter~\cite{MET:2012,ATLAS-CONF-2013-082}.
The contribution of the low-energy deposits from soft particles to the \met\ is further corrected to mitigate the effect of pileup on the \met\ reconstruction
performance~\cite{ATLAS-CONF-2014-019}.

%-------------------------------------------------------------------------------
\subsection{Event selection}
\label{sec:SignalSelection}

Events are required to contain at least three lepton candidates satisfying
the selection criteria described above.

In order to decrease the background from $ZZ$ processes, events containing
 four or more candidate leptons satisfying a looser \pt requirement of
\pt $> 7$~\GeV\ are discarded.

To ensure that the trigger efficiency is well determined, at least one of the candidate leptons
is required to have $p_{\mathrm{T}} > 25$~\GeV\ and to be geometrically matched to a lepton that triggered
the event.

The event must have at least one pair of leptons of the same flavor and opposite charge,
with an invariant mass that is consistent with the nominal \Zboson\ boson mass~\cite{Agashe:2014kda} within $10$~\GeV.
This pair is considered as  a \Zboson\ boson candidate.  If more than one pair
  is found, the pair whose invariant mass is closest to the nominal $Z$ boson mass is taken as the \Zboson\ boson candidate.
The third lepton is assigned to the $W$ boson.

To reduce the $Z+j$ background, the lepton assigned to the $W$ boson is required to satisfy more stringent
criteria than those required for the leptons attributed to the $Z$ boson. The \pt  threshold
for this lepton is increased to $20$~\GeV. In addition, electrons must satisfy tighter identification criteria  that include  requirements on the
 transverse impact parameter with respect to the primary vertex and on the number
 of hits in the innermost pixel layer in order to reject photon conversions.
In addition, the size of the lepton isolation cones is increased to $\Delta R = 0.3$ and
the sum of the \pt of the tracks in the isolation cone of the lepton must be less than $10\%$
of the lepton \pt.
Finally, the transverse mass of the $W$ candidate computed using the \met and the \pt
of the third lepton is required to be above $30$~\GeV.

To select VBS event candidates, in addition to the above-mentioned selection criteria, the presence
of at least two jets with \pt greater than $30$~\GeV\ with  an absolute value of $\eta$ less than $4.5$ is required.
The invariant mass of the two leading jets must be above $500$~\GeV\ and the angular distance between all selected leptons and jets is required to be greater than $0.3$.

For the search of aQGC, in addition to the selection criteria applied in the VBS analysis, it is required
that the difference in azimuthal angle
 between the reconstructed $W$ and $Z$
directions is greater than $2$~rad and that the scalar sum of the transverse momenta of the three charged leptons
associated with the $W$ and $Z$ bosons is greater than $250$~\GeV.

%-------------------------------------------------------------------------------
\section{Background estimation}
\label{sec:BackgroundEstimation}

 The background sources  are classified into two groups: events where at least one of the candidate leptons
 is not a prompt lepton (reducible background)
 and events where
 all candidates are prompt leptons (irreducible background).
 Candidates that are not prompt leptons are  called also ``misidentified'' or ``fake'' leptons.

 Events in the first group originate from  $Z+j$, $Z\gamma$, $t\bar{t}$, and $WW$ production processes.
  This background is  estimated with a data-driven method based on the inversion of a global matrix containing the
  efficiencies and the misidentification probabilities for prompt and fake leptons (see Section~\ref{sec:DataDriven}).
  In the inclusive analysis, this contribution represents about half of the total backgrounds.
About $2\%$ of this background contribution arises from events with two fake leptons.
The background from events with three fake leptons, e.g., from multijet processes, is negligible.

 The events contributing to the second group originate from $ZZ$, $t\bar{t} +V $, $VVV$ (where $V$ = $Z$ or $W$),  $t Z (j)$ events,  and
  DPS processes.
  The amount of irreducible  background is estimated using MC simulations
  due to the low cross sections of the corresponding processes and the statistical limitations of
  estimates using data-driven methods.
  In the inclusive analysis the dominant contribution in this second group  is from $ZZ$ production and
  represents about $70\%$ of the irreducible background.
  The MC-based estimation of the $ZZ$ background
   is validated by comparing data and MC simulation in properly defined control regions (see Section ~\ref{sec:ZZBKG}).

The main background in the VBS analysis originates from the processes defined as
$WZjj$-QCD in Section~\ref{sec:Signal} and amounts to $\sim 70\%$ of the total backgrounds.
The second most important background contribution arises from the $tZj$ process and amounts to $\sim 10\%$ of the total estimated background.
 Interference effects between the $WZjj$-QCD  and  $tZj$ background processes and the VBS signal ($WZjj$-EW) are expected to
  be negligible.
  The treatment of the $tZj$ background is further discussed in Section~\ref{sec:VBS}.
In the VBS analysis, background events due to misidentified leptons and due to $ZZ$ events amount to about $9\%$ and $7\%$ of the total background, respectively.

%-------------------------------------------------------------------------------
\subsection{Background from misidentified leptons ($Z+j$, $Z\gamma$, $t\bar{t}$, $WW$)}
\label{sec:DataDriven}

The matrix method~\cite{Aad:2014pda} is a data-driven method for the calculation of the
reducible background which exploits  the classification of the leptons as  loose (L) or tight (T) candidates
and  the probability that a fake lepton is misidentified as a loose or tight
lepton.

Three-lepton events in the $WZ$ data sample, selected as explained in Section~\ref{sec:SignalSelection},
 but relaxing some of the lepton identification criteria, are classified into eight categories.
Each category contains a number of events, $N_{\alpha \beta \gamma}$, where the first index refers always to
the $W$ lepton, the second to the $Z$ leading lepton, and the third to the $Z$ trailing lepton.
Each index can be L or T depending on whether the corresponding lepton met only the loose identification criteria or
satisfied the tight ones.
Loose leptons are leptons that survive the overlap removal criteria (as
described in Section~\ref{sec:ObjectReconstruction}) but do not meet the isolation criteria, while tight leptons are signal leptons
as defined in Sections~\ref{sec:ObjectReconstruction} and~\ref{sec:SignalSelection}.
These eight categories are called \textit{identification categories} here.
The number of events in each category, $N_{\alpha \beta \gamma}$, is measured directly in data.

 The same $WZ$ data sample of three-lepton events can be decomposed in eight \textit{true categories}
 according to the nature of each lepton as prompt or nonprompt.
 Each category contains a number of events, $N_{ijk}$, where each index, ordered as
 described above, can be  R or F depending on the kind of  corresponding lepton (prompt, R, or nonprompt, F).
 The number of events in each category $N_{ijk}$ is the result of the matrix method calculation.

The number of events,  $N_{\alpha \beta \gamma}$, in each identification category
is related to the number of events $N_{ijk}$ of the true categories
by an $8 \times 8$ matrix expressed in terms of
the probability that a prompt lepton is identified as a tight (loose) lepton, denoted here by $e$ (${\bar e} = 1- e$),
and  the probability that a fake lepton is misidentified as a tight (loose) lepton, denoted here by $f$ (${\bar f} = 1- f$).
The matrix reduces to a $7 \times 7$ matrix since the category $N_{\mathrm{FFF}}$ can be neglected, the number of events with three misidentified leptons being more than two orders of magnitude smaller than the number of those with only one misidentified lepton.
The value of $f$ is small, therefore terms with order higher than two in $f$ can be neglected.
It has been verified that these simplifications do not change the final result.

 The matrix is inverted to obtain the number of events with at least one misidentified lepton, which represents
 the amount of reducible background  in the $WZ$ sample, $N_{\textrm{reducible}}$
 \begin{equation}
N_{\textrm{reducible}}=  N^{\textrm{red.}}_{\mathrm{TTL}} F_3 + N^{\textrm{red.}}_{\mathrm{TLT}} F_2+ N^{\textrm{red.}}_{\mathrm{LTT}} F_1 - N^{\textrm{red.}}_{\mathrm{TLL}} F_2 F_3 - N^{\textrm{red.}}_{\mathrm{LTL}} F_1 F_3 - N^{\textrm{red.}}_{\mathrm{LLT}} F_1 F_2 \, ,
\label{eq:FinalMM3}
\end{equation}
  where $N^{\textrm{red.}}_{\alpha \beta \gamma}=N_{\alpha \beta \gamma}-N^{\textrm{irr.}}_{\alpha \beta \gamma}$, $F_i = \frac{f_i}{\bar f_i}$,
  and the index $i = 1, \, 2, \, 3$ refers to the $W$ lepton, the $Z$ leading lepton and the $Z$ trailing lepton, respectively.
  The value of $N_{\alpha \beta \gamma}$
   is obtained by counting the number of $WZ$ events in the selected data sample with leptons satisfying the loose or tight criteria.
 The  variable $N^{\textrm{irr.}}_{\alpha \beta \gamma}$ represents the number of events with three prompt leptons in the
 corresponding identification category $\alpha \beta \gamma$ and is estimated using MC simulation.
 The values of $F_i$ are measured  differentially as a function of the lepton transverse momentum,
using $W+j$ or $Z+j$  control samples taken from data for $F_1$ or for $F_2$ and $F_3$, respectively.
  The efficiencies $e (\bar{e})$ do not appear in Eq.~(\ref{eq:FinalMM3}) since they are included in the $N^{\textrm{irr.}}_{\alpha \beta \gamma}$ term.

 The control samples and the reducible background in the $WZ$ sample  are composed of
 events with misidentified leptons from light- or heavy-flavor jets and from photon conversions.
  The data-driven estimates of the $F_i$ factors  correspond to an average value weighted by the
  abundance of each kind of background  and may vary depending on the composition
  of the sample used to extract them.
 For this reason, data samples enriched in the different types of background have been used to verify
  that the background composition in the above-defined $W+j$ and $Z+j$
 control samples  is the same, within uncertainties, as in the signal region.

  Other methods to assess the reducible background  have been considered and
  provide results in good agreement with the matrix method estimation.

%-------------------------------------------------------------------------------
\subsection{Background from $ZZ$ processes}
\label{sec:ZZBKG}

The $ZZ$ background is estimated using MC simulation, as explained in Section~\ref{sec:BackgroundSimu}.
 The number of expected  $ZZ$ events from \POWHEG is scaled by $1.05$ to account for NNLO QCD
  and NLO EW corrections~\cite{Cascioli:2014yka, Bierweiler:2013dja, Baglio:2013toa}.
 In the VBS analysis,  the scale factor used for \SHERPA is taken to be $1.0$ since \SHERPA incorporates matrix element calculations
 up to three partons.

These estimations are validated by comparing the MC expectations with the event yield and several kinematic distributions of a data sample enriched
 in $ZZ$ events.
The $ZZ$ control sample is selected by requiring a $Z$ candidate  meeting all the analysis selection
 criteria accompanied by two additional leptons of the same flavor and opposite charge, satisfying the lepton criteria
 described in Section~\ref{sec:ObjectReconstruction}.
 The comparisons are performed  in the above-defined control region and in a subregion where at least two jets are present in addition.
In the first case, the data are compared with the predictions from \POWHEG and  \ggZZ Monte Carlo simulations, while in the second case \SHERPA and
\ggZZ Monte Carlo samples are used.
Overall the agreement between the data and the expectations is within one standard deviation of the experimental uncertainty.
The shapes of main kinematic variables are also found to be well described by the MC expectations.

% 
%-------------------------------------------------------------------------------
\section{Detector-level results}
\label{sec:RecoResults}

Table~\ref{tab:Results:YieldsSummary} summarizes the numbers of expected and observed events together with the
estimated background contributions in the inclusive analysis.
Only statistical uncertainties are quoted. 
Systematic uncertainties affecting the predicted yields include the theoretical uncertainty on the cross sections as discussed
in Section~\ref{sec:Theorypredictions}, and experimental uncertainties discussed in Section~\ref{sec:Systematics}.
Figure~\ref{fig:Results:WZControlPlots} shows, at detector level, the momentum and the invariant mass of the $Z$ candidate,
the transverse mass of the $W$ candidate and a transverse mass-like variable of the  $WZ$ system, $m_{\mathrm{T}}^{WZ}$, after applying all selection criteria.
The variable $m_{\mathrm{T}}^{WZ}$ is reconstructed as
\begin{equation}
m_{\mathrm{T}}^{WZ} = \sqrt{ \left( \sum_{\ell = 1}^3 p_{\mathrm{T}}^\ell + E_{\mathrm{T}}^{\mathrm{miss}} \right)^2
		- \left[ \left(\sum_{\ell = 1}^3 p_x^\ell + E_{x}^{\mathrm{miss}} \right)^2 + \left(\sum_{\ell = 1}^3 p_y^\ell + E_{y}^{\mathrm{miss}} \right)^2 \right]} \, .
\end{equation}
The expectations based on MC simulation are scaled to the integrated luminosity of the data using the predicted cross sections of
each sample.
The \powhegpythia MC prediction is used for the \wz\ signal contribution.
In Figure~\ref{fig:Results:WZControlPlots} it is scaled by a global factor of $1.17$ to match the measured inclusive \wz\ cross section of Section~\ref{sec:IntCrossSections}.
This scaling is only used for an illustrative purpose in this Figure and does not affect the measurements.
Table~\ref{tab:vbs_yields} shows the number of expected and observed events together with the
estimated background contributions for the VBS and aQGC analyses, respectively.
Figure~\ref{fig:Results:WZControlPlots}  indicates that the MC predictions provide a fair description of the shapes of the data distributions.

\begin{table}[!htbp]
\begin{center}
\begin{tabular}{l rrrr r} 
\hline 
Channel        & $e e e$	    & $\mu e e $	    & $e\mu\mu$ 	  & $\mu\mu\mu$ 	  & All \\ 
\hline 	
Data 	       & $ 406$ 	    & $ 483$		    & $ 539$		  & $ 663$		  & $2091$	  \\ 
\hline 
Total expected & $336.7 \pm 2.2$    & $410.8 \pm 2.4$	    & $469.1 \pm 2.1$	  & $608.2 \pm 3.5$	  & $1824.8 \pm 7.0$	  \\ 
\hline 
$WZ$  	       & $255.7 \pm 1.1$     & $337.2 \pm 1.0$	    & $367.0 \pm 1.1$	  & $495.9 \pm 2.3$	  & $1455.7 \pm 5.5$	  \\ 
Misid. leptons & $43.7 \pm 1.9$     & $32.2 \pm 2.1$	    & $50.2 \pm 1.7$	  & $52.8 \pm 2.6$	  & $178.9 \pm 4.2$	  \\ 
$ZZ$ 	       & $25.9 \pm 0.2$     & $26.7 \pm 0.3$	    & $36.1 \pm 0.3$	  & $39.5 \pm 0.3$	  & $128.2 \pm 0.6$	  \\ 
$t\bar{t} + V$ & $5.5 \pm 0.2$      & $6.7 \pm 0.2$	    & $7.2 \pm 0.3$	  & $9.1 \pm 0.3$	  & $28.5 \pm 0.5$	  \\ 
$tZ$ 	       & $4.2 \pm 0.1$      & $5.5 \pm 0.2$	    & $6.0 \pm 0.2$	  & $7.7 \pm 0.2$	  & $23.3 \pm 0.3$	  \\ 
DPS 	       & $1.2 \pm 0.1$      & $1.9 \pm 0.1$	    & $1.8 \pm 0.1$	  & $2.3 \pm 0.2$	  & $7.2 \pm 0.3$	  \\ 
$VVV$ 	       & $0.5 \pm 0.0$      & $0.7 \pm 0.0$	    & $0.8 \pm 0.0$	  & $0.9 \pm 0.0$	  & $3.0 \pm 0.1$	  \\ 
\hline 
\end{tabular} 
\end{center}
\caption{Numbers of observed and expected events after the \wz\ inclusive selection described in Section~\ref{sec:SignalSelection} in each of the considered channels and for
 the sum of all channels.  The expected number of \wz\ events from \powhegpythia and the estimated number of background events from other processes are detailed.
The sum of background events containing misidentified leptons is labeled ``Misid. leptons''.
Only statistical uncertainties are quoted.}
\label{tab:Results:YieldsSummary}
\end{table}

\begin{figure}[!htbp]
\begin{center}

\includegraphics[width=.4\textwidth]{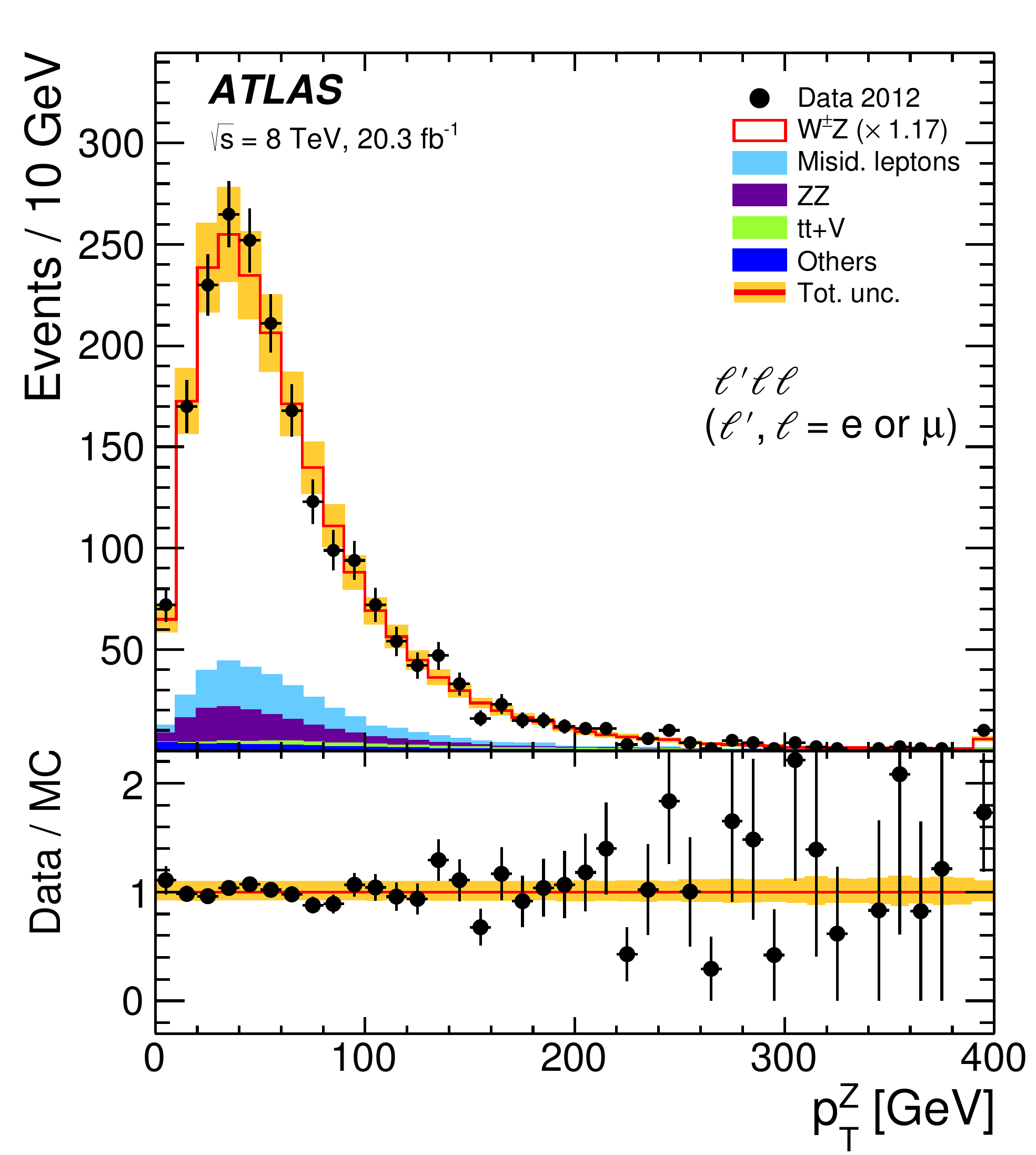}\put(-30,90){{(a)}}
\includegraphics[width=.4\textwidth]{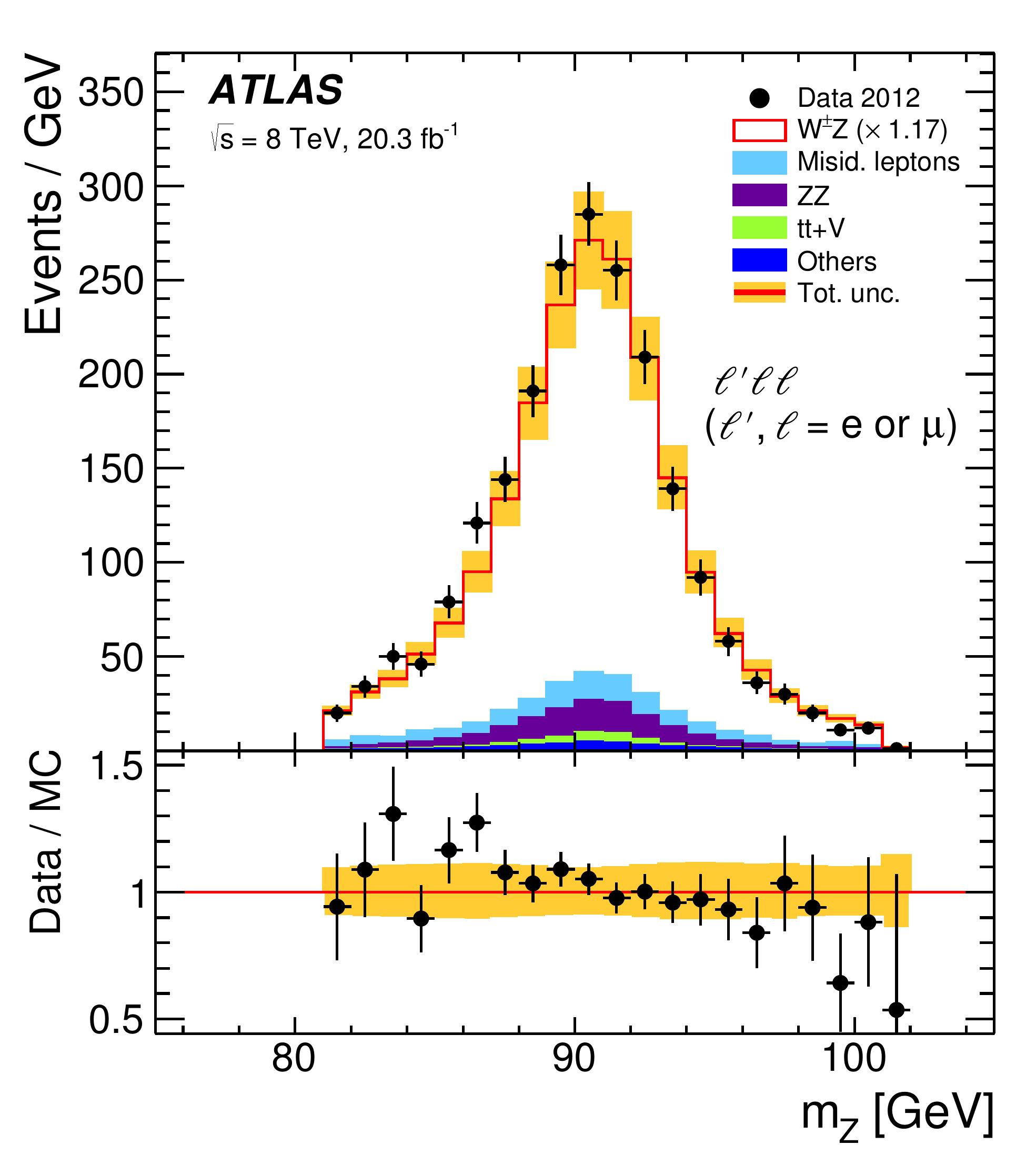}\put(-30,90){{(b)}}\\
\includegraphics[width=.4\textwidth]{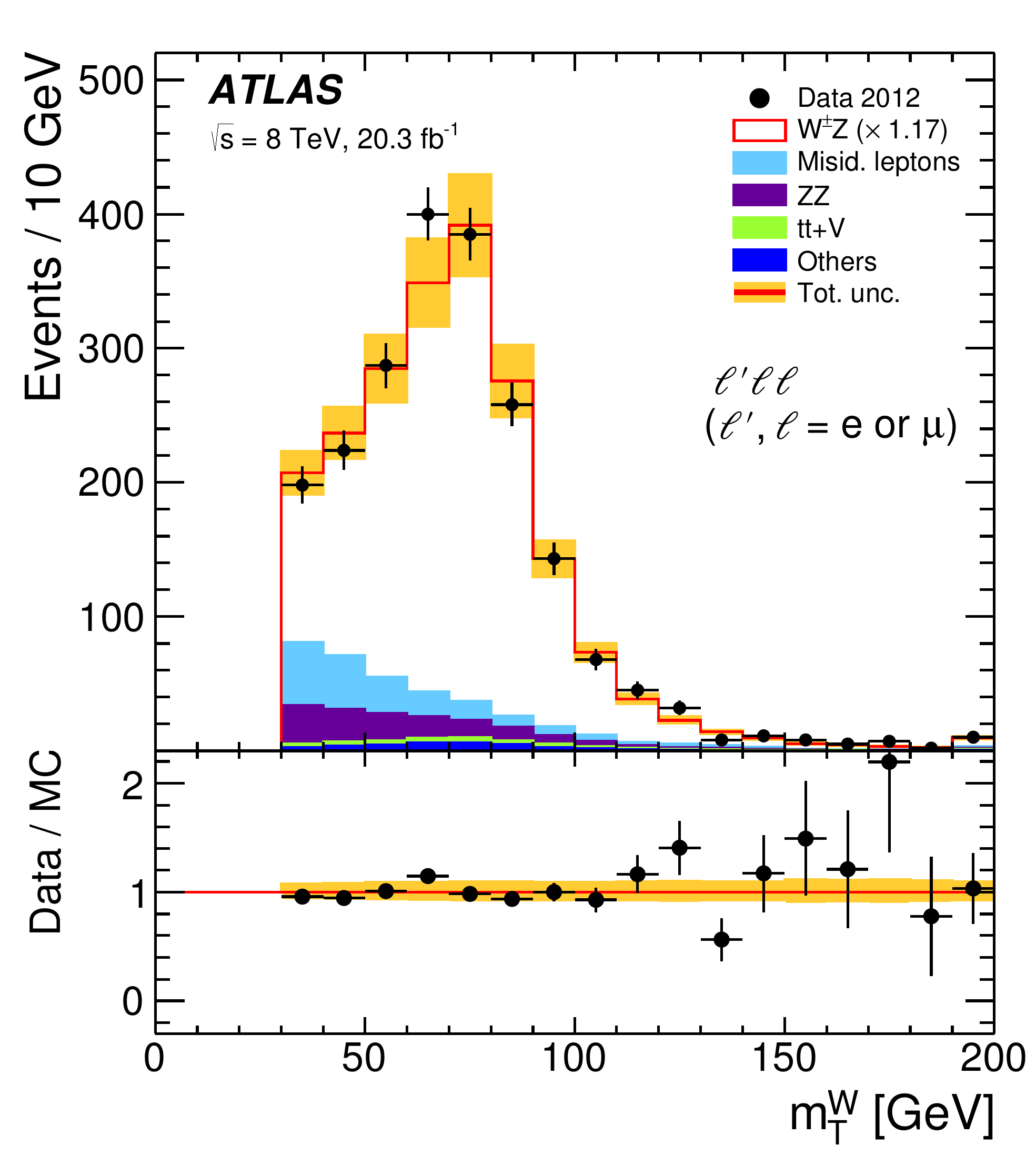}\put(-30,90){{(c)}}
\includegraphics[width=.4\textwidth]{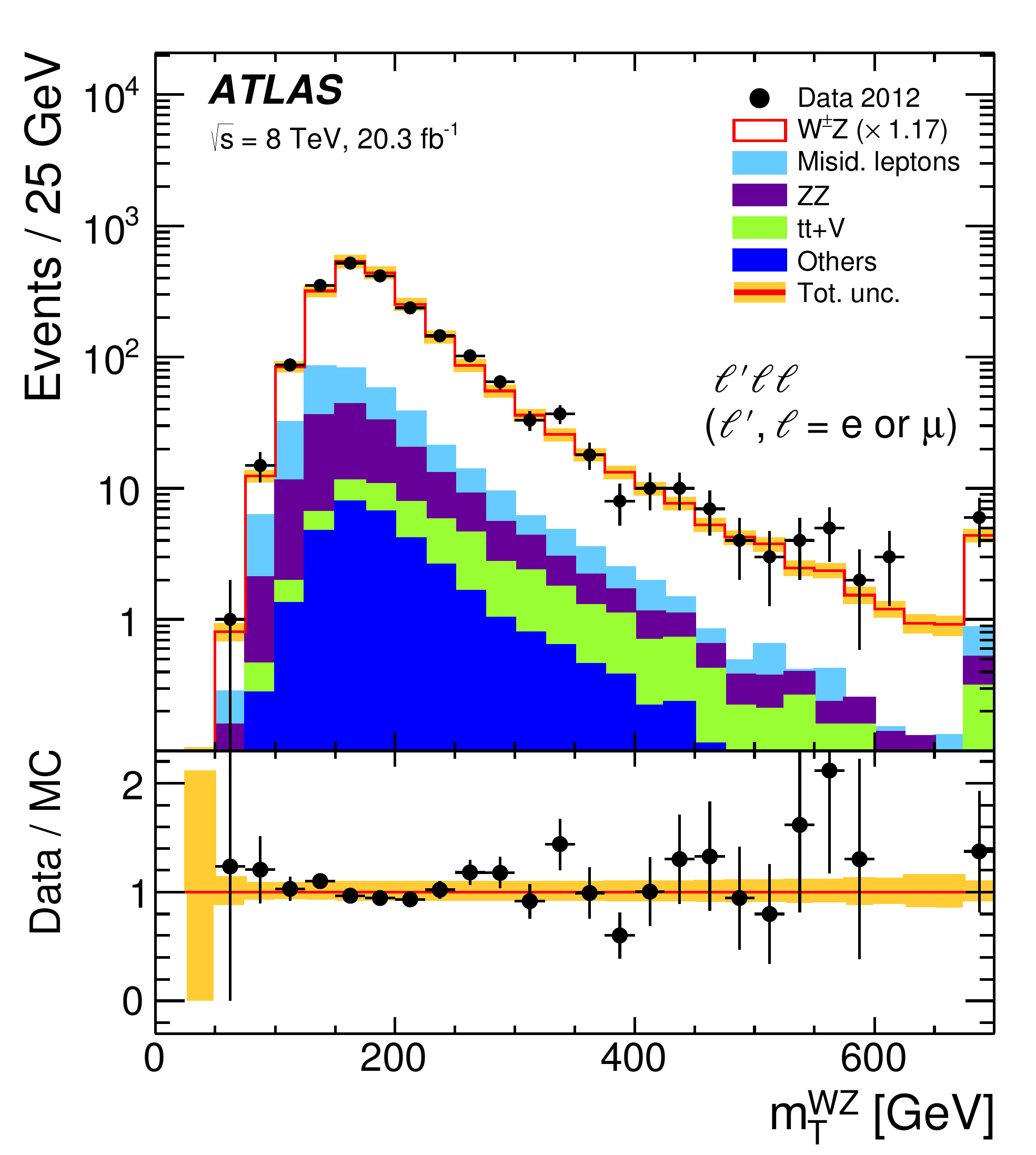}\put(-80,180){{(d)}}\\

\end{center}
 \caption{Distributions, summed over all channels, of the following kinematic variables: (a) the transverse
momentum of the reconstructed $Z$ boson $p_\textrm{T}^Z$, (b) the mass of the $Z$ $m_Z$, (c) the transverse mass of the reconstructed $W$ boson $m_\textrm{T}^W$ and (d)  the transverse-mass like variable for the $WZ$ system $m_\textrm{T}^{WZ}$.
The points correspond to the data and the histograms to the expectations of the different SM processes.
All Monte Carlo expectations are scaled to the integrated luminosity of the data using the predicted MC cross sections of each sample.
The sum of background events containing misidentified leptons is labeled ``Misid. leptons''.
The \powhegpythia MC prediction is used for the \wz\ signal contribution.
It is scaled by a global factor of $1.17$ to match the measured inclusive \wz\ cross section.
The open red histogram shows the total prediction and the shaded orange band its
estimated total uncertainty.
The last bin contains the overflow.
}
\label{fig:Results:WZControlPlots}
\end{figure}

\begin{table}[!htbp]
\begin{center}
\begin{tabular}{l r r}
\hline
  Selection                        & VBS 		&  aQGC 	\\
\hline
Data	 	          & $ 45$ 	  	&   $9$	        \\
\hline
Total Expected	  	  & $ 37.2\pm 1.1$  	&   $4.9 \pm 0.3$	\\
\hline
$WZjj$-EW	 	  & $ 7.4\pm 0.2$ 	&   $1.1 \pm 0.1$		\\
$WZjj$-QCD	 	  & $ 20.8\pm 0.8$ 	&   $2.8 \pm 0.3$		\\
$tZ$	 	          & $ 3.0\pm 0.1$ 	&  $0.3 \pm 0.0$		\\
Misid. leptons	 	  & $ 2.5\pm 0.6$ 	&  $0.1 \pm 0.1$		\\
$ZZ$ 	 	          & $ 1.9\pm 0.3$ 	&  $0.2 \pm 0.1$		\\
$t\bar{t} + V$	          & $ 1.6\pm 0.1$ 	&  $0.3 \pm 0.0$		\\
\hline
\end{tabular}
\end{center}
\caption{
Numbers of observed and expected events for the sum of all channels after the \wz\ VBS and aQGC selections described in Section~\ref{sec:SignalSelection}.
The expected number of $WZjj$-EW events from \sherpa and the estimated number of background events from other processes are detailed.
The sum of background events containing misidentified leptons is labeled ``Misid. leptons''.
Only statistical uncertainties are quoted.
}
\label{tab:vbs_yields}
\end{table}

%-------------------------------------------------------------------------------
\section{Corrections for detector effects and acceptance}
\label{sec:CrossSectionDefinition}

  For a given channel \wz\ $\rightarrow \ell^{'\pm} \nu \ell^+ \ell^-$, where $\ell$ and $\ell^{'}$ are either an electron or a
  muon,  the integrated fiducial cross section that includes the leptonic branching fractions of the
  $W$ and $Z$  is calculated as
 \begin{equation}\label{eq:sigma_fid}
\sigma^{\mathrm{fid.}}_{W^\pm Z \rightarrow \ell^{'} \nu \ell \ell}  = \frac {  N_{\textrm{data}} - N_{\textrm{bkg}} }
 {   \mathcal{L}  \cdot  C_{WZ} }  \times  \left( 1 - \frac {N_\tau} {N_{\textrm{all}} } \right) \, ,
 \end{equation}
 where  $N_{\textrm{data}}$ and  $N_{\textrm{bkg}}$ are the number of observed events and the estimated number of background events, respectively,
 $\mathcal{L}$ is the integrated luminosity, and $C_{WZ}$, obtained from simulation, is the ratio of the number of
 selected signal events at detector level to the number of events at particle level in the fiducial phase space. This factor corrects for detector efficiencies
 and for QED final-state radiation effects. The contribution from $\tau$ lepton decays, amounting approximately to $4\%$, is removed from the
 cross-section definition by introducing the term in parenthesis.
This term is computed using simulation, where $N_\tau$ is the number of selected
 events in which at least one of the bosons decays into a $\tau$ lepton and $N_{\textrm{all}}$ is the number of selected  $WZ$ events with
 decays into any lepton.

  The  $C_{WZ}$ factors for $W^- Z$,  $W^+ Z$, and $W^\pm Z$ inclusive processes computed with \powhegpythia for
  each of the four leptonic channels are shown in Table~\ref{tab:cwz}.

\begin{table}[!htbp]
\begin{center}
\begin{tabular}{cccc}
\hline
Channel &   $C_{W^-Z}$ &  $C_{W^{+}Z}$  & $C_{W^\pm Z}$\\
\hline

$eee$     & $0.412$ $\pm$ $0.002$ & $0.399$ $\pm$ $0.002$ & $0.404$ $\pm$ $0.001$ \\
$\mu{ee}$ & $0.532$ $\pm$ $0.002$ & $0.540$ $\pm$ $0.002$ & $0.537$ $\pm$ $0.001$ \\
${e}\mu\mu$  & $0.596$ $\pm$ $0.002$ & $0.572$ $\pm$ $0.002$ & $0.581$ $\pm$ $0.001$ \\
$\mu\mu\mu$ & $0.786$ $\pm$ $0.002$ & $0.789$ $\pm$ $0.002$ &  $0.788$ $\pm$ $0.002$\\
\hline
\end{tabular}
\end{center}
\caption{The $C_{WZ}$ factors for each of the $eee$, $\mu{ee}$, ${e}\mu\mu$, and $\mu\mu\mu$ inclusive channels.
The  \powhegpythia MC event sample with the ``resonant shape'' lepton assignment algorithm at particle level
is used. Only statistical uncertainties are reported. }
\label{tab:cwz}
\end{table}

   The total cross section is calculated as
\begin{equation}\label{eq:sigma_tot}
\sigma^{\mathrm{tot.}}_{W^\pm Z}  = \frac{  \sigma^{\mathrm{fid.}}_{W^\pm Z \rightarrow \ell^{'} \nu \ell \ell}  } { \mathcal{B}_W \, \mathcal{B}_Z  \, A_{WZ} } \, ,
 \end{equation}
where $ \mathcal{B}_W = 10.86 \pm 0.09 \, \%$ and $\mathcal{B}_Z = 3.3658 \pm 0.0023 \, \%$ are the $W$ and $Z$ leptonic branching fractions~\cite{Agashe:2014kda}, respectively, and
  $A_{WZ}$ is the acceptance factor calculated at particle level as the ratio of the number of events in the fiducial
 phase space to the number of events in the total phase space as defined in Section~\ref{sec:FiducialPS}.

   A single acceptance factor of $A_{WZ}$ = $0.395$ $\pm$ $0.001$~(stat.), obtained by averaging the acceptance factors computed
  in the  $\mu ee$ and $e \mu\mu$ channels, is used
  since it has been verified that interference effects related to the presence of
    identical leptons in the final state, as in the $eee$ and $\mu \mu \mu$ channels, are below $1\%$.
  The use of the  $\mu{ee}$ and ${e}\mu\mu$ channels for the computation of $A_{WZ}$
  avoids the ambiguity arising from the assignment at particle level
  of final-state leptons to the $W$ and $Z$ bosons.

   The differential detector-level distributions are corrected for detector resolution and for QED FSR effects using
   an iterative Bayesian unfolding method~\cite{D'Agostini:1994zf}, as implemented in the RooUnfold toolkit~\cite{RooUnfold}.
Three iterations were consistently used for the unfolding of each variable.
   The width of the bins in each distribution was chosen according to the experimental resolution
and to the statistical significance of the expected number of events in each bin.
For the data distributions used to extract the limits on anomalous gauge couplings,
  a dedicated bin optimization was performed using signal MC events, in order to reach the best sensitivity for
  the fitted parameters.
The fraction of signal MC events reconstructed in each bin is always greater than $50\%$ and around $60\%$ on average.

 Simulated signal events are used to obtain for each  distribution a response matrix that accounts
 for bin-to-bin migration effects between the reconstructed-level and particle-level distributions.
 In the inclusive measurements, the \powhegpythia signal sample is used since it provides
 a fair description of the data distributions.
  For the  jet multiplicity differential measurement  and in the VBS analysis, the  \SHERPA  signal sample
  is used for the computation of the response matrix since this sample includes up to three partons in the matrix element
 calculation and therefore better describes the jet multiplicity of data.
 To build the response matrix for the unfolding of the jet multiplicity, the \pt threshold of the particle level jets, as defined in Section~\ref{sec:FiducialPS}, is set to $25$~\GeV.
This threshold is similar to the one used in the recent measurement of the $WW$ cross section by the ATLAS Collaboration~\cite{WW_8TeV_atlas}.
A jet \pt threshold of $30$~\GeV, corresponding to the definition of the VBS phase space, is, however, used for the unfolding of the invariant mass spectrum of the two leading jets.

%-------------------------------------------------------------------------------
\section{Systematic uncertainties}
\label{sec:Systematics}

 The systematic uncertainties on the integrated and differential  cross sections are due to
  uncertainties of experimental
  and theoretical nature on the acceptance, on the correction procedure for detector effects, on the background estimation and on the luminosity.

 The systematic uncertainties on the $A_{WZ}$ and $C_{WZ}$ factors
 due to the theoretical modeling in the event generators are evaluated taking into account
 the uncertainties related to the choice of the PDF, QCD
 renormalization and factorization scales, and of the parton showering simulation.
 Uncertainties due to the choice of PDF are computed using the CT10 eigenvectors and the envelope
  of the differences between CT10, MSTW 2008, NNPDF 3.0, and ATLAS-epWZ12 PDF sets.
  QCD scale uncertainties are estimated by varying $\mu_{\mathrm{R}}$ and $\mu_{\mathrm{F}}$ by factors of two around the nominal scale $m_{WZ}/2$ with the constraint $0.5 \leq \mu_{\mathrm{R}} /\mu_{\mathrm{F}} \leq 2$.
Uncertainties due to the  choice of parton showering model are estimated by interfacing
  \POWHEG with either \PYTHIA or \HERWIG and comparing the results.
   These uncertainties of theoretical nature have no significant effect on the $C_{WZ}$ factors but affect the $A_{WZ}$ acceptance factor, where
   the dominant contribution originates from the PDF choice and is below $1.3\%$.

The uncertainty on the differential distributions arising from the MC modeling  of the response matrix
in the unfolding procedure  is estimated by  reweighting simulated events at particle level to the unfolded results obtained as described in Section~\ref{sec:CrossSectionDefinition}.
An alternative response matrix is defined using these reweighted MC events and is used to unfold \powhegpythia reconstructed MC events.
A systematic uncertainty is estimated by comparing this unfolded distribution to the original particle-level \powhegpythia prediction.

  The experimental systematic uncertainty on the $C_{WZ}$ factors and on the unfolding
  procedure  includes uncertainties on the electron  energy or muon momentum scale
  and resolution, on the \MET scale and resolution, on the jet energy scale and resolution,
  as well as uncertainties on the scale  factors applied to the simulation  in order to reproduce the
  trigger, reconstruction, identification, and isolation efficiencies measured in data.  The uncertainty
  associated with the pileup reweighting procedure is negligible.
  For the measurements of the $W$ charge-dependent cross sections,
  an uncertainty arising from the charge misidentification of leptons is also considered.
It affects only electrons and leads to uncertainties of $\sim 0.1\%$ on the integrated cross section combining all decay channels.
The systematic uncertainties on the measured cross section are determined by repeating
the analysis after applying appropriate variations for each source of systematic uncertainty to the simulated samples.

  The lepton energy or momentum scale corrections are obtained from a comparison of the
  $Z$ boson invariant mass distribution in data and simulations, while the uncertainties
  on the efficiency scale factors are derived from a comparison of tag-and-probe results in data and
  simulations~\cite{ElectronPerf:2014,ATLAS-CONF-2014-032,MuonPerf:2014}.
 Uncertainties on the jet energy scale are determined from a combination of methods based on simulation
 and in situ techniques~\cite{JetCleaning:2013,ATLAS-CONF-2015-002}.
 The uncertainty on the jet energy resolution is derived from a comparison of the resolutions obtained
 in data and in simulated dijet events~\cite{ATLAS-CONF-2015-017}.
 The uncertainty on the \MET is estimated by propagating the uncertainties on the objects and
 by applying energy scale and resolution uncertainties to the calorimeter energy clusters that
 are not associated with a jet or an electron. %
 The dominant contribution among the experimental systematic uncertainties
 in the $eee$ and $\mu ee$  channels derives from the electron identification efficiency,
  being at most  $2.9\%$, while  in the $e \mu\mu$ and
 $\mu\mu\mu$ channels it originates from the muon reconstruction efficiency   and is at most  $2.1\%$.

 The uncertainty on the amount of background from misidentified leptons is
 estimated taking into account the statistical uncertainties on the event yields in each identification
 category and on the $F_i$ factors (see Section~\ref{sec:DataDriven}).
  Uncertainties arising from the definition of the $W+j$ and $Z+j$ control
 samples and from their composition are also included.
 The former are evaluated by changing the control sample selection criteria and the latter by using
 a different way of computing the fake rate, which relies on a matrix method where the matrix is obtained
 using particle-level information.

An uncertainty of $7\%$ on the amount of $ZZ$ background is evaluated as the difference between the predicted and
  measured numbers of $ZZ$ events  in the defined control regions.
The uncertainty arising from other kinds of irreducible backgrounds is evaluated by propagating the uncertainty on
their MC cross section which are estimated to be $30\%$, $15\%$ and $50\%$ for $t\bar{t} + V$, $tZ$, and DPS processes, respectively.

The uncertainty on the unfolding procedure arising from the limited number of events in
 the simulation is estimated using  pseudoexperiments.

 The uncertainty on the integrated luminosity~\cite{LumiUncertainty}
 is applied to the signal normalization as well as to all background contributions that are estimated
  using MC simulations. It results in an effect of $2.2\%$ on the measured cross sections.

 The overall uncertainty on the single-channel \wz fiducial  cross section varies
 from  approximately  $6\%$ to $8\%$.
Table~\ref{tab:SystematicSummary} shows the statistical and main systematic
 uncertainties on the \wz fiducial cross section for each of the four channels and for their combination.

\begin{table}[!htbp]
\begin{center}
\begin{tabular}{l r r r r r}  
\hline 
 & $e e e$ & $\mu e e$ & $e \mu\mu$ & $\mu\mu\mu$ & combined \\ 
\hline 
Source & \multicolumn{5}{c}{Relative uncertainties [\%]}\\ 
\hline 
 $e$ energy scale& $0.8$ & $0.4$ & $0.4$ & $0.0$ & $0.3$ \\ 
 $e$ id. efficiency& $2.9$ & $1.8$ & $1.0$ & $0.0$ & $1.0$ \\ 
 $\mu$ momentum scale& $0.0$ & $0.1$ & $0.1$ & $0.1$ & $0.1$ \\ 
 $\mu$ id. efficiency& $0.0$ & $0.7$ & $1.3$ & $2.0$ & $1.4$ \\ 
 $E_{\mathrm{T}}^{\textrm{miss}}$ and jets& $0.3$ & $0.2$ & $0.2$ & $0.1$ & $0.3$ \\ 
 Trigger& $0.1$ & $0.1$ & $0.2$ & $0.3$ & $0.2$ \\ 
 Pileup& $0.3$ & $0.2$ & $0.2$ & $0.1$ & $0.2$ \\ 
 Misid. leptons background& $2.9$ & $0.9$ & $3.1$ & $0.9$ & $1.3$ \\ 
 $ZZ$ background& $0.6$ & $0.5$ & $0.6$ & $0.5$ & $0.5$ \\ 
 Other backgrounds& $0.7$ & $0.7$ & $0.7$ & $0.7$ & $0.7$ \\ 
%\hline 
 Uncorrelated& $0.7$ & $0.6$ & $0.5$ & $0.5$ & $0.3$ \\ 
\hline 
 Total systematics& $4.5$ & $2.6$ & $3.7$ & $2.5$ & $2.4$ \\ 
 Luminosity& $2.2$ & $2.2$ & $2.2$ & $2.2$ & $2.2$ \\ 
 Statistics& $6.2$ & $5.4$ & $5.3$ & $4.7$ & $2.7$ \\ 
\hline 
\hline 
 Total& $8.0$ & $6.3$ & $6.8$ & $5.7$ & $4.2$ \\ 
\hline 
\end{tabular} 
\end{center}
\caption{Summary of the relative uncertainties on the measured fiducial cross section $\sigma^{\mathrm{fid.}}_{W^\pm Z}$ for each channel and for their combination.
Uncertainties are given in percent.
The decomposition of the total systematic uncertainty into the main sources correlated between channels and a source uncorrelated between channels is indicated in the first rows.
}
\label{tab:SystematicSummary}
\end{table}

%-------------------------------------------------------------------------------
\section{Cross-section measurements}
\label{sec:CrossSections}
\subsection{Integrated cross sections}
\label{sec:IntCrossSections}

 The measured fiducial  cross sections in the four channels are combined
 using the measured total event yields and statistical procedure based on the minimization of a
 negative log-likelihood function that accounts for correlations between the sources of
  systematic uncertainty affecting each channel~\cite{Aad:2015rka}.
  The systematic uncertainties are included in the
 likelihood function as nuisance parameters.
  The combination of the \wz cross sections in the fiducial phase space yields a $p$-value of $48\%$,
 the combinations of $W^+ Z$ and $W^- Z$ cross sections yield $p$-values of
 $15\%$ and $26\%$, respectively.

  The \wz production cross section in the detector fiducial region resulting from the combination of the four channels including the $W$ and $Z$ branching ratio in a single leptonic channel
 with muons or electrons is
 \begin{eqnarray}
\sigma_{W^\pm Z \rightarrow \ell^{'} \nu \ell \ell}^{\mathrm{fid.}} =  35.1~\pm~ 0.9 \, \textrm{(stat.)}~\pm~0.8 \, \textrm{(sys.)}~\pm 0.8 \, \textrm{(lumi.) fb},
\end{eqnarray}
where the uncertainties correspond to statistical, systematic and luminosity uncertainties, respectively.
The measurement is to be compared to the SM expectation of $30.0  \pm 2.1$~fb from \powhegpythia, as discussed in Section~\ref{sec:Theorypredictions}.
 The measured \wz production cross sections are compared to the SM NLO prediction from \powhegpythia in Figure~\ref{fig:xsectionperchannel}
and all  results for $W^\pm Z$, $W^+Z$, and $W^-Z$ final states
 are reported in Table~\ref{tab:XSection:FiducialCSAll}.

The measured cross section is larger than the quoted SM prediction.
However, the SM prediction, which is at NLO accuracy in perturbative QCD, is highly sensitive to the choice of renormalization scale $\mu_{\mathrm{R}}$.
%
%For example, choosing a fixed renormalization scale of $\mu_{\mathrm{R}} = (m_W + m_Z)/2$ instead of a dynamic scale $\mu_{\mathrm{R}} = m_{WZ}$ increases the SM predicted cross section by $7\%$.
%
%
In addition, new perturbative effects appearing at NNLO could enhance the SM prediction compared to the NLO calculation.
Indeed, for the other diboson final states $ZZ$, $WW$, $Z\gamma$, and $W\gamma$ NNLO calculations have recently become available~\cite{Cascioli:2014yka,Grazzini:2015hta,Gehrmann:2014fva,Grazzini:2015nwa} and in all cases the NNLO
corrections were found to be positive and larger than the uncertainty on the NLO calculation estimated by the conventional independent up and down variations of $\mu_{\mathrm{R}}$ and $\mu_{\mathrm{F}}$ by a factor of two.

 %%%%%% TABLE: Cross section WZ
 \begin{table}[!htbp]
\begin{center}
\begin{tabular}{lccccc}

\hline
Channel   &  $\sigma^{\mathrm{fid.}}$ & $\delta_{\mathrm{stat.}}$ & $\delta_{\mathrm{sys.}}$ & $\delta_{\mathrm{lumi.}}$ &$\delta_{\mathrm{tot.}}$\\
          &  [fb] & [\%] &  [\%] & [\%] &[\%]\\
\hline
\multicolumn{6}{c}{~}\\ [-12.0pt]
\multicolumn{6}{c}{$\sigma^{\mathrm{fid.}}_{W^{\pm} Z \rightarrow \ell^{'} \nu \ell \ell}$}\\ [4.0pt]
\hline
$e^\pm{ee}$                                         &  38.1    &       6.2 &  4.5 &           2.2 &  8.0\\   
$\mu^\pm{ee}$                                      &  36.3    &      5.4 &  2.6 &           2.2 & 6.3\\ 
$e^\pm\mu\mu $                                 &  35.7   &       5.3 &  3.7 &            2.2 &  6.8\\ 
$\mu^\pm\mu\mu$                              &  33.3   &         4.7 &  2.5 &              2.2 &  5.7\\ 
\hline
Combined      & 35.1    &          2.7 &  2.4 &              2.2 &  4.2\\ 
\hline 
SM expectation  & 30.0    &          --- &  --- &            --- &  7.0\\ 
\hline 
\multicolumn{6}{c}{~}\\ [-12.0pt]
\multicolumn{6}{c}{$\sigma^{\mathrm{fid.}}_{W^{+} Z \rightarrow \ell^{'} \nu \ell \ell}$}\\ [4.0pt]
\hline
$e^+ee$                                            & 22.6    &          8.0 &             4.4 &             2.2 &  9.4\\   
$\mu^+{ee}$                                        & 23.9    &          6.5 &             2.5 &              2.2 & 7.3\\ 
$e^+\mu\mu$                                   & 19.9    &          7.2 &             3.5 &               2.2 & 8.3\\ 
$\mu^+\mu\mu$                               & 19.8    &           6.0 &             2.5 &               2.2 &  6.8\\ 
\hline
Combined          & 21.2    &           3.4 &             2.3 &                2.2 & 4.6\\ 
\hline 
SM expectation  & 18.8    &          --- &  --- &            --- &  6.8\\ 
\hline 
\multicolumn{6}{c}{~}\\ [-12.0pt]
\multicolumn{6}{c}{$\sigma^{\mathrm{fid.}}_{W^{-} Z \rightarrow \ell^{'} \nu \ell \ell}$}\\ [4.0pt]
\hline
$e^-ee$                                            & 15.4    &           9.8 &            5.0 &                2.3 &     11.2\\   
$\mu^-{ee}$                                        & 12.4   &            9.5 &            3.1 &                2.3 &     10.3 \\ 
$e^-\mu\mu$                                   & 15.7   &            8.0 &            4.2 &                2.3 &    9.2\\ 
$\mu^-\mu\mu$                                & 13.4   &            7.5 &            2.8 &               2.3&         8.3 \\  
\hline
Combined           & 14.0   &             4.3 &            2.8 &                2.3 &        5.6\\ 
\hline
SM expectation  & 11.1    &          --- &  --- &            --- &  8.9\\ 
\hline 
\end{tabular}
\caption{Fiducial integrated cross section in fb, for $W^\pm Z$, $W^+ Z$, and $W^- Z$ production, measured in each of the $eee$, $\mu{ee}$, ${e}\mu\mu$, and $\mu\mu\mu$ channels and all four channels combined.
The statistical ($\delta_{\mathrm{stat.}}$), total systematic ($\delta_{\mathrm{sys.}}$), luminosity ($\delta_{\mathrm{lumi.}}$), and total ($\delta_{\mathrm{tot.}}$) uncertainties are given in percent.
}
\label{tab:XSection:FiducialCSAll}
\end{center}
\end{table}

The ratio of $ W^+Z$ to $W^-Z$ production cross sections is also measured in the fiducial phase space and yields
\begin{eqnarray}
\frac{\sigma_{W^{+}Z \rightarrow \ell^{'} \nu \ell \ell}^{\textrm{fid.}}}{\sigma_{W^{-}Z \rightarrow \ell^{'} \nu \ell \ell}^{\textrm{fid.}}} & = & 1.51\pm 0.08 \,\textrm{(stat.)} \pm 0.01 \,\textrm{(sys.)} \pm 0.01 \,\textrm{(lumi.) } \nonumber.
\end{eqnarray}
Most of the systematic uncertainties cancel in the ratio and the measurement is dominated by the statistical uncertainty.
The measured cross-section ratios, for each channel and for their combination, are compared in Figure~\ref{fig:WPMXSection:WpWmRatio} to the SM expectation of $1.69 \pm 0.07$, calculated with \powhegpythia and the CT10 PDF set.
The use of the ATLAS-epWZ12 PDF set instead of CT10 changes the SM prediction to $1.63$, indicating the sensitivity of the ratio $\sigma_{W^{+}Z}^{\textrm{fid.}} / \sigma_{W^{-}Z}^{\textrm{fid.}}$ to the PDFs.
The total uncertainty of the present measurement is of the same order of magnitude as the estimated uncertainties in the PDF and the SM prediction.

%%%%%% FIGURE: Ratio Cross section
\begin{figure}[!htbp]
\begin{center}
\includegraphics[width=11cm]{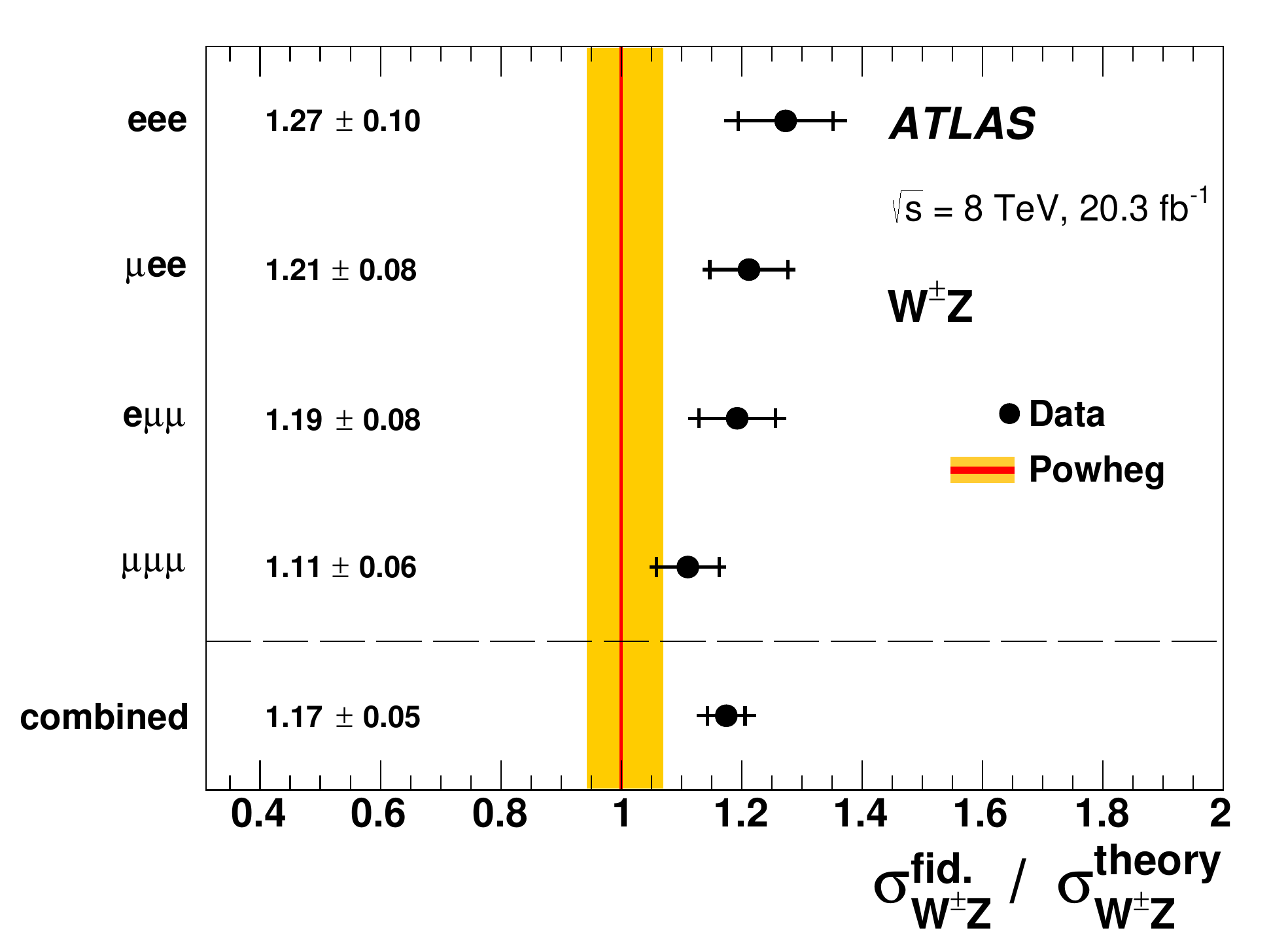}
\caption{Ratio of the measured \wz\ integrated cross sections in the fiducial phase space to the NLO SM prediction from \powhegpythia using the CT10 PDF set and renormalisation and factorisation scales $\mu_R = \mu_F = m_{WZ}/2$, in each of the four channels and for their combination.
The inner and outer error bars on the data points represent the statistical and total uncertainties, respectively.
The shaded orange band represents the uncertainty associated with the SM prediction.}
\label{fig:xsectionperchannel}
\end{center}
\end{figure}

%%%%%% FIGURE: Ratio Cross section f Wp and Wm
\begin{figure}[!htbp]
\begin{center}
\includegraphics[width=11cm]{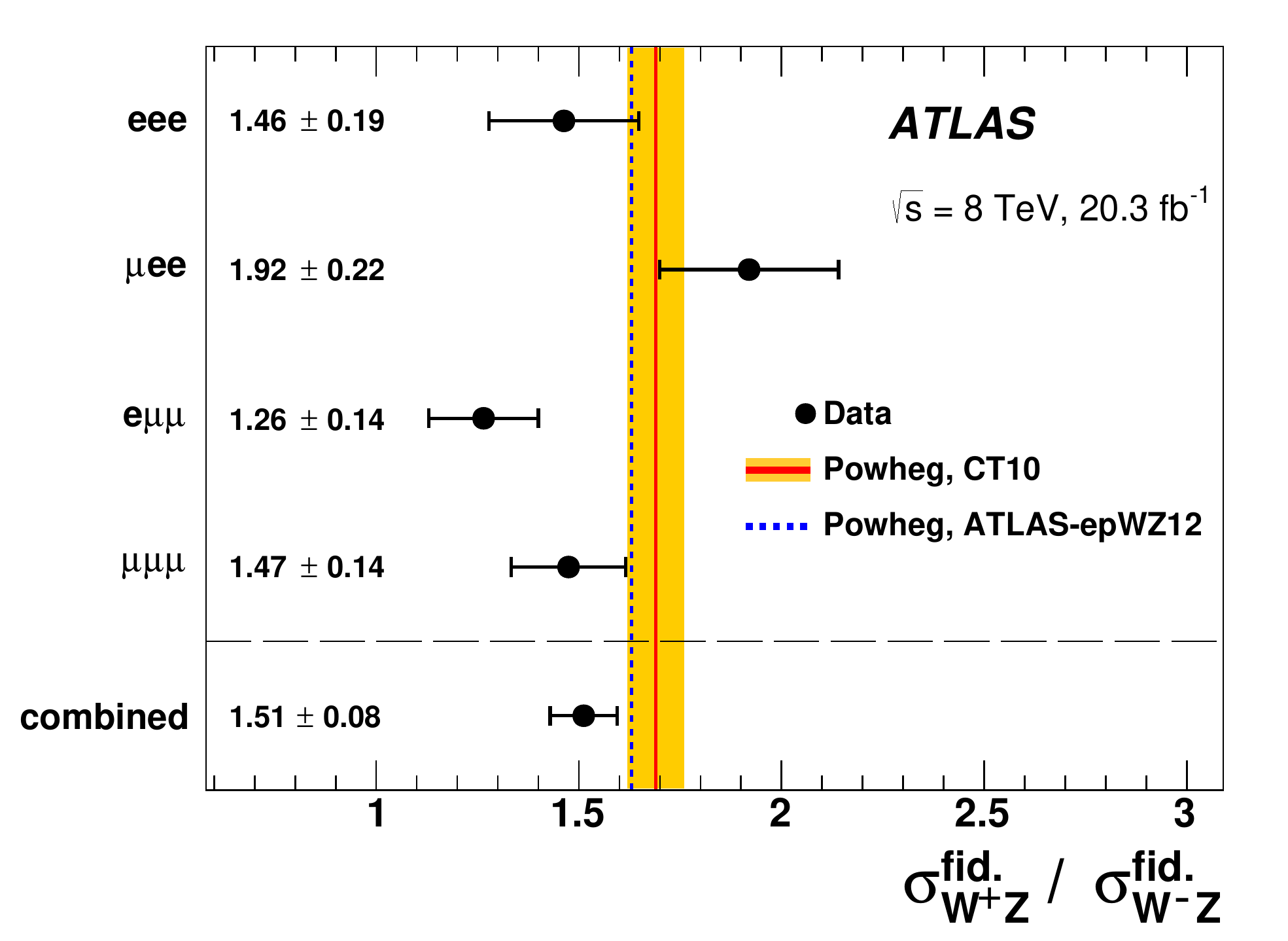}
\caption{Measured ratios $\sigma^{\mathrm{fid.}}_{W^{+}Z} / \sigma^{\mathrm{fid.}}_{W^{-}Z}$ of $W^{+}Z$ and $W^{-}Z$ integrated cross sections in the fiducial phase space in each of the four channels and for their combination.
The error bars on the data points represent the total uncertainties, dominated by statistical uncertainties.
The NLO SM prediction from \powhegpythia using the CT10 PDF set and renormalisation and factorisation scales $\mu_R = \mu_F = m_{WZ}/2$ is represented by the red line and the shaded orange band for the associated uncertainty.
The \powhegpythia prediction using the ATLAS-epWZ12 PDF set is also displayed as the dashed line.
}

\label{fig:WPMXSection:WpWmRatio}
\end{center}
\end{figure}

Finally, the combined fiducial cross section is extrapolated to a total phase space, defined by requiring
 that the invariant mass of the lepton  pairs associated with the $Z$ boson decay be in the range~
 $66 < m_{Z} < 116$~\GeV.
 The result is
\begin{eqnarray}
\sigma_{W^{\pm}Z}^{\textrm{tot.}} & = & 24.3 \pm 0.6 \,\textrm{(stat.)} \pm 0.6 \,\textrm{(sys.)} \pm 0.4 \,\textrm{(th.)} \pm 0.5 \,\textrm{(lumi.) \, pb} \, \nonumber,
\end{eqnarray}
where besides the statistical and systematic uncertainties a theory uncertainty (th.) has been included from the propagation of the theoretical uncertainty on $A_{WZ}$ to the total cross section.
The measurement is to be compared to the SM expectation calculated  with \powhegpythia of $21.0$ $\pm$ $1.6$ pb.

%-------------------------------------------------------------------------------
\subsection{Differential cross sections}
\label{sec:DiffCrossSections}

For the measurements of the differential distributions, all four decay channels, $eee$, $e\mu\mu$, $\mu ee$,
and $\mu\mu\mu$, are added together.
The resulting distributions are unfolded with a response matrix computed
using a \powhegpythia MC signal sample that includes all four topologies and divided by four such
that cross sections refer to final states where the $W$ and $Z$ decay in a single leptonic channel with muons or electrons.

\begin{figure}[!htbp]
\begin{center}
\includegraphics[width=.49\textwidth]{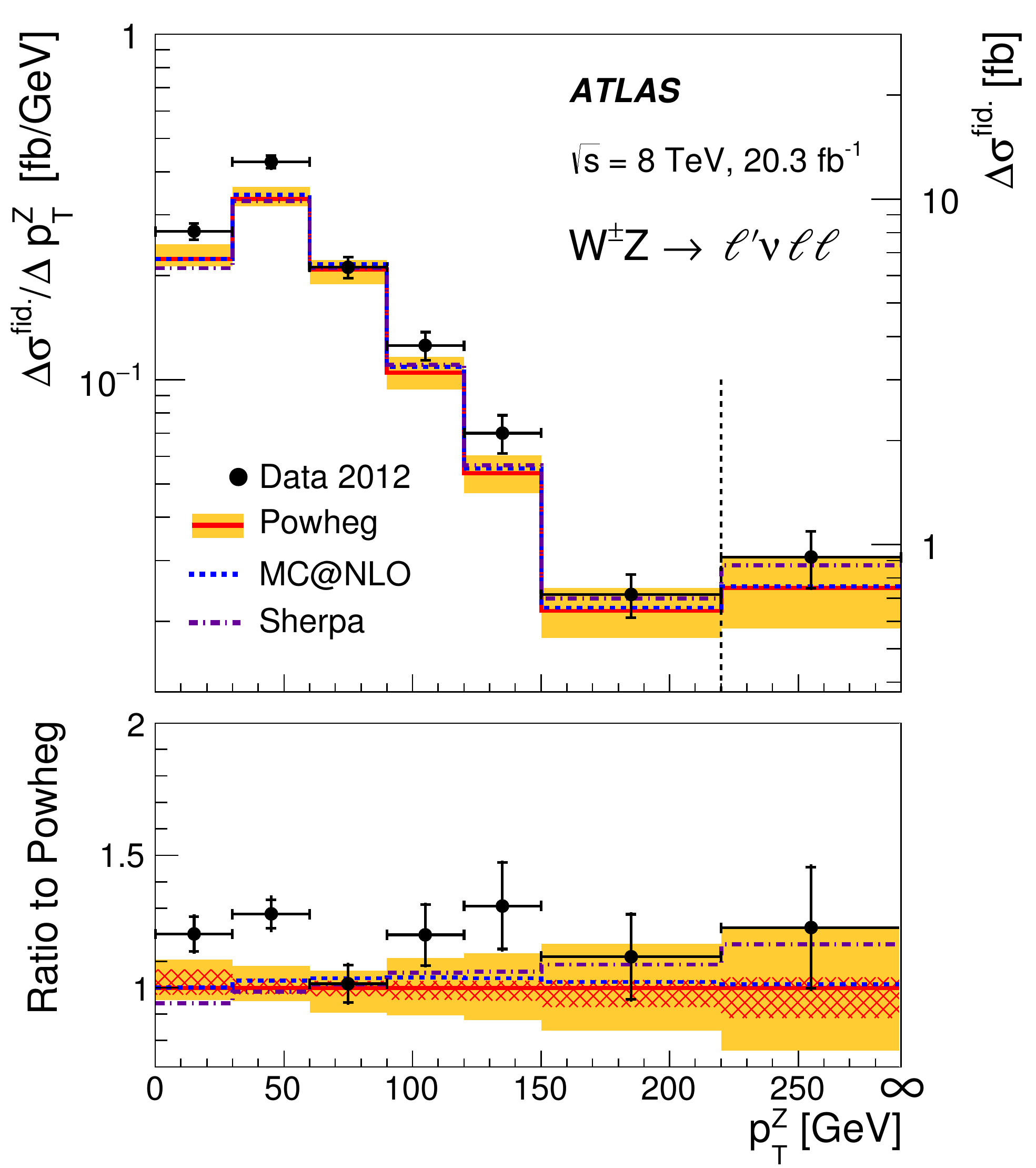}\put(-50,170){{(a)}}
\includegraphics[width=.49\textwidth]{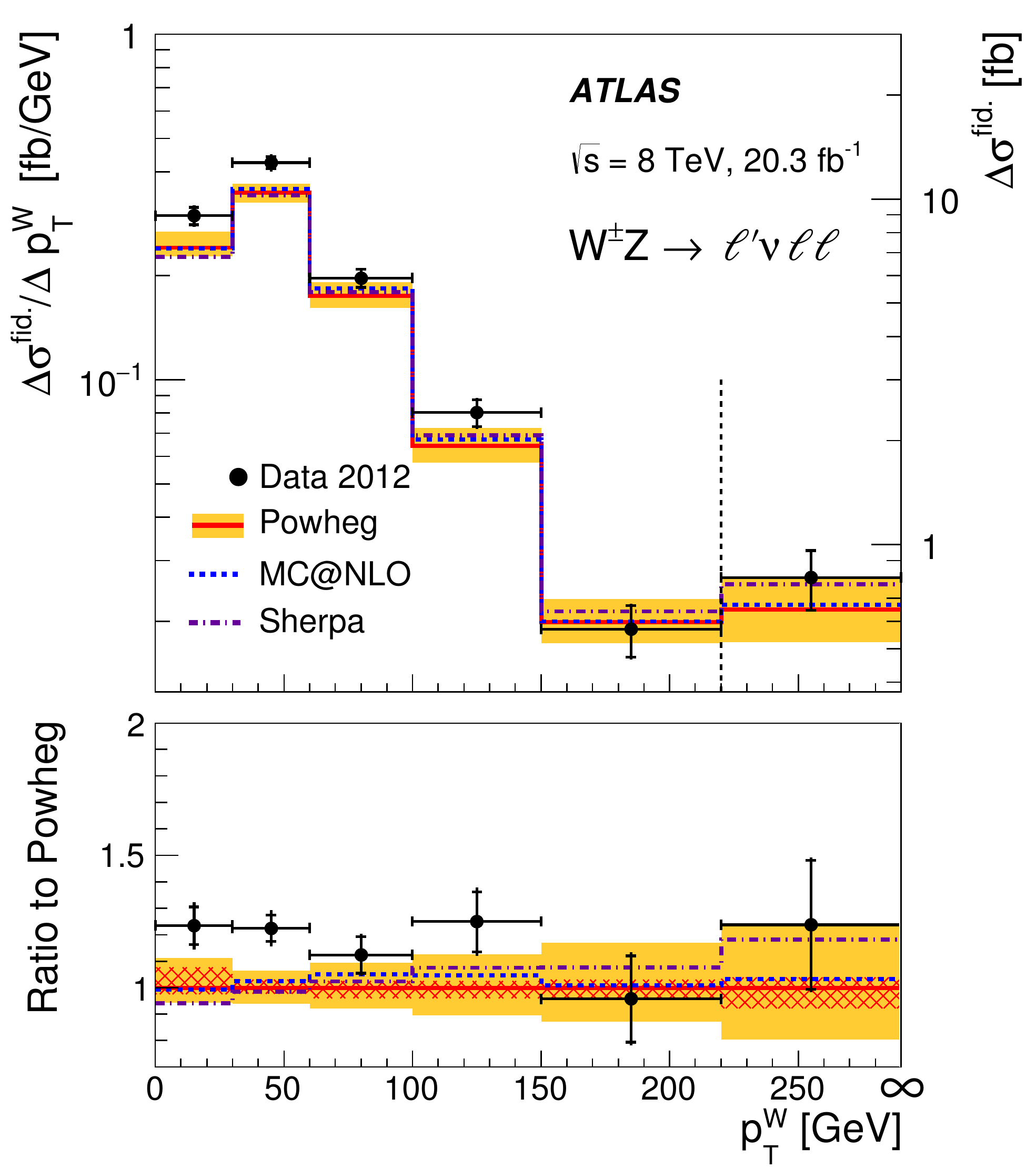}\put(-50,170){{(b)}}
\caption{The measured $W^{\pm}Z$ differential cross section in the fiducial phase space as a function of (a) $p_\textrm{T}^Z$  and (b) $p_\textrm{T}^W$.
The inner and outer error bars on the data points represent the statistical and total uncertainties, respectively.
The measurements are compared to the prediction from \powhegpythia (red line, see text for details).
The orange band represents its total theoretical uncertainty and the hatched red area the part of the theoretical uncertainty arising from the PDF and parton shower uncertainties.
The predictions from the \mcatnlo and \sherpa MC generators are also indicated by dashed and dotted-dashed lines, respectively.
The \sherpa prediction is rescaled to the integrated cross section predicted by \powhegpythia.
The right $y$-axis refers to the last cross-section point, separated from the others by a vertical dashed line, as this last bin is integrated up to the maximum value reached in the phase space. }
\label{fig:DiffXSection:pTZW}
\end{center}
\end{figure}

%%%%%%%%%  MTWZ
\begin{figure}[!htbp]
\begin{center}
\includegraphics[width=0.5\textwidth]{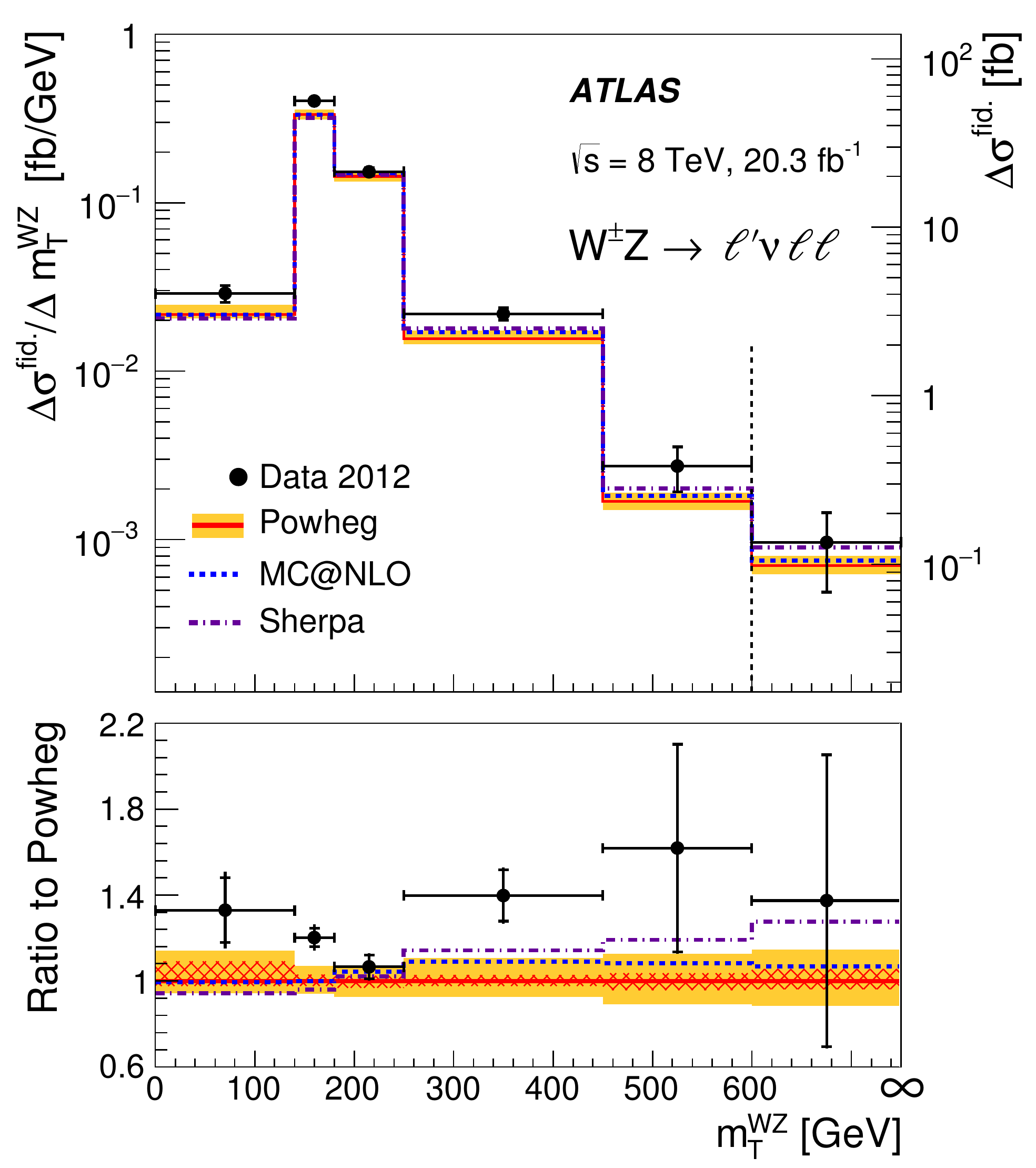}
\caption{The measured $W^{\pm}Z$ differential cross section in the fiducial phase space as a function of $m_\textrm{T}^{WZ}$.
The inner and outer error bars on the data points represent the statistical and total uncertainties, respectively.
The measurements are compared to the prediction from \powhegpythia (red line, see text for details).
The orange band represents its total theoretical uncertainty and the hatched red area the part of the theoretical uncertainty arising from the PDF and shower uncertainties.
The predictions from the \mcatnlo and \sherpa MC generators are also indicated by dashed and dotted-dashed lines, respectively.
The \sherpa prediction is rescaled to the integrated cross section predicted by \powhegpythia.
The right $y$-axis refers to the last cross-section point, separated from the others by a vertical dashed line, as this last bin is integrated up to the maximum value reached in the phase space.}
\label{fig:DiffXSection:mTWZ}
\end{center}
\end{figure}

 The \wz production cross section is  measured as a function of the transverse momentum
 of the $Z$ and $W$ boson, $p_\textrm{T}^Z$ and $p_\textrm{T}^W$ (Figure~\ref{fig:DiffXSection:pTZW}),
 as a function of the transverse mass of the \wz system $m_\textrm{T}^{WZ}$ (Figure~\ref{fig:DiffXSection:mTWZ}), as a function of
 the \pt of the neutrino associated with the decay of the $W$ boson, $p_\textrm{T}^\nu$,
  and as a function of the absolute difference between the rapidities of the $Z$ boson and the lepton
  from the decay of the $W$ boson, $|y_Z - y_{\ell,W}|$ (Figure~\ref{fig:DiffXSection:yZlW_all}).

The differential cross sections as a function of the transverse momenta of the neutrino or of the lepton from the $W$ decay are interesting because of their sensitivity to the polarization of the $W$ boson.
Experimentally, given the fiducial phase space of the measurement, the $p_\textrm{T}^\nu$ observable has the advantage of probing lower transverse momenta than
 the transverse momentum of the lepton from the $W$ boson decay, $p_\textrm{T}^{\ell,W}$,  which is restricted to values above $20$~\GeV.
Therefore, despite the worse experimental resolution for the reconstruction of $p_\textrm{T}^\nu$ compared to  $p_\textrm{T}^{\ell,W}$, $p_\textrm{T}^\nu$ could be more sensitive to polarization effects.

In order to derive the $p_\textrm{T}^\nu$ from data events, the assumption is made that the whole $E_\textrm{T}^\textrm{miss}$ of events arises from the neutrino of the $W$ boson decay.
Using MC samples, this assumption was verified to be valid for SM $WZ$ events.
The observed $E_\textrm{T}^\textrm{miss}$ distribution is therefore unfolded to $p_\textrm{T}^\nu$ using $WZ$ MC events.

Previously, no observable related to decay angles of final-state particles had been measured for $WZ$ events.
The rapidity correlations between the $W$ and $Z$ decay products have been found to be useful tools in searching for the approximately zero $WZ$ helicity amplitudes expected at LO in the SM or for aTGC~\cite{Baur:1994aj,Accomando:2005xp}.
These rapidity correlations are also sensitive to QCD corrections, PDF effects, and polarization effects of the $W$ and $Z$ bosons.
The rapidity difference between the $W$ and $Z$ bosons, $|y_Z - y_W|$, is a boost-invariant substitute for the center-of-mass scattering angle $\theta$ of the $W$ with respect to the direction of the incoming quark.
Since the rapidity of the $W$ boson cannot be uniquely reconstructed due to the presence of the neutrino, the rapidity of the lepton from the $W$ boson decay is used.
 Therefore the rapidity difference  $|y_Z - y_{\ell,W}|$ is measured instead of $|y_Z - y_W|$.

%%%%%%%%%  YZ-etaL
\begin{figure}[!htbp]
\begin{center}
\includegraphics[width=0.49\textwidth]{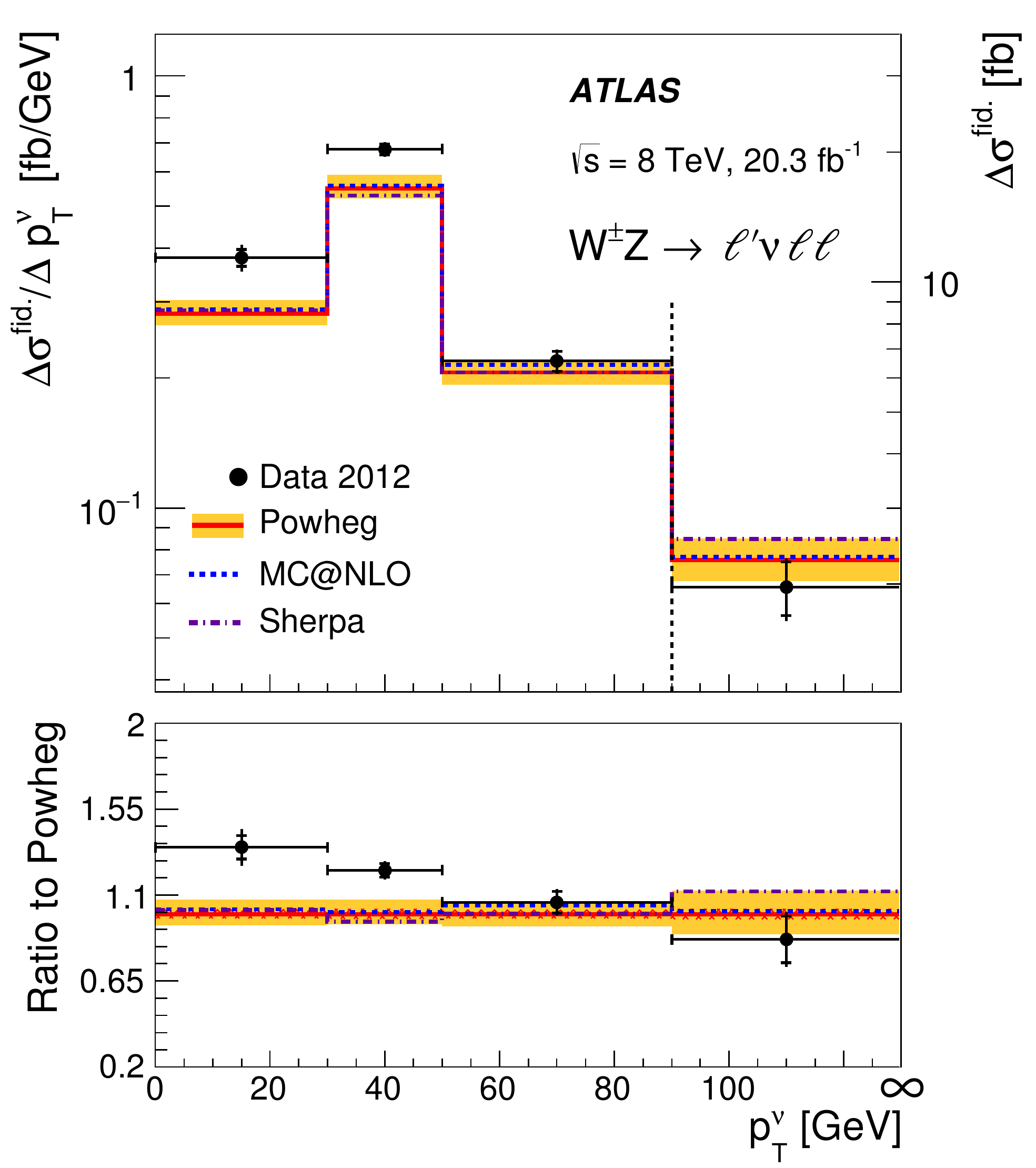}\put(-50,170){{(a)}}
\includegraphics[width=0.49\textwidth]{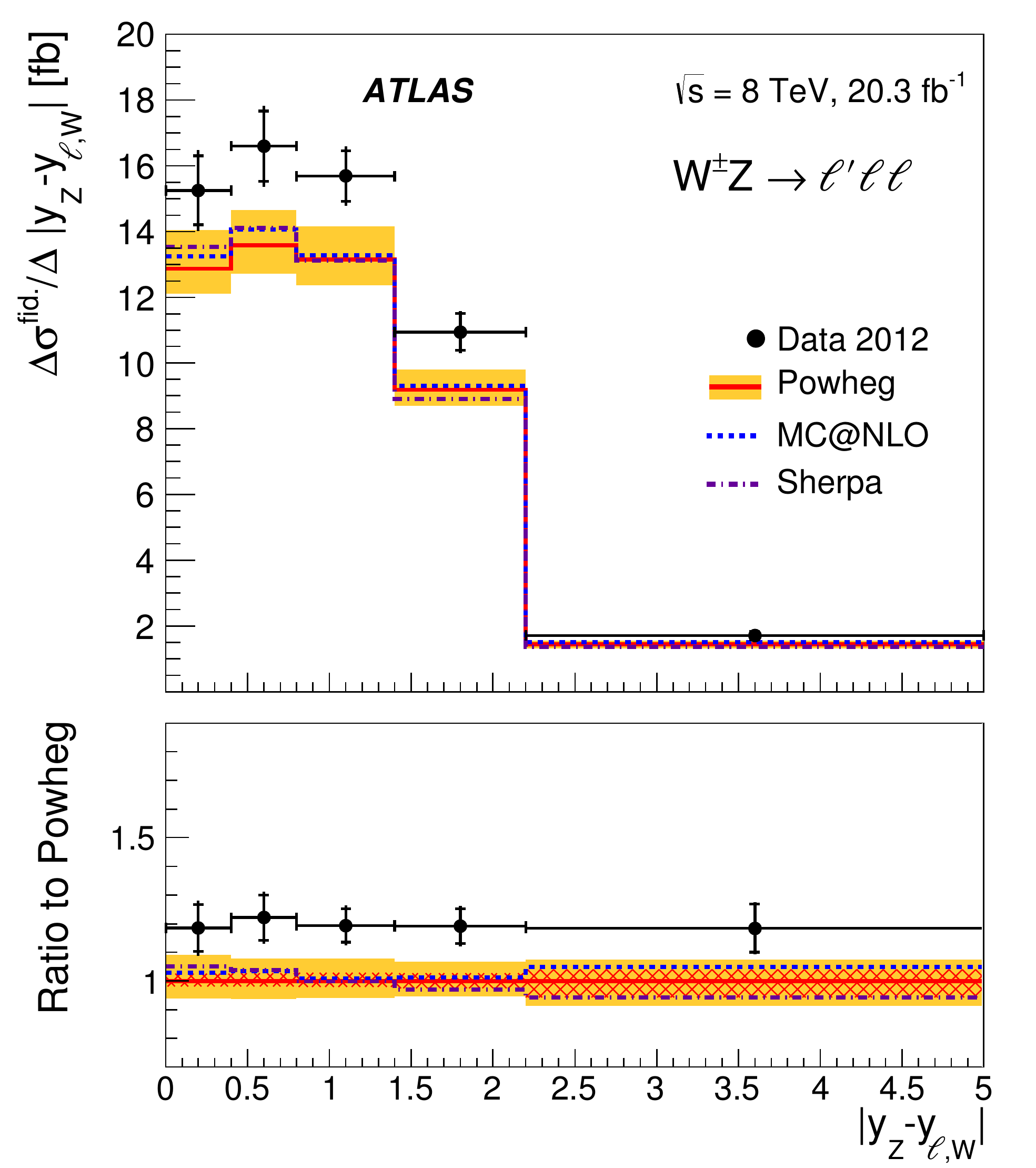}\put(-50,200){{(b)}}
\caption{
The measured $W^{\pm}Z$ differential cross section in the fiducial phase space as a function of (a) $p_{\textrm {T}}^\nu$  and (b) $|y_Z - y_{\ell,W}|$.
The inner and outer error bars on the data points represent the statistical and total uncertainties, respectively.
The measurements are compared to the prediction from \powhegpythia (red line, see text for details).
The orange band represents its total theoretical uncertainty and the hatched red area the part of the theoretical uncertainty arising from the PDF and shower uncertainties.
The predictions from the \mcatnlo and \sherpa MC generators are also indicated by dashed and dotted-dashed lines, respectively.
The \sherpa prediction is rescaled to the integrated cross section predicted by \powhegpythia.
The right $y$-axis in Figure 6(a) refers to the last cross-section point, separated from the others by a vertical dashed line, as this last bin is integrated up to the maximum value reached in the phase space.
}
\label{fig:DiffXSection:yZlW_all}
\end{center}
\end{figure}

  The $W^+Z/W^- Z$ ratio of the production cross sections is also measured as a function
  of  $p_\textrm{T}^Z$,  $p_\textrm{T}^W$, $m_\textrm{T}^{WZ}$, $p_\textrm{T}^\nu$, and $|y_Z - y_{\ell,W}|$ and presented in Figures~\ref{fig:DiffXSection:pTZW_ratio},~\ref{fig:DiffXSection:mTWZ_ratio}, and~\ref{fig:DiffXSection:yZlW_ratio}.

 The measured  differential cross sections are compared to the predictions from the \powhegpythia
  MC generator, which uses the CT10 PDF set and  dynamic QCD scales  of $\mu_\textrm{F}$ = $\mu_\textrm{R}$ = $m_{WZ}/2$.
 The theoretical uncertainties on the differential predictions from \powhegpythia arise from the choice of PDF set and QCD
 scales and are evaluated as explained in Section~\ref{sec:Theorypredictions}.
The total uncertainty on the theoretical predictions is estimated
as the linear sum of the PDF, parton shower, QCD scale and EW correction uncertainties, following the recommendations of Ref.~\cite{Dittmaier:2011ti}.
The measured cross-section distributions are also compared to predictions from the \mcatnlo and \SHERPA MC event generators.

Fair agreement of the shapes of measured distributions of inclusive cross sections and  $W^+Z/W^- Z$ cross section ratios with the different MC predictions is observed.
However, the precision of SM predictions of \wz\ production is limited to NLO and LO accuracy for perturbative QCD and EW effects, respectively.
New effects of higher perturbative orders could therefore potentially affect the present SM predictions, beyond the presently estimated theoretical uncertainties.
From the $p_\textrm{T}^\nu$ differential cross section in Figure~\ref{fig:DiffXSection:yZlW_all}(a) we observe that the global excess of the measured integrated cross section compared to the \powhegpythia prediction seems to be related to the region with $p_\textrm{T}^\nu < 50$~\GeV, this difference being more pronounced for $p_\textrm{T}^\nu < 30$~\GeV\ for $W^-Z$ events as seen in the first bin of Figure~\ref{fig:DiffXSection:yZlW_ratio}(a).

The exclusive multiplicity of jets unfolded at particle level
is presented in Figure~\ref{fig:unfoldResultNjetMjj}.
This distribution uses the same jet definition as for the VBS analysis (see Section~\ref{sec:ObjectReconstruction}) but with a lower jet
\pt threshold of $25$~\GeV\ at detector and at particle level.
The measurement is compared with predictions from \SHERPA and \powhegpythia.
The \SHERPA prediction provides a good description of the measured jet multiplicity while this is not the case for \powhegpythia and \mcatnlo.
Moreover, the ratio of $0$-jet to $1$-jet event cross sections predicted by \powhegpythia is lower than predicted by \SHERPA and than measured in data.
Finally, the measured $W^{\pm}Z$ differential cross section as a function of the invariant mass, $m_{jj}$, of the two leading  jets with $p_T > 30$~\GeV\ is presented in Figure~\ref{fig:unfoldResultMjj_notNorm}.
The measurement is better described by the \SHERPA prediction, which includes the sum of $WZjj$-QCD and $WZjj$-EW contributions.
The contribution of $WZjj$-EW events, which is increasing at higher $m_{jj}$ is exemplified in the figure.

%%%%%%% ratio ptZ, ptW
\begin{figure}[!htbp]
\begin{center}
\includegraphics[width=0.45\textwidth]{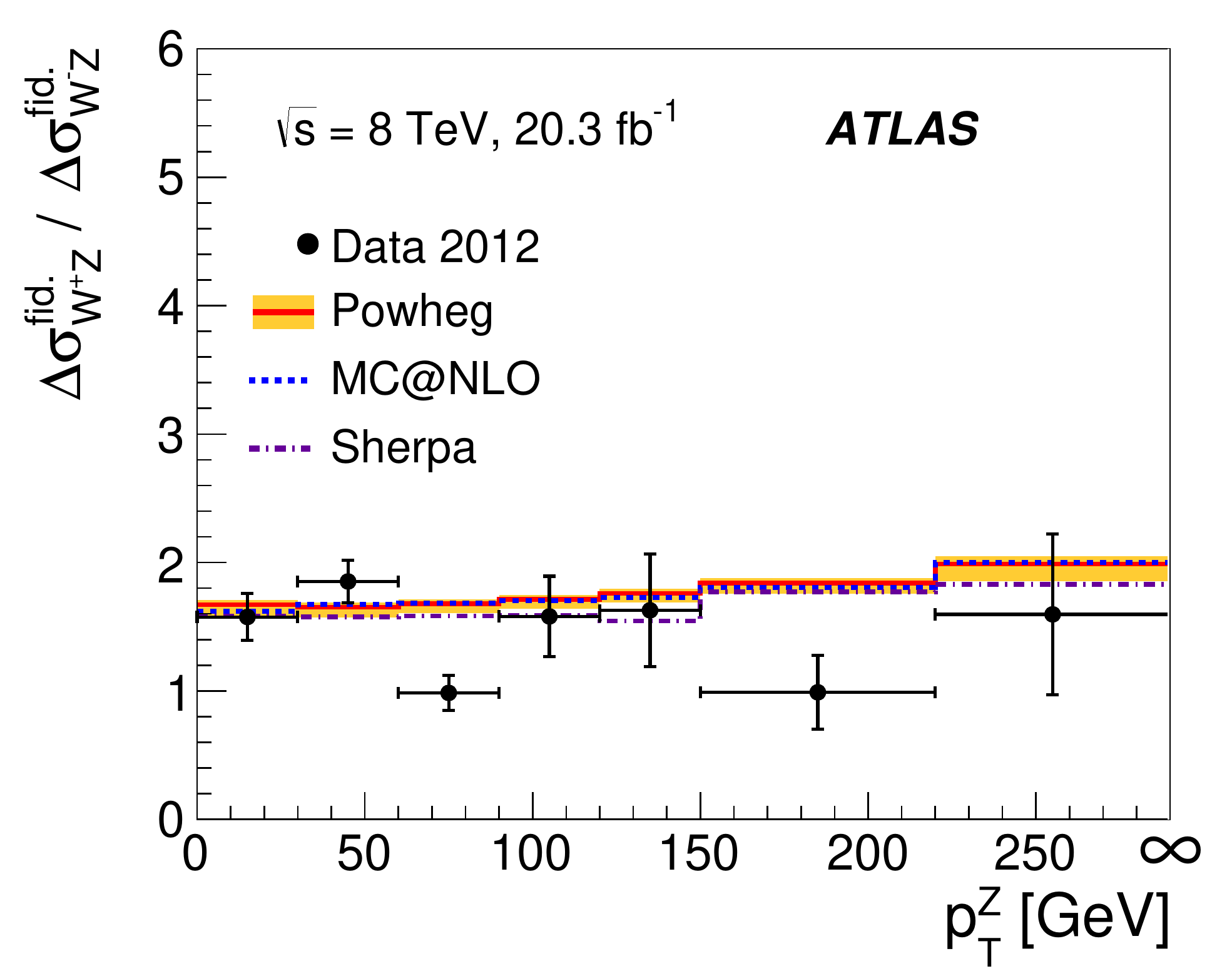} \put(-35,120){{(a)}}
\includegraphics[width=0.45\textwidth]{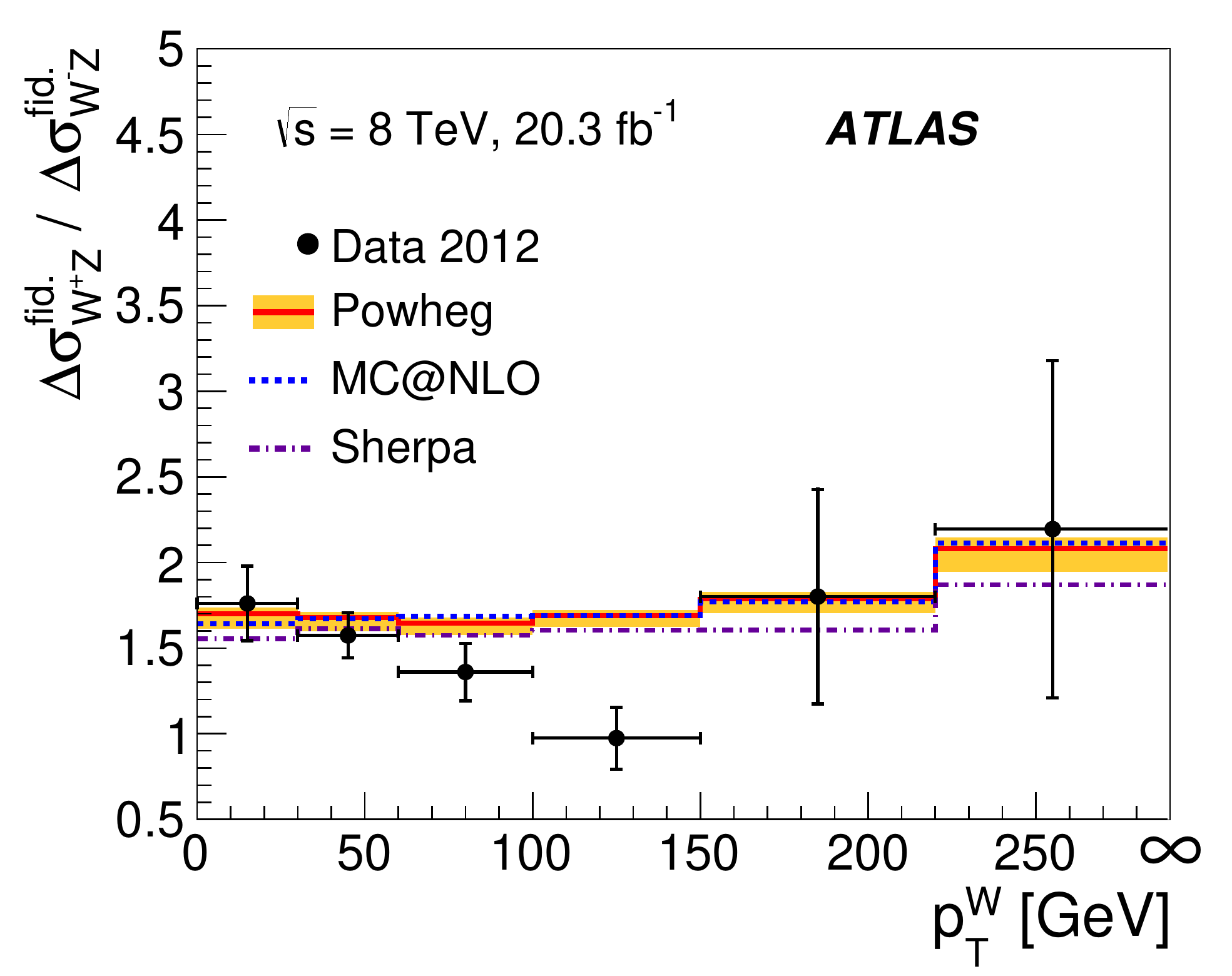} \put(-35,120){{(b)}}
\caption{The ratio of the $W^{+}Z$ and $W^{-}Z$ differential cross sections in the fiducial phase space as a function of (a) $p_\textrm{T}^Z$ and (b) $p_\textrm{T}^W$.
The inner and outer error bars on the data points represent the statistical and total uncertainties, respectively.
The measurements are compared to the prediction from \powhegpythia (red line, see text for details).
The orange band represents its total theoretical uncertainty, which is dominated by the PDF uncertainty.
The predictions from the \mcatnlo and \sherpa MC generators are also indicated by dashed and dotted-dashed lines, respectively.}
\label{fig:DiffXSection:pTZW_ratio}
\end{center}
\end{figure}

%%%%%% ratio MWZ
\begin{figure}[!htbp]
\begin{center}
\includegraphics[width=0.55\textwidth]{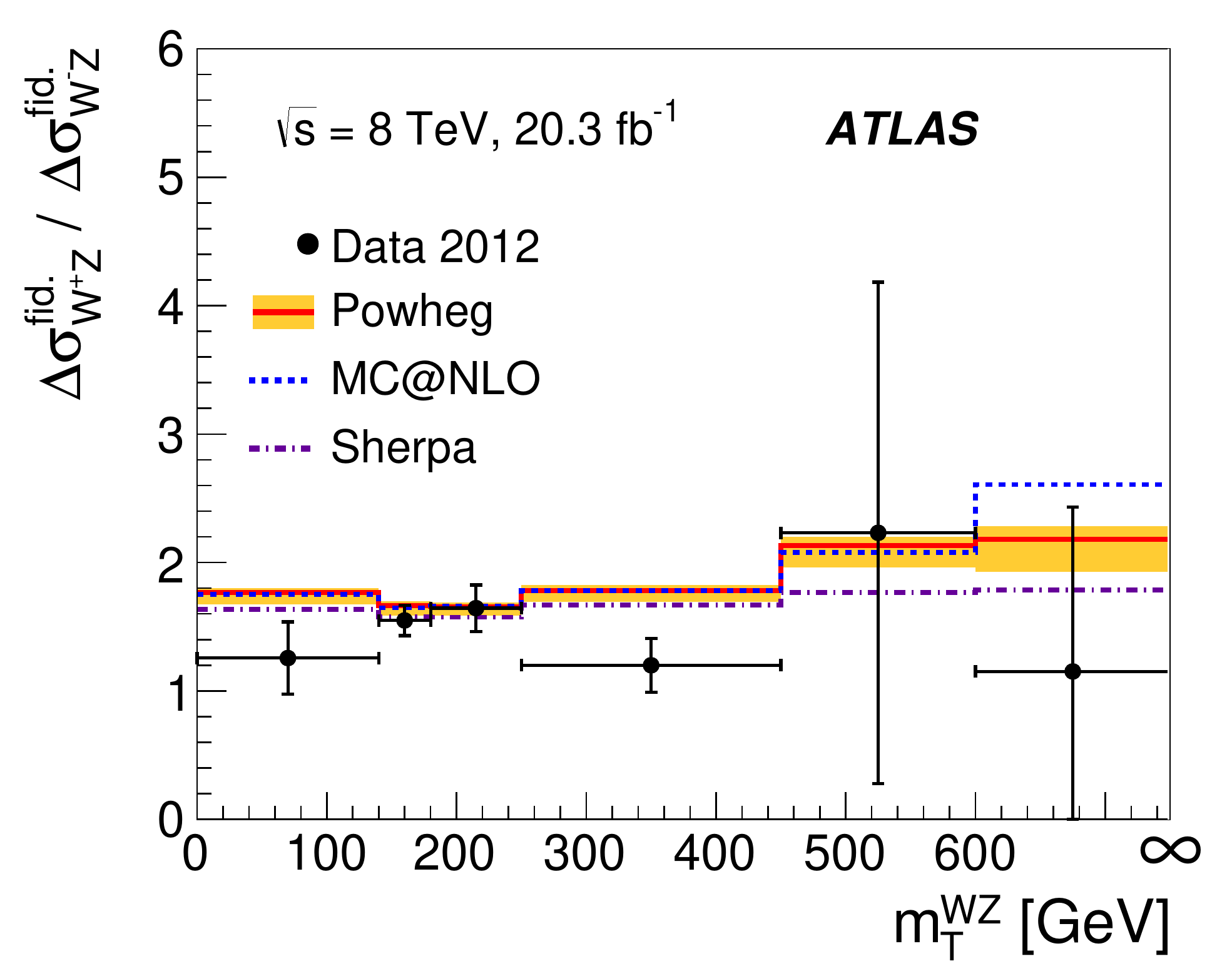}
\caption{The ratio of the $W^{+}Z$ and $W^{-}Z$ differential cross sections in the fiducial phase space as a function of $m_\textrm{T}^{WZ}$.
The inner and outer error bars on the data points represent the statistical and total uncertainties, respectively.
The measurements are compared to the prediction from \powhegpythia (red line, see text for details).
The orange band represents its total theoretical uncertainty, which is dominated by the PDF uncertainty.
The predictions from the \mcatnlo and \sherpa MC generators are also indicated by dashed and dotted-dashed lines, respectively.}
\label{fig:DiffXSection:mTWZ_ratio}
\end{center}
\end{figure}

%%%%%% ry
\begin{figure}[!htbp]
\begin{center}
\includegraphics[width=0.45\textwidth]{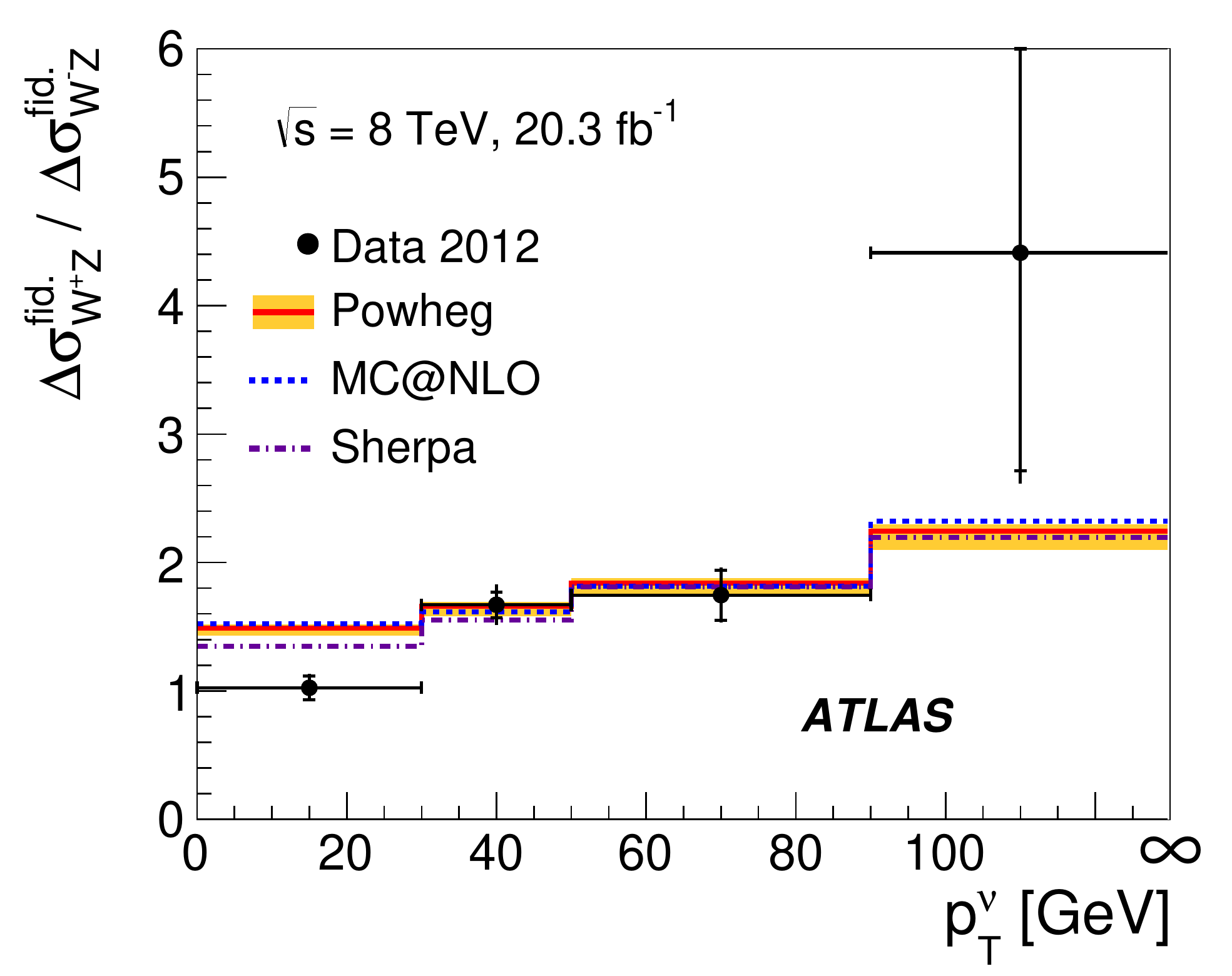}\put(-30,140){{(a)}}
\includegraphics[width=0.45\textwidth]{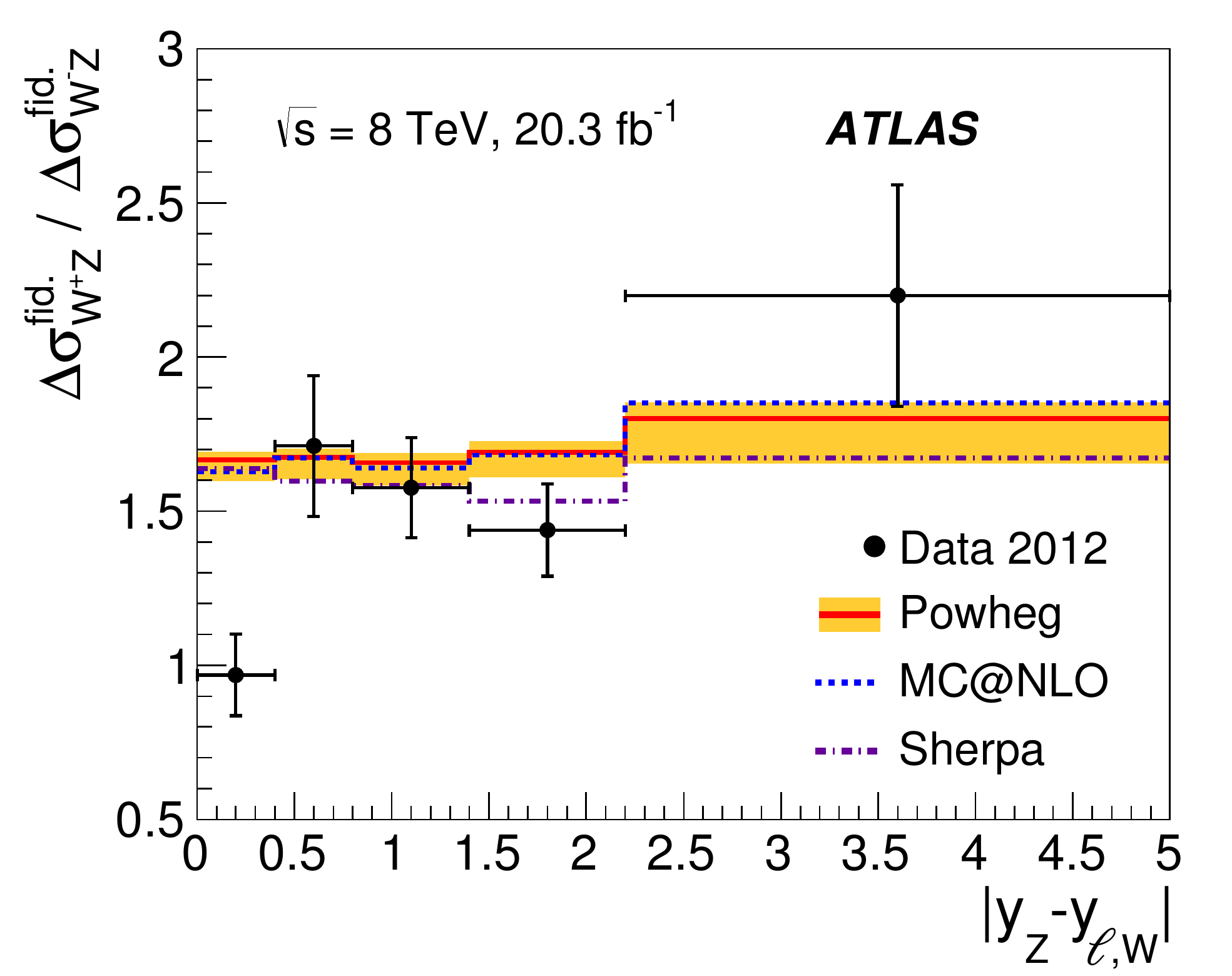} \put(-30,120){{(b)}}
\caption{The ratio of the $W^{+}Z$ and $W^{-}Z$ differential cross sections in the fiducial phase space as a function of (a) $p_{\textrm{T}}^\nu$
and (b) $|y_Z - y_{\ell,W}|$.
The inner and outer error bars on the data points represent the statistical and total uncertainties, respectively.
The measurements are compared to the prediction from \powhegpythia (red line, see text for details).
The orange band represents its total theoretical uncertainty, which is dominated by the PDF uncertainty.
The predictions from the \mcatnlo and \sherpa MC generators are also indicated by dashed and dotted-dashed lines, respectively.
}
\label{fig:DiffXSection:yZlW_ratio}
\end{center}
\end{figure}

%%%%%  Njet and MJJ
\begin{figure}[!htbp]
\begin{center}
\includegraphics[width=0.5\textwidth]{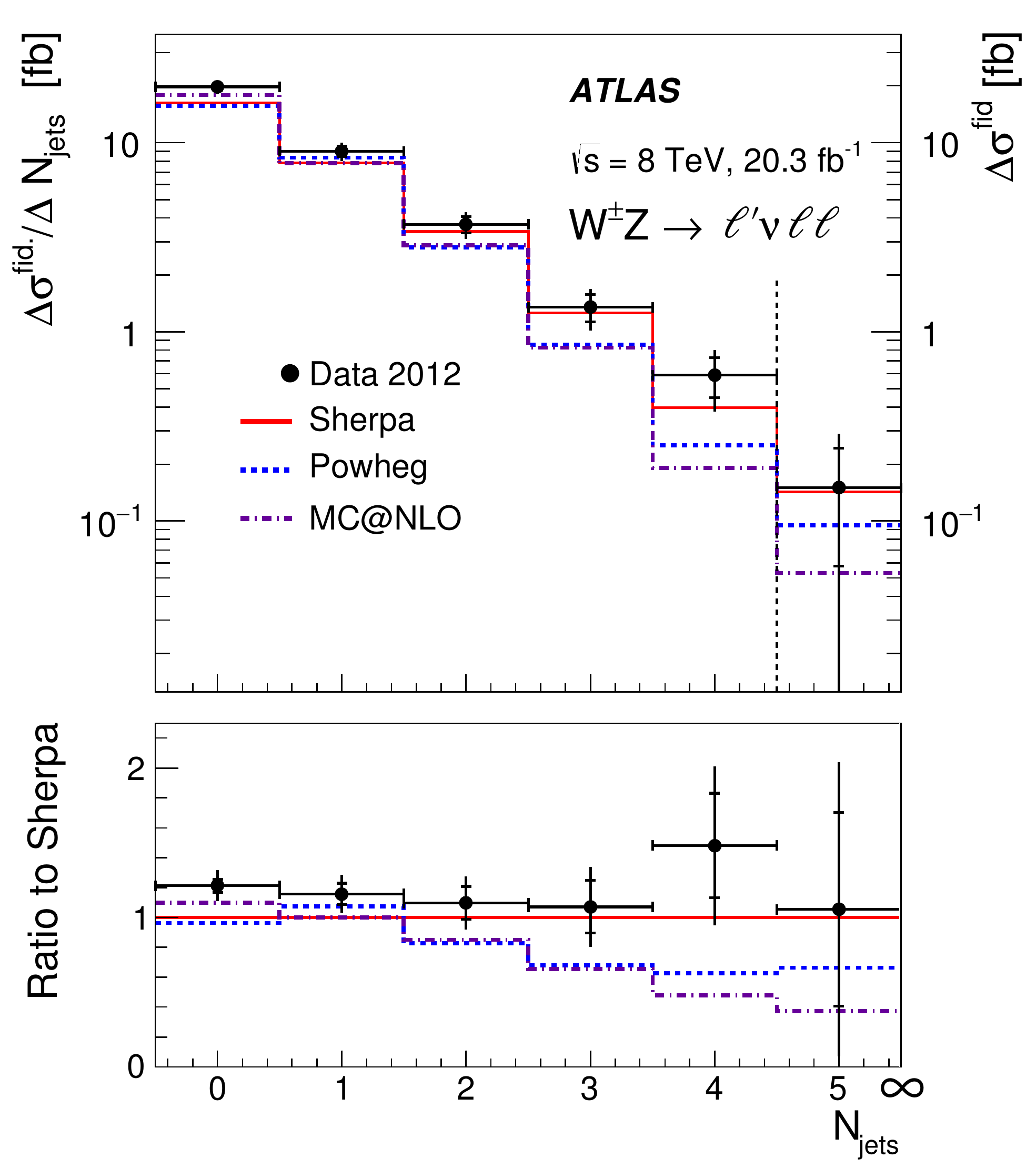}
\caption{The measured $W^{\pm}Z$ differential cross section in the fiducial phase space as a function of the exclusive jet multiplicity of jets with $p_\textrm{T} > 25$~\GeV.
The inner and outer error bars on the data points represent the statistical and total uncertainties, respectively.
The measurements are compared to the prediction from \sherpa (red line), \powhegpythia (dashed blue line)  and \mcatnlo (dotted-dashed violet line).
The \sherpa prediction is rescaled to the integrated cross section predicted by \powhegpythia.
The right $y$-axis refers to the last cross section point, separated from the others by a vertical dashed line, as this last bin is integrated up to the maximum value reached in the phase space.
}
\label{fig:unfoldResultNjetMjj}
\end{center}
\end{figure}

\begin{figure}[!htbp]
\begin{center}
\includegraphics[width=0.5\textwidth]{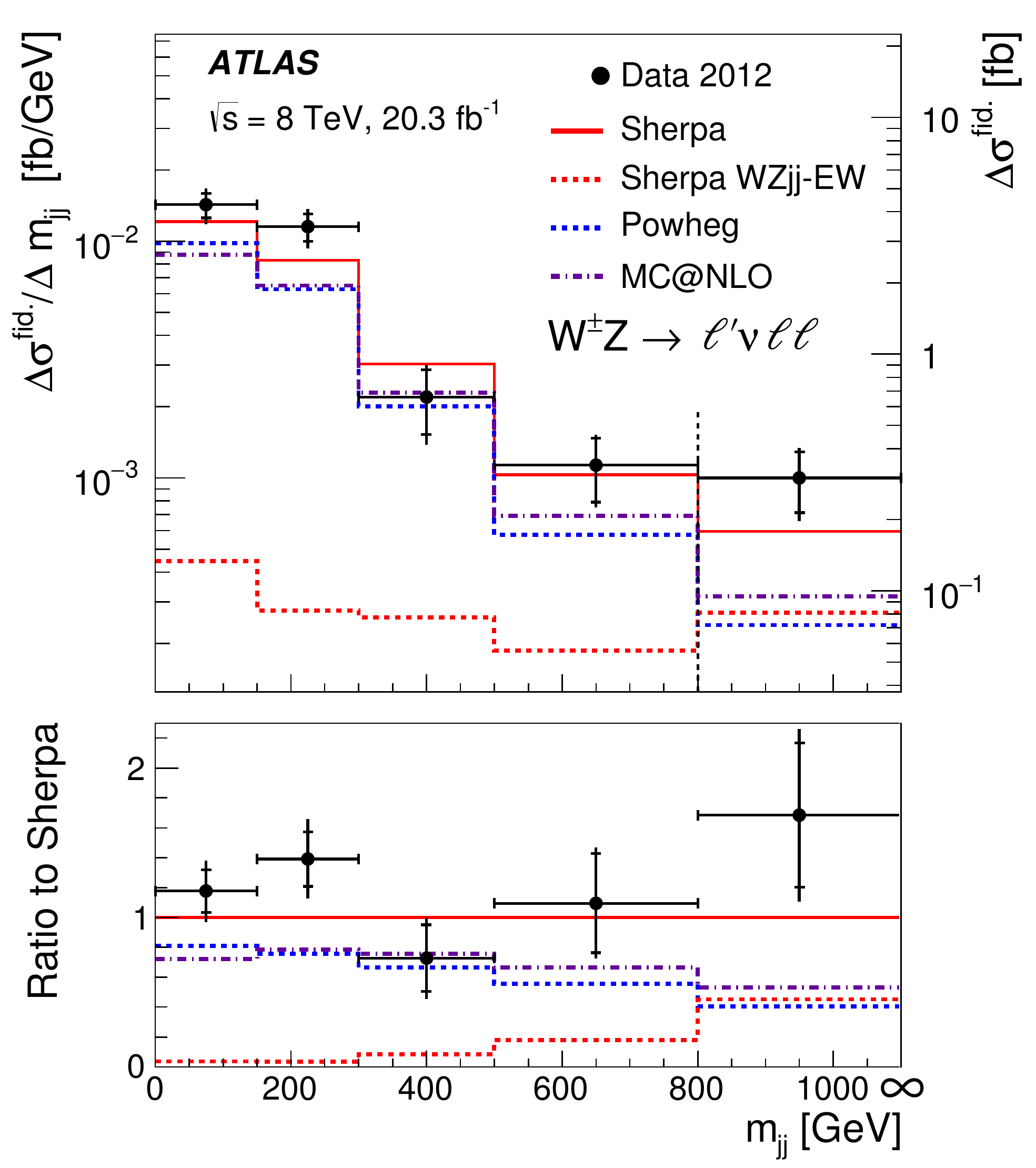}
\caption{The measured $W^{\pm}Z$ differential cross section as a function of the invariant mass of the two leading  jets with $p_\textrm{T} > 30$~\GeV.
The inner and outer error bars on the data points represent the statistical and total uncertainties, respectively.
The measurements are compared to the prediction from \sherpa (red line), which includes both the $WZjj$-QCD and $WZjj$-EW processes, \powhegpythia (dashed blue line) and \mcatnlo (dotted-dashed violet line).
The part of the \sherpa prediction corresponding to $WZjj$-EW events is also represented by a dashed red line.
In the bottom panel the dashed red line therefore corresponds to the $WZjj$-EW fraction of the total \sherpa prediction.
The right $y$-axis refers to the last cross section point, separated from the others by a vertical dashed line, as this last bin is integrated up to the maximum value reached in the phase space.
}
\label{fig:unfoldResultMjj_notNorm}
\end{center}
\end{figure}

%-------------------------------------------------------------------------------
\subsection{Limits on vector boson scattering production}
\label{sec:VBS}

This part of the analysis aims to study $WZjj$-EW production, which includes VBS and $tZj$ processes.
The latter process results from a t-channel exchange of a $W$ boson  between a $b$ and a $u$-quark giving a final state with a $t$-quark, a $Z$ boson and a light quark jet, but does not exhibit diagrams with gauge boson couplings.
 Its contribution in the \SHERPA $WZjj$-EW sample is disentangled  from the VBS part, considered in this paper
 as the signal, with a splitting procedure relying on the presence of $b$-quarks at generator level.
Interference effects between the signal and the $tZj$ process are expected to be  negligible.
Since the $b$-tagged sample, enriched in $tZj$ events, still  contains a small fraction
  ($\sim 5\%$) of signal events from the scattering of the initial-state $b$-quark, two results
 with or without subtraction of the $tZj$ contribution, are provided.

Given a too-low expected statistical significance for a cross-section measurement,
the experimental result is reported as an upper limit at $95\%$ CL on the
 fiducial cross section multiplied by the $W$ and $Z$ branching ratios in a single leptonic channel
 with muons or electrons.
 Observed and expected upper limits are calculated using the numbers of observed and expected events, the estimated number of background events (see Table~\ref{tab:vbs_yields}),
  the luminosity of the data sample, the detector and reconstruction efficiencies of $\sim 67\%$, and are presented in Table~\ref{tab:vbs_xsec_results}.
Similar upper limits on the $\sigma^{\mathrm{fid.}}_{W^{\pm} Zjj \textrm{-EW} \rightarrow \ell^{'} \nu \ell \ell}$ production cross section are measured in the aQGC phase-space and shown in Table~\ref{tab:vbs_xsec_results}.
The measured upper cross-section limits are within $1$~$\sigma$ and $2$~$\sigma$ uncertainty on the expected limit for the VBS and aQGC phase-space measurements, respectively.
In the VBS phase-space, the number of observed data events corresponds to a cross section for $WZjj$-EW production of $0.29 \, ^{+0.14}_{-0.12} \, \textrm{(stat.)} \, ^{+0.09}_{-0.1} \, \textrm{(sys.)}$~fb, to be compared to the SM expectation of $0.13 \pm 0.01$~fb from VBFNLO.

\begin{table}
\begin{center}
\begin{tabular}{rcc}
\hline
  \multicolumn{3}{c}{$95\%$ CL upper limit  on $\sigma^{\mathrm{fid.}}_{W^{\pm} Zjj \textrm{-EW} \rightarrow \ell^{'} \nu \ell \ell}$ [fb]}\\[1ex]
\hline
  & VBS only & VBS + $tZj$ \\
\hline
  \multicolumn{3}{c}{VBS phase space}\\
\hline
Observed & $0.63$  & $0.67$ \\
Expected  & $0.45$  & $0.49$ \\
$\pm 1 \sigma$ Expected  & $[0.28 \,; 0.62]$  & $[0.33 \,; 0.67]$ \\
$\pm 2 \sigma$ Expected  & $[0.08 \,; 0.80]$  & $[0.19 \,; 0.84]$ \\
\hline
  \multicolumn{3}{c}{aQGC phase space}\\
\hline
Observed & $0.25$  & $0.25$ \\
Expected  & $0.13$  & $0.13$ \\
$\pm 1 \sigma$ Expected  & $[0.08 \,; 0.20]$  & $[0.08 \,; 0.20]$ \\
$\pm 2 \sigma$ Expected  & $[0.04 \,; 0.28]$  & $[0.06 \,; 0.28]$ \\
\hline
\end{tabular}
\end{center}
\caption{Observed and expected upper limits at $95\%$ CL in fb on the  fiducial cross section $\sigma^{\mathrm{fid.}}_{W^{\pm} Zjj \textrm{-EW} \rightarrow \ell^{'} \nu \ell \ell}$~, multiplied by the $W$ and $Z$ branching ratios in a single leptonic channel
 with muons or electrons in the VBS and aQGC fiducial phase space. Values obtained with or without subtraction of the $tZj$ contribution are presented.
The $1$~$\sigma$ and $2$~$\sigma$ uncertainty intervals around the expected limits are also indicated.}
\label{tab:vbs_xsec_results}
\end{table}

%-------------------------------------------------------------------------------
\section{Anomalous triple gauge couplings}
\label{sec:aTGC}

To extract the aTGC, two model-independent parameterizations of possible effects beyond the SM
are followed.
The first makes use of an effective Lagrangian describing the $WWZ$ vertex and
includes only terms that separately conserve the charge conjugation (C) and parity (P)
quantum numbers~\cite{Hagiwara, EllisonWudka}.
The deviation of the vector boson $WWZ$ couplings from the SM predicted values
are introduced as dimensionless anomalous couplings $\Delta \kappa^Z$, $\Delta g^Z_1$, and $\lambda^Z$.

Without effects not described by the SM, the anomalous terms cause a violation of the unitarity
bound in the interaction amplitudes.
To prevent this violation, the anomalous couplings are introduced as
form factors dependent on the partonic center-of-mass energy, $\hat{s}$:
 $\alpha(\hat{s}) = \alpha(0)/ (1+ \hat{s}/ \Lambda_{\mathrm{co}}^2)^2 $, where $\alpha(0)$ is the generic anomalous coupling value
 at low energy  and $\Lambda_{\mathrm{co}}$ is a cutoff scale at which physics effects beyond the SM should manifest.

The second parameterization is based on an effective field theory (EFT) in which the particle content of the SM is not changed and
 the theory is extended by adding to the SM Lagrangian a linear combination of operators of mass dimension higher than four~\cite{Buchmuller:1985jz, EFT}.
 The dimension-six operators are expected to be dominant. There are three independent dimension-six C- and P-conserving
 operators that affect the electroweak vector boson self-interactions and that can lead to anomalous triple vector boson couplings.
  The corresponding new terms in the Lagrangian are

\begin{align}
\mathcal{O}_{WWW} &= {\frac{c_{WWW}}{\Lambda^2} } \mathrm{Tr}[{W_{\mu\nu}W^{\nu\rho}W^{\mu}_{\rho}}] \, ,\nonumber \\
\mathcal{O}_{W} &=  {\frac{c_{W}}{\Lambda^2} }  \left(D_{\mu} \Phi \right)^{\dagger}W^{\mu\nu}\left(D_{\nu} \Phi \right) \, ,\nonumber \\
\mathcal{O}_{B} &=  {\frac{c_{B}}{\Lambda^2} } \left(D_{\mu} \Phi \right)B^{\mu\nu}\left(D_{\nu} \Phi \right) \, ,
\end{align}

where $W_{i j}, W^{ij}, W^i_j (i= \mu, \nu, j= \nu, \rho ) $, and $B^{\mu \nu}$ are built from the SM electroweak gauge boson fields,
 $D_i (i=\mu,\nu)$ are the covariant derivatives as introduced in the SM, and $\Phi$ is the Higgs doublet field.
The dimensionless coefficients $c_i (i= WWW, W, B)$  and $\Lambda$ represent the strength of the new couplings
 and the energy scale of new physics, respectively. This approach does not require the introduction of arbitrary
 form factors to restore unitarity.

The effective field theory allows the anomalous couplings to be reinterpreted in
terms of the EFT parameters, $c_i/\Lambda^2 (i= WWW, W, B)$~\cite{EFT1}.
For this reason the two parameterizations can be considered equivalent.
They are both used in this analysis because the first allows a comparison with previous analyses and the second is a
 flexible way of parameterizing effects beyond the SM in a model-independent way.
Therefore, the free parameters considered in this analysis are
  $\Delta \kappa^Z$, $\Delta g^Z_1$, and $\lambda^Z$ or $c_i/\Lambda^2 (i= WWW, W, B)$.

\begin{figure}[!htbp]
\begin{center}
 \includegraphics[width=0.5\textwidth]{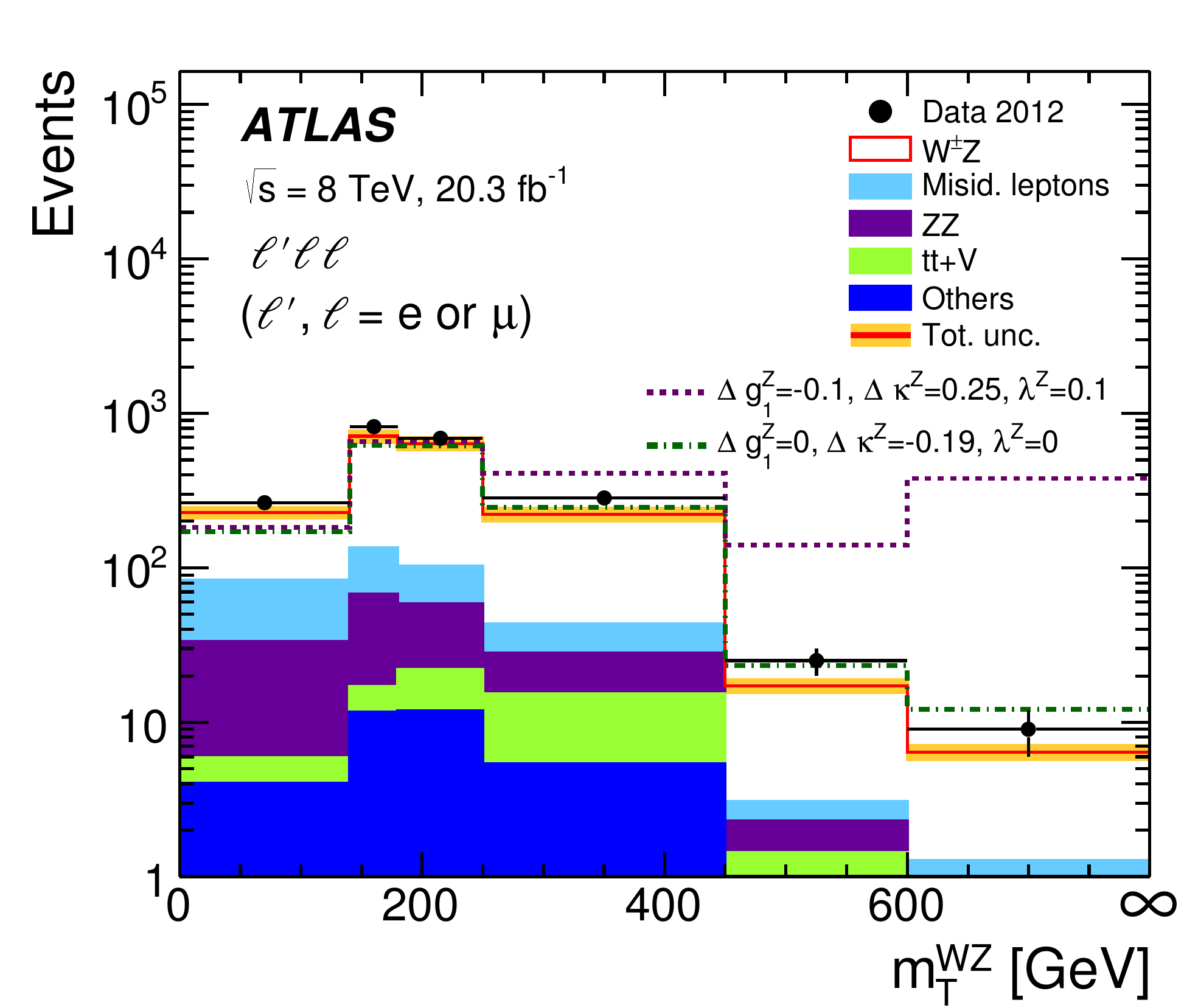}
 \caption{Distribution of $m_\textrm{T}^{WZ}$ in the sum of all channels with the same binning as used for the calculation of limits on aTGC.
The points correspond to the data and the histograms to the expectations of the different SM processes.
All Monte Carlo expectations are scaled to the integrated luminosity of the data using the predicted MC cross sections of each sample.
The \powhegpythia MC prediction is used for the SM \wz\ signal contribution.
The open red histogram shows the total prediction and the shaded orange band its
estimated total uncertainty.
The last bin contains the overflow.
Two predictions with nonzero values of some of the anomalous coupling parameters are also represented by the dashed and dotted-dashed lines, respectively.
 }
 \label{fig:mTWZ_and_aTGCs}
 \end{center}
 \end{figure}

The presence of aTGC would affect the \wz integrated cross section and manifest itself as an increased yield of events at
high values of $p_\textrm{T}^Z$ or $m_\textrm{T}^{WZ}$.
Limits on the aTGC are extracted from the $m_\textrm{T}^{WZ}$ differential distribution at detector level, as presented in Figure~\ref{fig:mTWZ_and_aTGCs}.
The $m_\textrm{T}^{WZ}$ distribution is expected to be less sensitive to higher-order QCD and EW effects in perturbation theory (as discussed in Section~\ref{sec:Theorypredictions}).
For this reason it has smaller theoretical uncertainties than the $p_\textrm{T}^Z$ distribution  at high values
 and provides more stringent expected limits, as proven by a dedicated MC study.

The MC event generator \mcatnlo is used to generate \wz events
and to compute, for each event, a set of weights that are employed to reweight the SM sample to any chosen value of the
anomalous couplings, or EFT coefficients.
With this procedure, expected $m_\textrm{T}^{WZ}$ distributions are obtained for different values of the anomalous
couplings, or EFT coefficients. This reweighting procedure is validated by comparing the SM sample reweighted to a
given set of aTGC values with a sample generated using the same set of aTGC values. A global systematic
uncertainty of $10\%$ across all $m_\textrm{T}^{WZ}$ bins was included in the aTGC limit extraction procedure to account for
the reweighting method.

Frequentist confidence intervals on the anomalous coupling are computed by forming a profile likelihood test
that incorporates the observed and expected numbers of signal events  in each bin of the $m_\textrm{T}^{WZ}$ distribution
for different values of the anomalous couplings. The systematic uncertainties are included in the likelihood function
as nuisance parameters.

Table~\ref{table:limObs} presents the observed and expected one-dimensional intervals at $95\%$ CL on
$\Delta \kappa^Z$, $\Delta g^Z_1$, and $\lambda^Z$  with the cutoff scale $\Lambda_{\mathrm{co}} =2$~\TeV,  $\Lambda_{\mathrm{co}} = 15$~\TeV\ and $\Lambda_{\mathrm{co}} = \infty$ (no cutoff).
Each limit is obtained by setting the other two couplings to the SM value.
The  $\Lambda_{\mathrm{co}}$ value of $15$~\TeV\ is the largest form factor scale that can preserve unitarity for all aTGC in this analysis.

\begin{table}[!htbp]
	 \centering
       	 \begin{tabular}{ccccc}
	 \hline
	 \noalign{\smallskip}
	 $\Lambda_{\mathrm{co}}$  & Coupling & Expected  & Observed \\
	 	 \noalign{\smallskip}
		% &  $\Lambda = 100 $ TeV &  $\Lambda = 100 $ TeV \\
		% \noalign{\smallskip}

	 \hline
  	\noalign{\smallskip}                              %  Expected    & Observed
	\multirow{3}{*}{$2$ TeV} & $\Delta g^Z_1$                   &    [$-0.023$ ; $0.055$]  & [ $-0.029$ ; $0.050$] \\
	\noalign{\smallskip}
       &  $\Delta \kappa^Z$                                                 &    [$-0.22$ ; $0.36$]  & [ $-0.23$ ; $0.46$] \\     
        	 \noalign{\smallskip}
       & $\lambda^Z$                                                        &    [$-0.026$ ; $0.026$]  & [ $-0.028$ ; $0.028$] \\
         \noalign{\smallskip}
         \hline
	\noalign{\smallskip}
         	\multirow{3}{*}{ $15$ TeV} & $\Delta g^Z_1$           &    [$-0.016$ ; $0.033$]  & [ $-0.019$ ; $0.029$] \\
	\noalign{\smallskip}
       &  $\Delta \kappa^Z$                                 &    [$-0.17$ ; $0.25$]  & [ $-0.19$ ; $0.30$] \\
        	 \noalign{\smallskip}
       & $\lambda^Z$                                        &    [$-0.016$ ; $0.016$]  & [ $-0.017$ ; $0.017$] \\
         \noalign{\smallskip}
         \hline
	\noalign{\smallskip}
	\multirow{3}{*}{ $ \infty$} & $\Delta g^Z_1$                 &    [$-0.016$ ; $0.032$]  & [$-0.019$ ; $0.029$] \\
	\noalign{\smallskip}
       &  $\Delta \kappa^Z$                                 &    [$-0.17$ ; $0.25$]    & [$-0.19$ ; $0.30$] \\
        	 \noalign{\smallskip}
       & $\lambda^Z$                                        &    [$-0.016$ ; $0.016$]  & [$-0.016$ ; $0.016$] \\
         \noalign{\smallskip}
	 \hline
	 \end{tabular}

	 \caption{Expected and observed one-dimensional $95\%$ CL intervals on the anomalous coupling parameters.
}
\label{table:limObs}
\end{table}

Expected and observed $95\%$ CL limit contours in the planes ($\Delta \kappa^Z$, $\Delta g^Z_1$), ($\Delta g^Z_1$, $\lambda^Z$), and ($\Delta \kappa^Z$,  $\lambda^Z$) are shown in Figure~\ref{fig:tgc_2D}.
For each of the contours, the third parameter is set to the SM value and the limits are derived without any cut-off.

\begin{figure}[!htbp]
\begin{center}

 \includegraphics[width=0.33\textwidth]{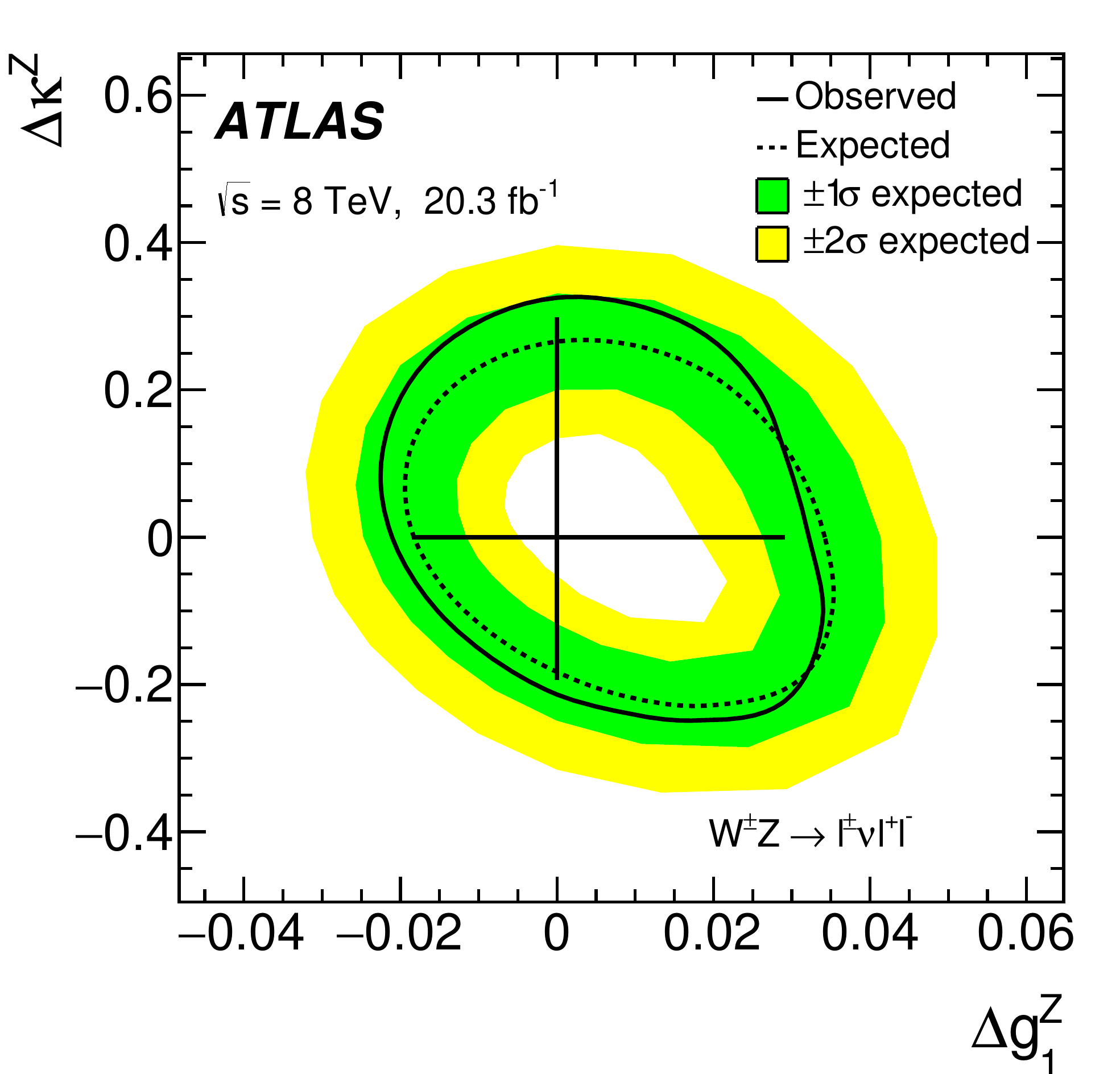}
 \includegraphics[width=0.33\textwidth]{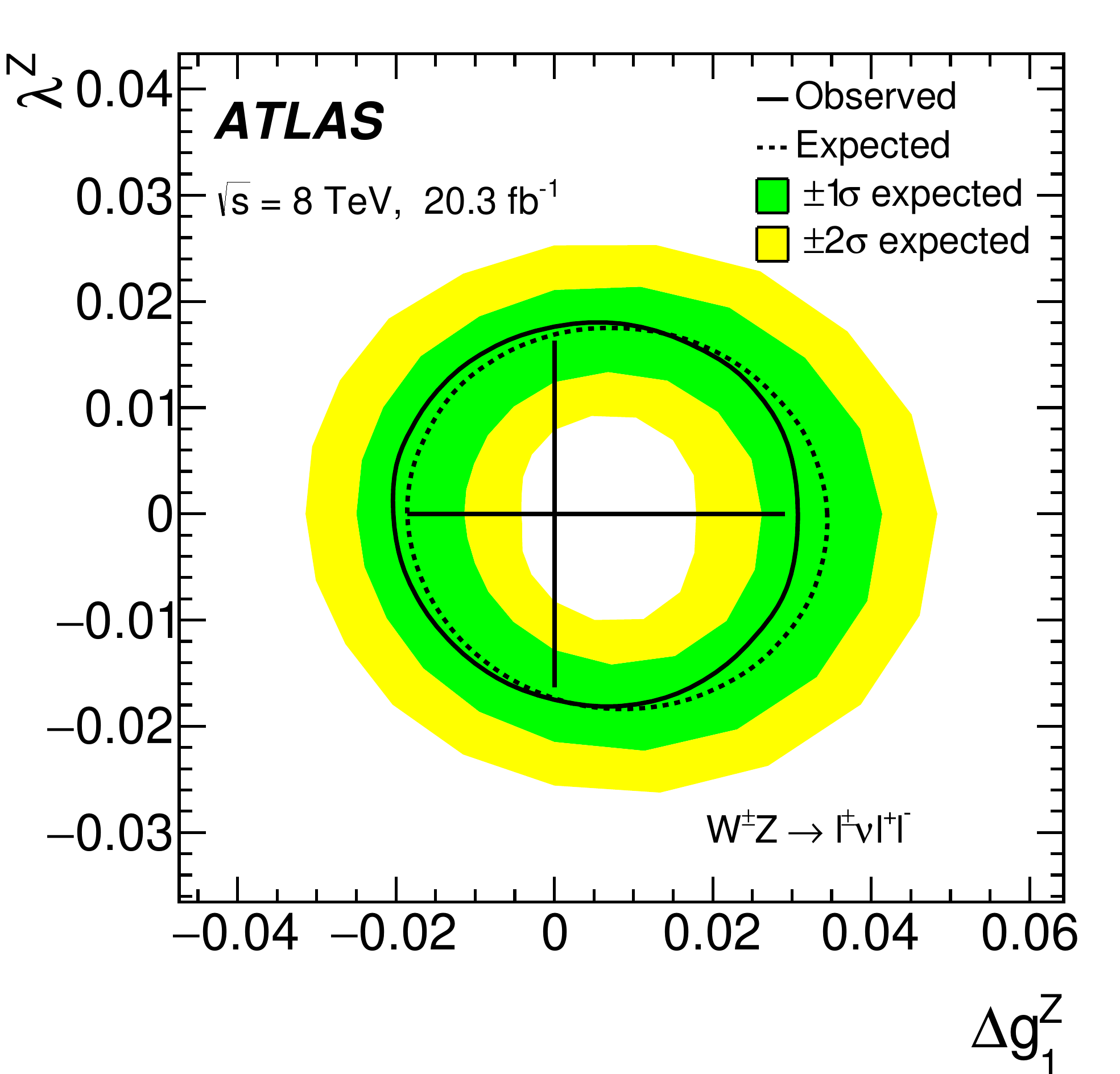}
 \includegraphics[width=0.33\textwidth]{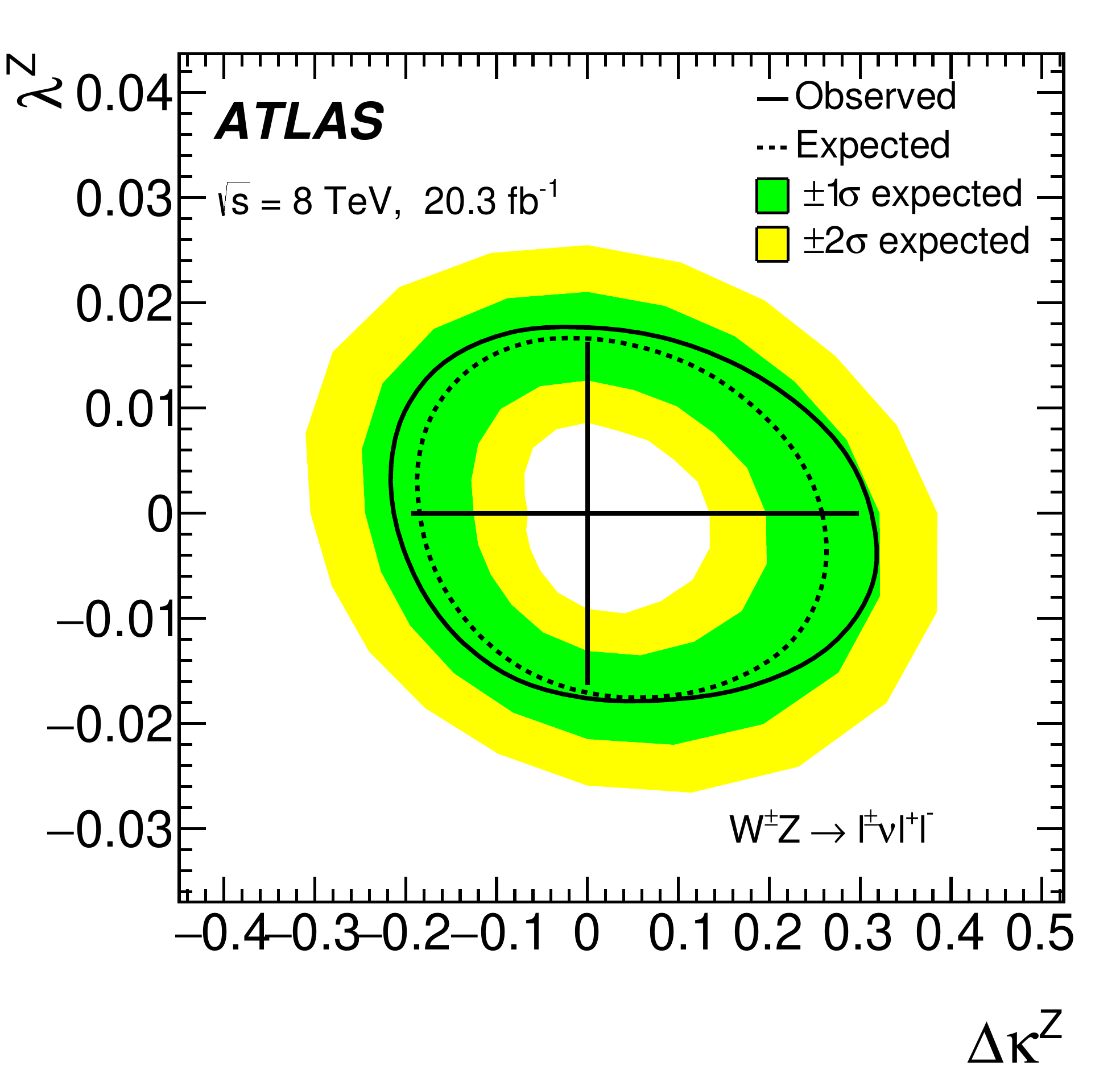}
 \end{center}
 \caption{
Expected and observed $95\%$ CL limit contours for $\Lambda_{\mathrm{co}} \rightarrow \infty$ in the planes ($\Delta \kappa^Z$, $\Delta g^Z_1$),
 ($\Delta g^Z_1$, $\lambda^Z$), and ($\Delta \kappa^Z$,  $\lambda^Z$).
The solid and dashed lines in the figures represent the observed and expected limits, respectively.
The regions outside the black contours are excluded.
The green and yellow bands correspond to the $1$~$\sigma$ and $2$~$\sigma$ uncertainty on the expected limit, respectively.
The vertical and horizontal lines represent the $95\%$ CL one-dimensional limits calculated separately.
}
 \label{fig:tgc_2D}
 \end{figure}

In Figure~\ref{fig:aTGC_summary_LHC_Tevatron} the present observed limits are compared
to limits previously obtained using $WZ$ events produced in $p\bar{p}$ collisions at the Tevatron~\cite{Abazov:2010qn,CDF:2012WZ} and by ATLAS  with $\sqrt{s} =7$~\TeV\ $pp$ collisions~\cite{Aad:2012twa}.
The new limits improve previous constraints by factors of $1.5$ to $2.5$ and are now the most stringent model-independent limits on $WWZ$ anomalous couplings.

\begin{figure}[!htbp]
\begin{center}
 \includegraphics[width=0.7\textwidth]{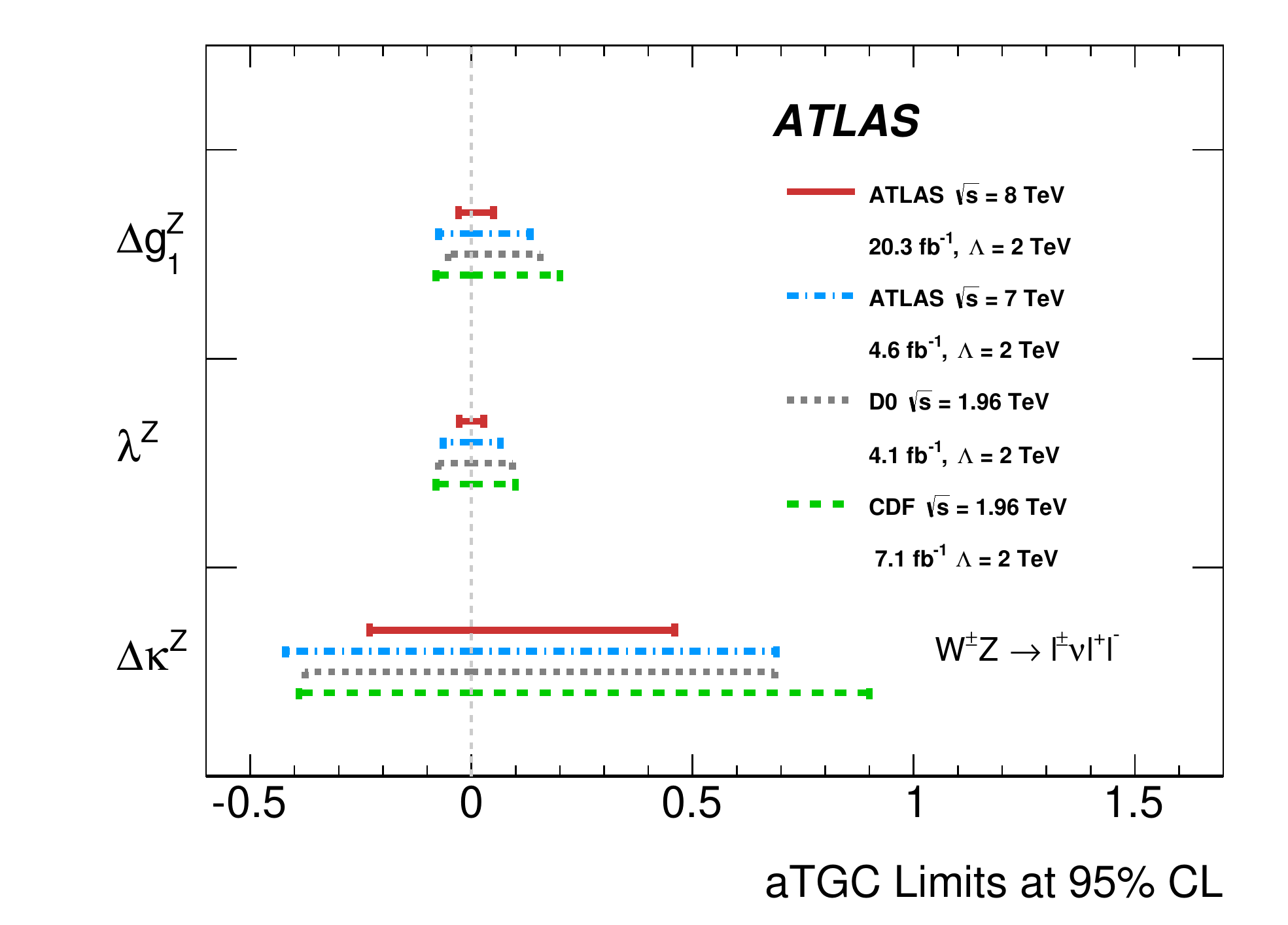}
 \end{center}
 \caption{Comparison of one-dimensional limits at $95\%$ CL on the anomalous coupling parameters using a cutoff scale of $\Lambda_{\mathrm{co}} = 2$~\TeV\ and obtained from the analysis of $W^\pm Z$ events by the ATLAS~\protect\cite{Aad:2012twa}, D0~\protect\cite{Abazov:2010qn}, and CDF experiments~\protect\cite{CDF:2012WZ}.
}
 \label{fig:aTGC_summary_LHC_Tevatron}
 \end{figure}

Table~\ref{table:EFTlimObs} presents the observed and expected one-dimensional intervals at $95\%$ CL on $c_{WWW}/\Lambda^2$, $c_{B}/\Lambda^2$, and $c_{W}/\Lambda^2$.
The sensitivity of the \wz final state to the EFT parameter $c_{B}/\Lambda^2$  is much weaker.

%%%%%%%%%%%%%%%%%%%%%%%%%%%%%%%%%%%%%%%%%%%%%%%%%%%%%%%%%%%%%%%%%%%%%%%%%%%%%%%%%%%%%%%%%%%%%%%%%%%%%%%%%%%%%%%%%%%%%%%%%%%
%%%%%%%%%%%%%%%%%%%%%%%%%%%%%%%%%%%%%%%%%%%%%%%%%%%%%%%%%%%%%%%%%%%%%%%%%%%%%%%%%%%%%%%%%%%%%%%%%%%%%%%%%%%%%%%%%%%%%%%%%%%

\begin{table}[!htbp]
	 \centering
	 \begin{tabular}{ccccc}
	 \hline
	 \noalign{\smallskip}
	 EFT coupling & Expected [TeV$^{-2}$]  & Observed [TeV$^{-2}$]  \\
	 	 \noalign{\smallskip}
         \hline
	\noalign{\smallskip}    
	$c_{W}/\Lambda^2$            &    [$-3.7$ ; $7.6$]   & [$-4.3$ ; $6.8$]  \\
	\noalign{\smallskip}
        $c_{B}/\Lambda^2$                                 &    [$-270$ ; $180$]  & [$-320$ ; $210$]  \\
	\noalign{\smallskip}
         $c_{WWW}/\Lambda^2$                               &    [$-3.9$ ; $3.8$]  & [$-3.9$ ; $4.0$]  \\
	\noalign{\smallskip}
     	 \hline
	 \end{tabular}
	 \caption{One-dimensional intervals at $95\%$ CL on the EFT parameters expected and observed in data.
}
\label{table:EFTlimObs}
\end{table}

%-------------------------------------------------------------------------------
\section{Anomalous quartic gauge Couplings}
\label{sec:aQGC}

To extract limits on aQGC, the EFT approach introduced in the previous section is used.
Several ways of parameterizing possible deviations with respect to the SM exist.
In this analysis, the choice is to express the deviation using two parameters $\alpha_4$ and $\alpha_5$ following existing notations~\cite{Appelquist:1980, Longhitano:1980, Longhitano:1981, Alboteanu:2008my}.
They are the coefficients of the two linearly independent
 dimension-four operators  contributing to the quartic gauge couplings beyond the SM.

 The \WHIZARD event generator
  is used to compute the ratio in the aQGC fiducial phase space, at particle level, of 
   the expected fiducial cross section for different values of $\alpha_4$ and $\alpha_5$, to the SM cross section.
\WHIZARD includes a unitarization scheme  in order to ensure the unitary of the scattering amplitude, which would be violated for values of the  quartic gauge couplings different from the SM value.

\begin{figure}[!htbp]
\begin{center}
 \includegraphics[width=0.4\textwidth]{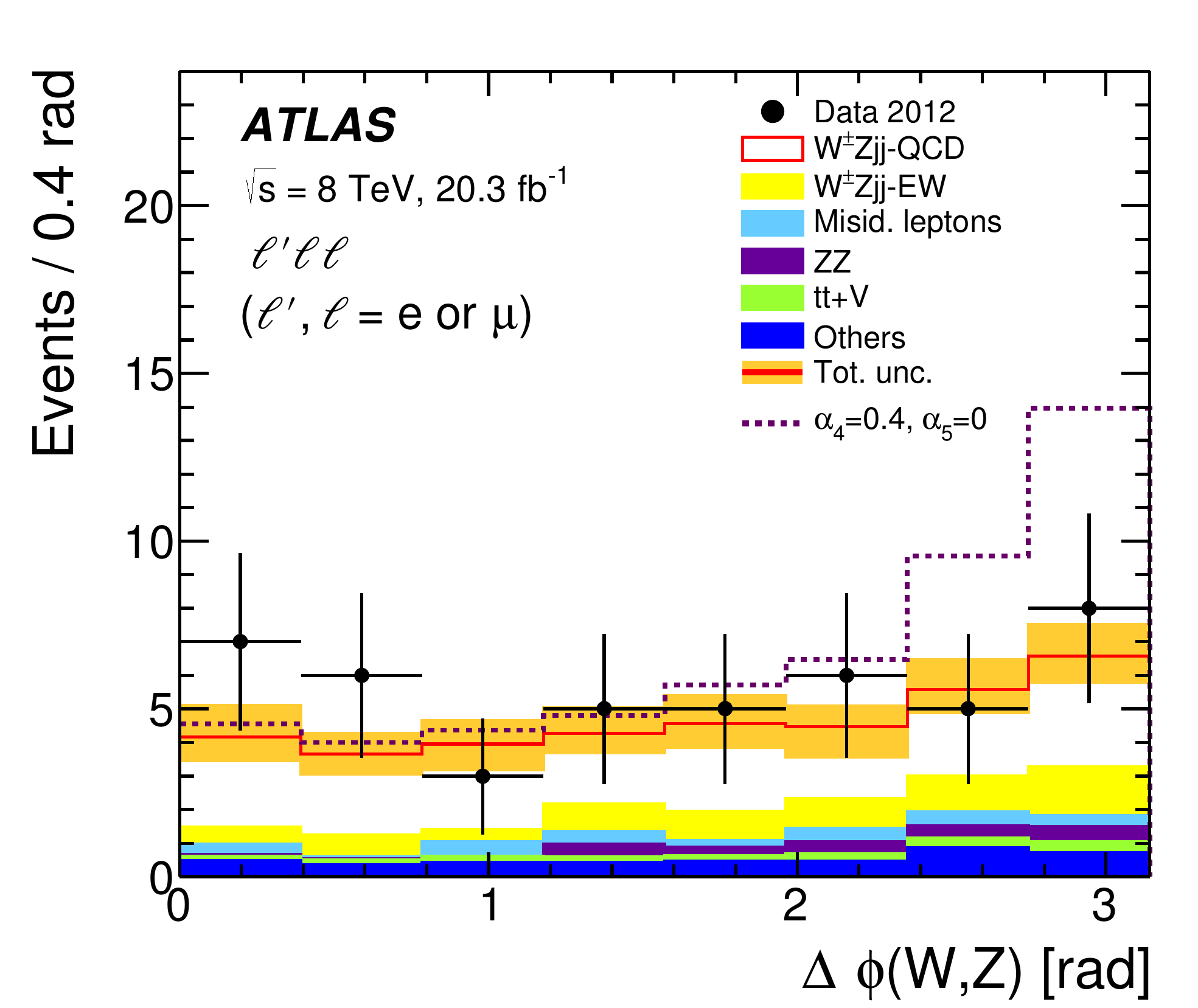}\put(-90,120){{(a)}}
  \includegraphics[width=0.4\textwidth]{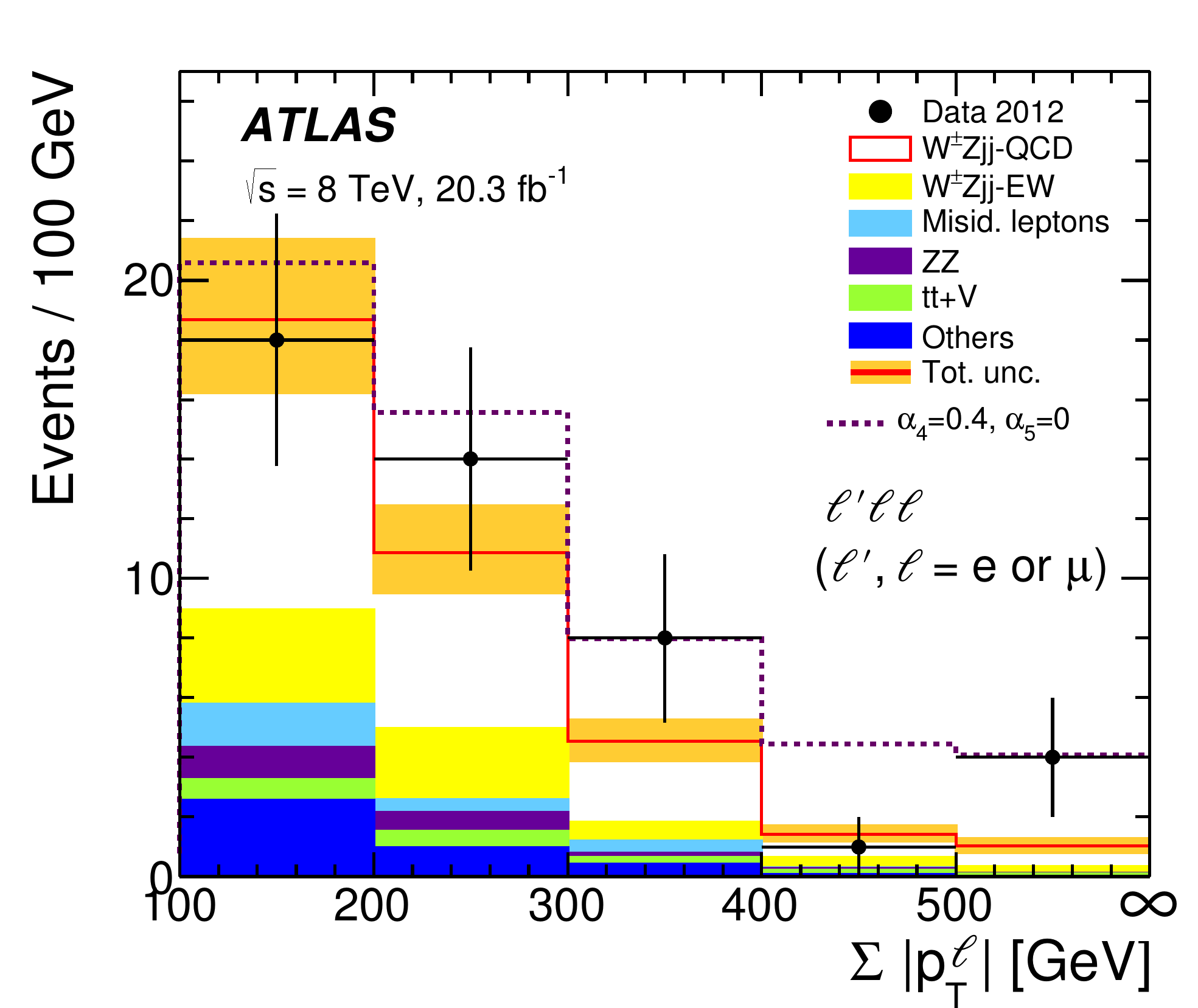}\put(-85,120){{(b)}}
\caption{Distribution of the difference in azimuthal angle between the reconstructed $W$ and $Z$ bosons, (a) $| \Delta \phi(W,Z)|$ and of the scalar sum of the transverse momenta of the three  charged leptons associated with the $W$ and $Z$ bosons, (b) $\sum |p_{\textrm T}^\ell |$, for the sum of all channels, in the VBS phase space.
All Monte Carlo expectations are scaled to the integrated luminosity of the data using the predicted MC cross sections of each sample.
The \SHERPA MC prediction is used for the SM $WZjj$-QCD and $WZjj$-EW predictions.
The open red histogram shows the total prediction and the shaded orange band its
estimated total uncertainty.
The prediction with nonzero values of one of the aQGC parameters is also represented by the dashed  line.
 }
 \label{fig:dist_aQGCs}
 \end{center}
 \end{figure}

These ratios are multiplied by the SM  fiducial cross section estimated with \SHERPA to obtain  the predicted fiducial
 cross sections as a function of $\alpha_4$ and $\alpha_5$.
The \SHERPA MC generator is used as  reference SM generator  for sake of consistency with the VBS
 cross-section limit measurement of Section~\ref{sec:VBS} and with a previous search for aQGC
 using $W^\pm W^\pm jj$ events~\cite{PhysRevLett.113.141803}.
 The expected fiducial cross sections include only the VBS part of the $WZjj$-EW process.

Distributions for the variables $| \Delta \phi(W,Z) |$ and $\sum |p_{\textrm T}^\ell |$ that are used to select events in the aQGC fiducial phase space are shown in Figure~\ref{fig:dist_aQGCs} for events passing the VBS phase space selection.
The change of the shape of these distributions when one of the aQGC parameters has a nonzero value is also shown.
After correcting for the selection efficiency, the measured fiducial cross section in the aQGC
 phase space is used to set limits on the aQGC.
The selection efficiency is estimated to be $\approx 70\%$ and found to be constant over the considered
 $\alpha_4$ and $\alpha_5$ values, within the MC statistical uncertainties.
Limits are obtained as for the aTGC limits of Section~\ref{sec:aTGC} from a profile likelihood  method that incorporates the
 systematic uncertainties.
The expected and observed two-dimensional limit contours at $95\%$ CL on $\alpha_4$ and $\alpha_5$ are shown in Figure~\ref{fig:aqgc_confidenceregions_2D}.
The present limit is compared to the expected limit obtained by the ATLAS Collaboration using $W^\pm W^\pm jj$ events~\cite{PhysRevLett.113.141803}.
This analysis of $W^\pm Z jj$ events probes a domain of the ($\alpha_4$, $\alpha_5$) parameter space
that could not excluded by the analysis of $W^\pm W^\pm jj$ events.

The limits on ($\alpha_4$, $\alpha_5$) coefficients used in \WHIZARD can be translated to limits on the ($f_{S,0}/\Lambda^4$, $f_{S,1}/\Lambda^4$) coefficients of the $\mathcal{O}_{S,0}$ and $\mathcal{O}_{S,1}$ operators of Ref.~\cite{Eboli:2006wa} using the following conversion for the $WWZZ$ vertex~\cite{Degrande:2013rea}:
\begin{equation}
\frac{f_{S,0(1)}}{\Lambda^4} = \alpha_{4(5)} \times \frac{16}{v^4} \, ,
\end{equation}
where $v = 246.22$~\GeV\ is the Higgs vacuum expectation value.
Assuming $\Lambda$ = 1~\TeV\ and that this conversion also holds for the $K$-matrix unitarization,
a value of $\alpha_{4(5)} = 0.5$ corresponds to $f_{S,0(1)} = 2177$ for $W^\pm Z jj$ events.

\begin{figure}[!htbp]
	\centering
		\includegraphics[width=0.8\textwidth]{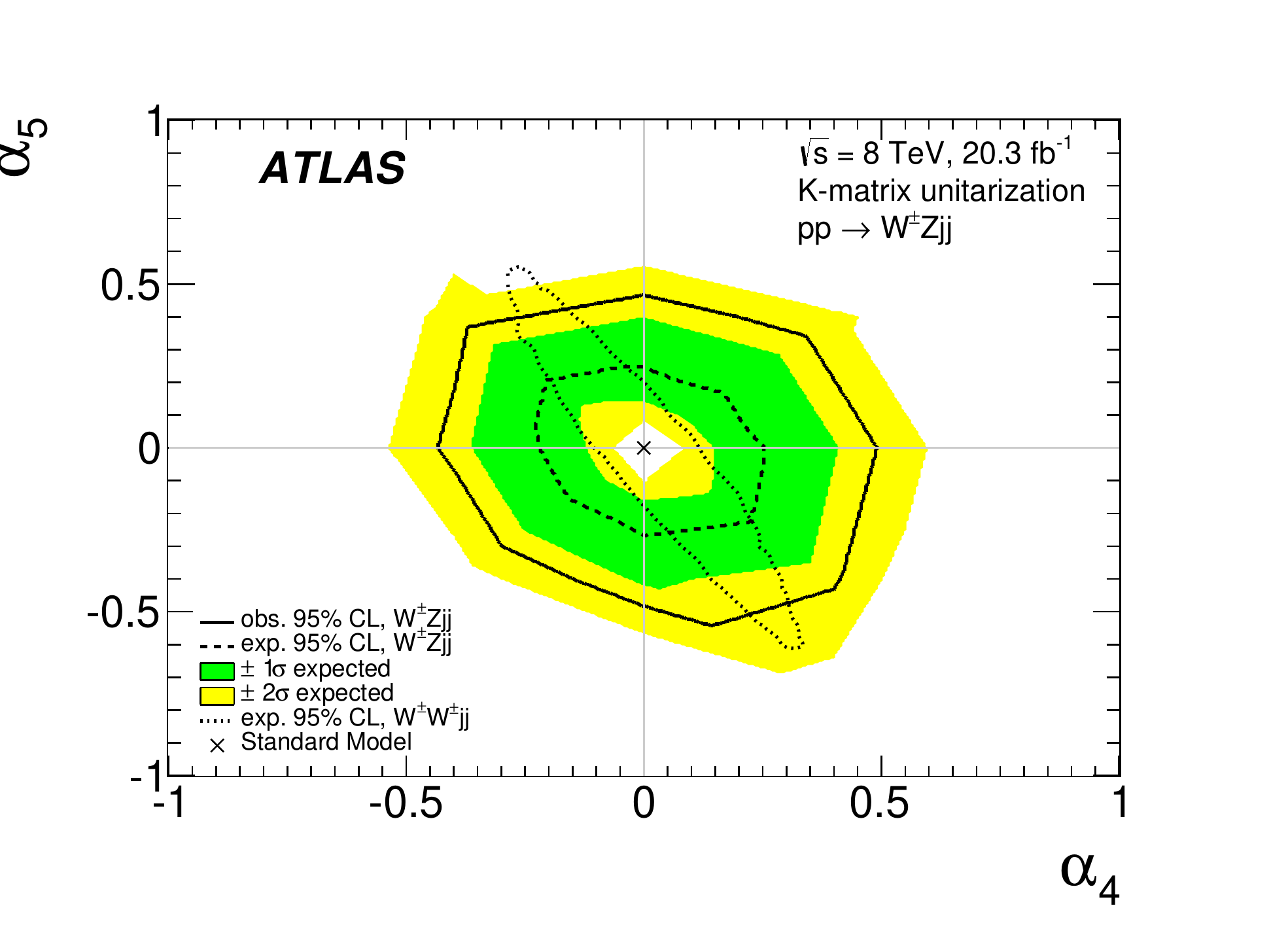}
	\caption{Expected and observed $95\%$ CL limit contours on aQGC parameters $\alpha_4$ and $\alpha_5$.
The solid and dashed lines in the figures represent the observed and expected limits, respectively.
The regions outside the black contours are excluded.
The green and yellow bands correspond to the $1$~$\sigma$ and $2$~$\sigma$ uncertainty on the expected limit, respectively.
The expected exclusion contour from an analysis by the ATLAS Collaboration using $W^\pm W^\pm jj$ events~\protect\cite{PhysRevLett.113.141803} is indicated by the dotted line.
}
\label{fig:aqgc_confidenceregions_2D}
\end{figure}

%-------------------------------------------------------------------------------
\section{Conclusion}
\label{sec:Conclusion}
Measurements of \wz\ production using $\sqrt{s} = 8$~\TeV\ $pp$ collisions at the LHC are presented.
The data were collected with the ATLAS detector and correspond to an integrated luminosity of $20.3$~fb$^{-1}$.
The measurements use leptonic decay modes of the gauge bosons to electrons or muons and are performed in a fiducial phase space approximating the detector acceptance.
The measured inclusive cross section in the fiducial region for one leptonic decay channel is
 $\sigma_{WZ^\pm \rightarrow \ell^{'} \nu\ \ell \ell} =  35.1~\pm$~0.9 (stat.)~$\pm~0.8$ (sys.)~$\pm~0.8$ (lumi.)~fb,
to be compared to next-to-leading-order Standard Model expectation of  $30.0  \pm 2.1$~fb.
With a total experimental relative uncertainty of $4.2\%$, a precision better than presently available from theoretical predictions is reached.
The measured cross section is found to be slightly larger than the NLO SM prediction.
A comparison to a prediction incorporating full NNLO QCD effects would therefore be of very interesting.

Furthermore, the \wz\ production cross section is measured as a function of each of several kinematic variables and compared to SM predictions of the \powhegpythia, \mcatnlo, and \SHERPA Monte Carlo event generators.

The ratio of the cross sections for $W^+Z$ and $W^-Z$ production is measured.
Integrated over the detector fiducial phase space it is $\sigma_{W^{+}Z \rightarrow \ell^{'} \nu\ \ell \ell}^{\textrm{fid.}}/\sigma_{W^{-}Z \rightarrow \ell^{'} \nu\ \ell \ell}^{\textrm{fid.}} = 1.51 \pm 0.11$
to be compared to the NLO SM expectation of $1.69 \pm 0.07$.
The differential evolution of this cross-section ratio as a function of each of a few kinematic variables is also measured and compared to available SM predictions.

The transverse mass spectrum of the \wz  system is used to search for anomalous triple gauge boson couplings and limits on $\Delta k^Z$, $\Delta g^Z_1$ and $\lambda^Z$ are derived.
With an improvement by a factor of about two compared to previously existing constraints, these are the most stringent model-independent limits on $WWZ$ anomalous couplings to date.
Results are also interpreted as limits on the $c_{W}/\Lambda^2$, $c_{B}/\Lambda^2$,
and $c_{WWW}/\Lambda^2$ coefficients of the EFT parameterization.

Finally, events with a $W$ and a $Z$ boson associated with two or more forward jets have been analyzed and an upper limit at $95\%$ CL on the \wz scattering cross section of $0.63$~fb for one leptonic decay channel has been established.
Limits on anomalous  quartic gauge boson couplings have also been extracted.

For \wz\ production, the measurements presented here are the most precise and complete to date and have the potential to further constrain existing Standard Model theoretical predictions, which are presently only available at next-to-leading order in QCD.
%-------------------------------------------------------------------------------
\section*{Acknowledgments}
%-------------------------------------------------------------------------------

% Acknowledgements for papers with collision data
% Version 7-Feb-2016

% Standard acknowledgements start here
%----------------------------------------------
We thank CERN for the very successful operation of the LHC, as well as the
support staff from our institutions without whom ATLAS could not be
operated efficiently.

We acknowledge the support of ANPCyT, Argentina; YerPhI, Armenia; ARC, Australia; BMWFW and FWF, Austria; ANAS, Azerbaijan; SSTC, Belarus; CNPq and FAPESP, Brazil; NSERC, NRC and CFI, Canada; CERN; CONICYT, Chile; CAS, MOST and NSFC, China; COLCIENCIAS, Colombia; MSMT CR, MPO CR and VSC CR, Czech Republic; DNRF and DNSRC, Denmark; IN2P3-CNRS, CEA-DSM/IRFU, France; GNSF, Georgia; BMBF, HGF, and MPG, Germany; GSRT, Greece; RGC, Hong Kong SAR, China; ISF, I-CORE and Benoziyo Center, Israel; INFN, Italy; MEXT and JSPS, Japan; CNRST, Morocco; FOM and NWO, Netherlands; RCN, Norway; MNiSW and NCN, Poland; FCT, Portugal; MNE/IFA, Romania; MES of Russia and NRC KI, Russian Federation; JINR; MESTD, Serbia; MSSR, Slovakia; ARRS and MIZ\v{S}, Slovenia; DST/NRF, South Africa; MINECO, Spain; SRC and Wallenberg Foundation, Sweden; SERI, SNSF and Cantons of Bern and Geneva, Switzerland; MOST, Taiwan; TAEK, Turkey; STFC, United Kingdom; DOE and NSF, United States of America. In addition, individual groups and members have received support from BCKDF, the Canada Council, CANARIE, CRC, Compute Canada, FQRNT, and the Ontario Innovation Trust, Canada; EPLANET, ERC, FP7, Horizon 2020 and Marie Sk{\l}odowska-Curie Actions, European Union; Investissements d'Avenir Labex and Idex, ANR, R{\'e}gion Auvergne and Fondation Partager le Savoir, France; DFG and AvH Foundation, Germany; Herakleitos, Thales and Aristeia programmes co-financed by EU-ESF and the Greek NSRF; BSF, GIF and Minerva, Israel; BRF, Norway; the Royal Society and Leverhulme Trust, United Kingdom.

The crucial computing support from all WLCG partners is acknowledged
gratefully, in particular from CERN and the ATLAS Tier-1 facilities at
TRIUMF (Canada), NDGF (Denmark, Norway, Sweden), CC-IN2P3 (France),
KIT/GridKA (Germany), INFN-CNAF (Italy), NL-T1 (Netherlands), PIC (Spain),
ASGC (Taiwan), RAL (UK) and BNL (USA) and in the Tier-2 facilities
worldwide.
%----------------------------------------------

%-------------------------------------------------------------------------------
% \clearpage
% \appendix
% \part*{Appendix}
% \addcontentsline{toc}{part}{Appendix}
%-------------------------------------------------------------------------------

% In a paper, an appendix is used for technical details that would otherwise disturb the flow of the paper.
% Such an appendix should be printed before the Bibliography.

%-------------------------------------------------------------------------------
% If you use biblatex and either biber or bibtex to process the bibliography
% just say \printbibliography here
\printbibliography
% If you want to use the traditional BibTeX you need to use the syntax below.
%\bibliographystyle{bibtex/bst/atlasBibStyleWoTitle}
%\bibliography{mydocument,bibtex/bib/atlas-paper}
%-------------------------------------------------------------------------------

\newpage 
% ATLAS Collaboration author list
% Data extracted on 23-Jan-2016 for paper reference STDM-2014-02
% \documentclass[11pt]{article}
% \usepackage{a4wide}\begin{document}
\begin{flushleft}
{\Large The ATLAS Collaboration}

\bigskip

G.~Aad$^\textrm{\scriptsize 86}$,
B.~Abbott$^\textrm{\scriptsize 113}$,
J.~Abdallah$^\textrm{\scriptsize 151}$,
O.~Abdinov$^\textrm{\scriptsize 11}$,
B.~Abeloos$^\textrm{\scriptsize 117}$,
R.~Aben$^\textrm{\scriptsize 107}$,
M.~Abolins$^\textrm{\scriptsize 91}$,
O.S.~AbouZeid$^\textrm{\scriptsize 137}$,
H.~Abramowicz$^\textrm{\scriptsize 153}$,
H.~Abreu$^\textrm{\scriptsize 152}$
% ...
\bigskip
\\
% ...
$^{11}$ Institute of Physics, Azerbaijan Academy of Sciences, Baku, Azerbaijan\\
% ...
$^{86}$ CPPM, Aix-Marseille Universit{\'e} and CNRS/IN2P3, Marseille, France\\
% ...
$^{91}$ Department of Physics and Astronomy, Michigan State University, East Lansing MI, United States of America\\
% ...
$^{107}$ Nikhef National Institute for Subatomic Physics and University of Amsterdam, Amsterdam, Netherlands\\
% ...
$^{113}$ Homer L. Dodge Department of Physics and Astronomy, University of Oklahoma, Norman OK, United States of America\\
% ...
$^{117}$ LAL, Univ. Paris-Sud, CNRS/IN2P3, Universit{\'e} Paris-Saclay, Orsay, France\\
% ...
$^{137}$ Santa Cruz Institute for Particle Physics, University of California Santa Cruz, Santa Cruz CA, United States of America\\
% ...
$^{151}$ Institute of Physics, Academia Sinica, Taipei, Taiwan\\
$^{152}$ Department of Physics, Technion: Israel Institute of Technology, Haifa, Israel\\
$^{153}$ Raymond and Beverly Sackler School of Physics and Astronomy, Tel Aviv University, Tel Aviv, Israel\\
% ...
\end{flushleft}

%\end{document}
% Created with xml2latex.py

\end{document}